\documentclass[fleqn,usenatbib]{mnras}

\usepackage[T1]{fontenc}
\usepackage{ae,aecompl}

\usepackage{graphicx}	
\usepackage{amssymb}	

\newcommand{\hyperfit}{{\sc HyperFit}}	
\newcommand{\profound}{{\sc ProFound}}	

\newcommand{\prospect}{{\sc ProSpect}}
\newcommand{\shark}{{\sc Shark}}
\newcommand{\stingray}{{\sc Stingray}}
\newcommand{\viperfish}{{\sc Viperfish}}

\newcommand{\lambdar}{{\sc LAMBDAR}}

\newcommand{\magphys}{{\sc MagPhys}}
\newcommand{\R}{{\sc R}}

\newcommand{\msol}{M$_{\odot}$}

\title[ProSpect: Generating Spectral Energy Distributions]{ProSpect: Generating Spectral Energy Distributions with Complex Star Formation and Metallicity Histories}

\author[A.~S.~G. Robotham et al.]{
A.~S.~G. Robotham,$^{1,2}$\thanks{E-mail: aaron.robotham@uwa.edu.au}
S. Bellstedt,$^{1}$
C. del P. Lagos,$^{1,2}$
J.~E. Thorne,$^{1}$
L.~J. Davies,$^{1}$
S.~P. Driver,$^{1}$
\newauthor
M. Bravo$^{1}$
\\\\
$^{1}$ICRAR, M468, University of Western Australia, Crawley, WA 6009, Australia\\
$^{2}$ARC Centre of Excellence for Astrophysics in Three Dimensions (ASTRO3D)\\
}

\date{Accepted XXX. Received YYY; in original form ZZZ}

\pubyear{2019}

\begin{document}
\label{firstpage}
\pagerange{\pageref{firstpage}--\pageref{lastpage}}
\maketitle

\begin{abstract}
{ We introduce \prospect{}, a generative galaxy spectral energy distribution (SED) package that encapsulates the best practices for SED methodologies in a number of astrophysical domains. \prospect{} comes with two popular families of stellar population libraries (BC03 and EMILES), and a large variety of methods to construct star formation and metallicity histories. It models dust through the use of a Charlot \& Fall attenuation model, with re-emission using Dale far-infrared templates. It also has the ability to model AGN through the inclusion of a simple AGN and hot torus model. Finally, it makes use of MAPPINGS-III photoionisation tables to produce line emission features. We test the generative and inversion utility of \prospect{} through application to the \shark{} galaxy formation semi-analytic code, and informed by these results produce fits to the final ultraviolet to far-infrared photometric catalogues produces by the Galaxy and Mass Assembly Survey (GAMA). As part of the testing of \prospect{}, we also produce a range of simple photometric stellar mass approximations covering a range of filters for both observed frame and rest frame photometry.}
\end{abstract}





\section{Introduction}

A huge amount of effort has been invested in creating theoretical spectral energy distributions (SED) for stars and galaxies over the last 50 years \citep{tins68, bc03, emiles}. To fully capture the complexities of creating a galaxy SED would { require} solving a number of ongoing problems in astronomy, e.g.\ the evolution or otherwise of the initial mass function \citep[IMF; see][]{krou01}; the full and accurate mapping of stellar isochrones over a fine resolution and high dynamic range of metallicities \citep{bert94, gira00}; the proper treatment of stellar binary evolution \citep{eldr09}; the accurate production of stellar atmospheres over a suitably dense grid of temperatures and metallicities \citep{kuru92, pick98, lebo03, ivan19};  the treatment of dust for a broad range of geometries and galaxies \citep{cf00, tray20}; and the correct parameterisation for galaxy star formation history \citep{conr13, mitc13}.

Regardless of these numerous limitations, in practice we have witnessed a huge breadth of utility in creating and using galaxy SEDs. In particular it has become routine to use physically motivated SED models to infer properties of galaxies, e.g.: stellar mass, recent star formation rate and dust masses and luminosities \citep{dacu08, noll09}. Given the almost limitless complexity that could be applied to the problem of SED fitting, there is a huge scope for a range of different approaches that cover the restrictive (in terms of modelling assumptions) and computationally cheap \citep{bell01, zibe09, tayl11}, through to the highly flexible and computationally expensive \citep{fioc19}.

This work is interested in combining the best methods, codes, tables, philosophies, practices and practicalities to produce software (\prospect) that whilst known to be imperfect, is at least useful, simple to use, and rapid to run. The original use-case for \prospect{} was to produce useful and self-consistent FUV--FIR (0.2--1,000 $\mu$m) outputs for current and future semi-analytic models \citep{lago18, lago19}. This naturally required it to be written in a generative and functional manner. However, due to this generative structure it is deliberately simple to execute \prospect{} in a Bayesian mode that achieves model fits to observational data by varying combinations of parameters. This allows for the inversion of interesting galaxy properties given broad-band FUV--FIR data, i.e.\ the sort of data being routinely collected by large modern surveys \citep[e.g.][]{lisk15}. 

{ This paper serves as a broad introduction to the new \R{} software package \prospect. Section 2 discusses the conceptual implementation of the generative SED mode of \prospect. Section 3 discusses the main use cases for \prospect, with a discussion of its use for producing SEDs out of semi-analytic models (SAMs) and for inverting observations to extract model parameters. Section 4 explores using \prospect{} to extract stellar masses via approximate means. Section 5 presents the initial observational application of \prospect{} using data from the Galaxy Mass And Assembly survey \citep[GAMA;][]{lisk15}. The exploitation of the novel characteristics of galaxy properties that can be extracted with \prospect{} will be explored in detail in future papers (Bellstedt et al. in prep.; Thorne et al. in prep.).}

To be consistent with the default SAMs used in this work, \prospect{} by default assumes a Planck 2015 cosmology \citep{planck15}. Therefore, in this paper we assume an H$_0$ = 67.8 (kms/s)/Mpc, $\Omega_M$ = 0.308 and $\Omega_\Lambda = 0.692$ Universe. The cosmology assumed is only relevant in aspects of SED generation, or fitting as a function of redshift. Intrinsic properties are unaffected by the choice of cosmology, so the majority of the results presented in this paper are insensitive to it.


\section{Methods}
\label{sec:methods}

\prospect{} aims to generate reasonable quality (sub 0.01 mag accuracy) SEDs that can be reliably used to estimate the broad band photometric properties of galaxies from the FUV--FIR that have known star formation and gas metallicity histories. It is written in the free and open source \R{} language under a LGPL-3 license and is available to install and use immediately from GitHub\footnote{\url{https://github.com/asgr/ProSpect}} and via an interactive web tool\footnote{\url{http://prospect.icrar.org}}.

{This paper does not serve as a full pedagogical guide, so potential users are encouraged to read the full 77-page manual that comes with the package. Every function in the package also comes with one or more working examples in the help files to assist users in gaining confidence. We include one short example of using \prospect{} in a generative mode in Appendix \ref{sec:simple_example}. Longer form examples (including parameter fitting) and more complicated use cases are available as online vignettes\footnote{\url{https://rpubs.com/asgr}}.}

In basic terms it uses a similar strategy to published codes \citep[e.g.\ MAGPHYS and CIGALE;][]{dacu08, noll09, boqu19} to create radiation from an episode of star formation, attenuate it and re-emit it at longer wavelengths. Once this intrinsic spectrum has been made, we place it at a target redshift and pass it through a set of desired filters. When doing this we assume our fiducial galaxy has the most recent period of star formation (< 10 Myrs) embedded in birth clouds, and outside of this we have a screen-like inter-stellar medium. We also allow for the presence of an accretion disk AGN that may have its light first attenuated by a dusty torus, and then further attenuated by a screen-like ISM.

\begin{figure*}
\begin{center}
\includegraphics[width=18cm]{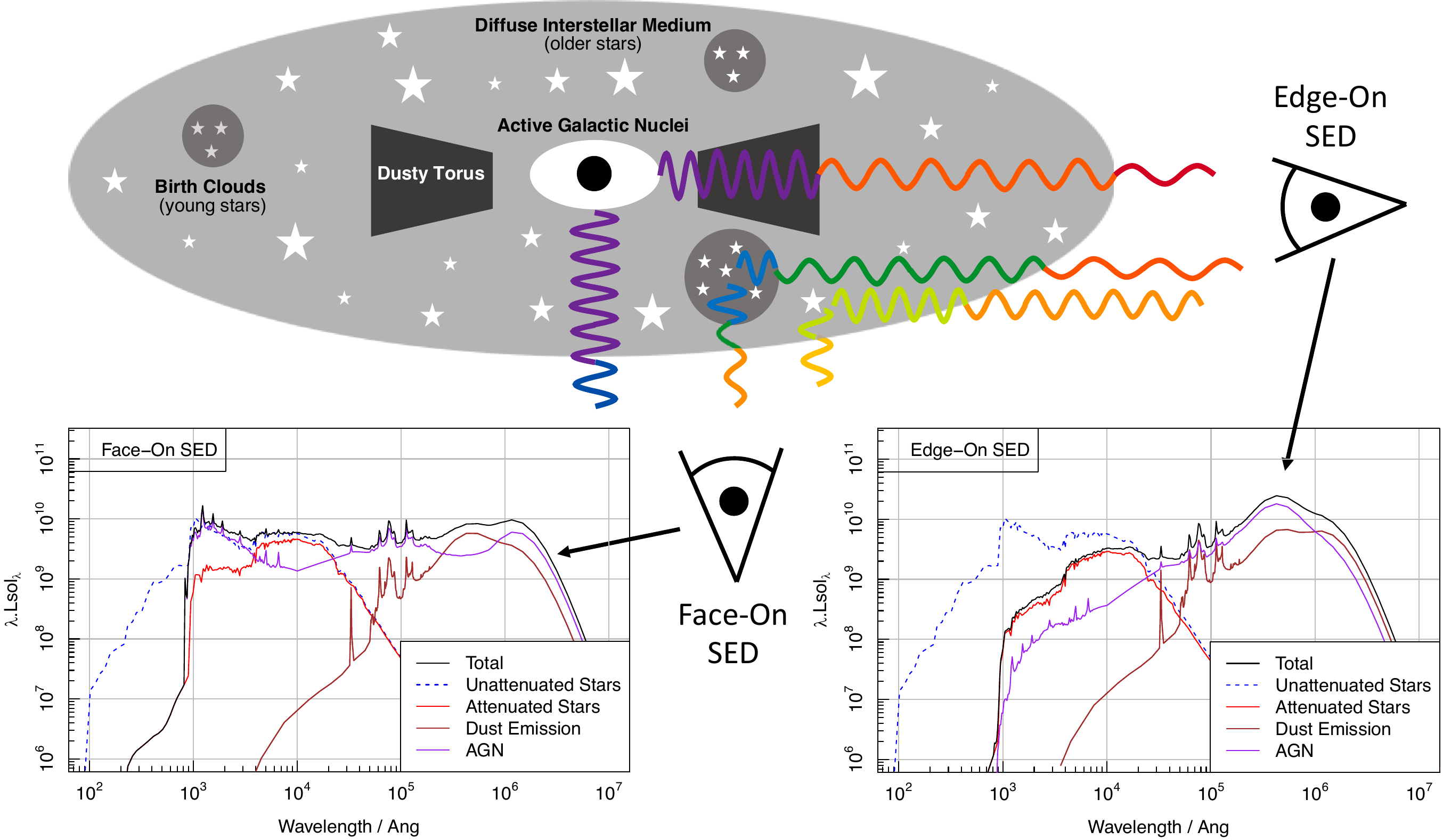}
\caption{Schematic view of the impact optical depth has on the observed SED. Here a galaxy with a constant star formation history (1 \msol/yr for 13 Gyrs) and a moderately powerful AGN ($10^{44}$ erg/s) is viewed face-on (light rays coming out of the bottom of the galaxy schematic) and edge-on (light rays coming out the right of the schematic). The respective SEDs are shown in the bottom-left and bottom-right. The general effect is that the edge on view is significantly more attenuated (a lot less blue flux is observed in the output), and there is significantly more flux on the far-infrared that is re-emitted by dust. In this example the face-on SED is dominated by AGN light at blue and cold far-infrared wavelengths, but in the edge-on view the AGN instead dominates at hotter dust temperatures, having been strongly attenuated by its own dusty torus.}
\label{fig:ProSpect_schematic}
\end{center}
\end{figure*}

A simplified schematic of how we produce and attenuate these different components is shown in Figure \ref{fig:ProSpect_schematic}. AGN and young stars can be attenuated by their own dust torus or birth cloud respectively, and this light can be further attenuated by ISM dust. Older stars in comparison are only attenuated by ISM dust. In all cases it is possible to adjust the optical depth of the different dust components, allowing a large amount of flexibility in SED generation even for the same intrinsic star formation history, e.g.\ the difference between the highly attenuated edge-on view of a galaxy and the barely attenuated face-on view shown in the bottom SED panels of Figure \ref{fig:ProSpect_schematic}. { Note that the birth cloud and ISM dust are in practice treated as screen-like with a mean optical depth; i.e.\ internally there is no attempt to model the full geometry of galaxy components \citep[such more complicated galaxy models are discussed in][]{fioc19}.}

In this initial incarnation of \prospect{} we have mostly focussed on generating broad band photometry over the range available to the GAMA survey \citep[FUV--FIR;][]{driv16}, where this has also been the SED focus of the recently developed semi-analytic galaxy formation code \shark{} \citep{lago18, lago19}. \shark{} produces star formation and star forming gas metallicity histories (SFH and ZH respectively from here) for bulge and disk components separately, with the bulge further sub-divided into merger driven and disk instability formation. Combining this detailed component-wise modelling with physically motivated SED generation (i.e.\ \prospect) offers a powerful predictive tool.

Extensions beyond the FUV--FIR range are planned for the future (e.g.\ adding X-ray and radio continuum modelling), but this paper will focus exclusively on the aforementioned FUV--FIR range. Since the design goal of \prospect{} is to generate reasonably accurate broad band SEDs given a target star formation and gas metallicity history, a number of pragmatic design choices were made early on in its development. The most significant of these, in terms of having a likely impact on the possible output SEDs, are:

\begin{itemize}

\item a choice of \citet{bc03} (BC03 from here) or \citet{emiles} (EMILES from here) simple stellar population (SSP) libraries, but in both cases a fixed \citet{chab03} initial mass function (C03-IMF from here);

\item the option to add an AGN component to the stellar population of arbitrary luminosity;

\item a free form variant of the \citet{cf00} dust attenuation prescription (CF00 from here) for light that operates separately on birth clouds, the inter-stellar medium and the AGN dust torus (optionally);

\item forced energy balance when re-emitting attenuated stellar light;

\item re-emission of attenuated light using the \citet{dale14} library of far-infrared templates (D14 from here) that operates separately on birth clouds, the inter-stellar medium and the AGN dust torus (optionally);

\item the option to define or derive both the metallicity and star formation history, with reasonable functional forms included by default.

\end{itemize}

In \prospect{} a composite stellar population is constructed as the weighted sum of many simple (or single-aged) stellar populations (SSPs), where the weights are informed by the star formation history and/or metallicity. In the standard mode of operation the following assumptions are made for weighting the SSP models:

\begin{itemize}

\item the star formation rate is constant in the time between adjacent SSPs;

\item at the age of a given SSP the target metallicity is approximated by logarithmically weighting the coarse available grid of SSP metallicities (i.e. a target $Z=0.1$ would be achieved by the average flux of a $Z=0.05$ and $Z=0.2$ metallicity SSPs if these are the closest available).

\end{itemize}

{ If the SFH is changing rapidly between adjacent SSPs the age weighting assumed in the above default mode will lose some accuracy. It is possible in this case to compute a more accurate weighting integral across the coarse SSP age bins, however this adds significant computational cost (factors of a few). In practice, the simplistic behaviour assumed above is accurate to less than 0.01 mag even when processing highly variable SFHs (as are produced by the SAM software \shark, seen later in this work). For this reasons we believe it is reasonable to run \prospect{} using the default weighting mode in almost all standard applications.} In the regime where the star formation history is somewhat smooth and no expensive SFH integration is required, the series of operations made in \prospect{} boils down to a series of large, but computationally highly efficient and parallelisable, matrix operations.

To achieve computationally efficient SED generation we have embedded the above SSPs (BC03 and EMILES) and dust libraries (D14) into the \prospect{} package and formatted them into a consistent binary structure, which means in its most efficient mode they are already in wired memory for each generation of SED creation. This pre-loaded library of SSPs is then multiplied through by the appropriate age weights required to simulate a target SFH, with further weighting calculated to interpolate between the available discrete metallicity isochrones.

{ To further increase processing speed, the SSP spectra that are used to compute the broad band photometry can optionally be sparse sampled; i.e.\ only use every $N^{th}$ spectral data point and assume the spectra behaves in a linear manner between these sparsely sampled points. \prospect{} uses a default sparsity factor of 5; beyond this the photometric magnitudes start to become appreciably changed by more than 0.01 mag.} On top of this, the per band filter responses can be pre-compiled into interpolation functions that are then used to process the spectra. This saves about 30\% of the computation time required when running \prospect{} in its faster mode (where all necessary data are explicitly passed in rather than being dynamically loaded).

To aid usability, it is possible to use \prospect{} in a more interactive mode, where the various required libraries are dynamically loaded as necessary rather than the user supplying them explicitly. This reduces the code required to generate a quick SED to few statements, assuming the user is happy with the default options for the SFH (constant over cosmic time), and dust attenuation and re-emission (reasonable typical values). The difference between running in the most computationally efficient mode (with more code required to pass data into functions) versus the simplest (in terms of code simplicity) is vast, with run times varying by nearly a factor of 100 between only processing the Sloan filters \citep{fuku96} in the most efficient manner and processing all of the default filters (97 in total) with dynamically loaded data (roughly 5 ms versus 0.5 s on a modern MacBook Pro).

{ \prospect{} comes pre-loaded with a useful and easily accessible set of 97 filter response curves that cover the Galex FUV through to mm wavelengths. On top of this, 347 EAZY filters \citep{bram08} are included in a loadable \R list structure that requires the user to identify the bands desired. It is also possible to process user definable filter responses that are not included in the base package, making the photometric outputs highly generic and somewhat future proof (at least in regards to adding filters for new telescopes). All filters are expected by default to be in the photon counting standard common to optical and NIR telescopes. \prospect{} also has the functionality to use filters specified via energy counting transmission (more common for, e.g.\, FIR telescopes), however it is not possible to mix the two paradigms within a particular SED generation. For this reason, the recommendation is that all filters are provided in a photon counting form (where it is quite trivial for a user to convert between the two modes) since this matches the format of the filters that are already provided.}

{ Compared to alternative SED codes, \prospect{} is quite unique in terms of how easy (for the user) and rapid (for the computer) it is to generate a galaxy spectrum (or a large number of photometric magnitudes) for a given star formation and metallicity history. There is a huge amount of flexibility in how the star formation and metallicity histories can be described, where a number of useful forms have been explored by the authors and are discussed in the following sections of this paper.}

{ Regarding the use of a C03-IMF, it is usually appropriate to convert generative fluxes to a different IMF by a uniform scaling factor. In detail this will not be correct since top or bottom heavy IMFs will produce more flux at the bluer or redder ends of the spectrum respectively. However, the assumption of a non-variable C03-IMF does not appear to be the limiting factor in photometric accuracy with current SED software, at least for broad band energy production.}

\subsection{Star Formation Histories}
\label{sec:SFH}

{ The majority of current SED codes give access to a limited set of unrealistic and/or unphysical star formation histories. In comparison \prospect{} is almost entirely flexible in how it can describe star formation histories, with users also able to specify new forms that are not included in the base package.} These can be either discrete outputs (of e.g.\ a semi analytic galaxy formation model, or of individual particles in a hydrodynamic simulation as discussed in Harbourne et al., in press) or functional forms with arbitrary complexity. The assumption is that when used in a purely generative mode to produce SEDs for simulation outputs the SFH will tend to be in the former state of discrete values of star formation at various age intervals, but when being used as part of an inversion process to infer the SFH of a particular galaxy in a Bayesian framework it will be the latter functional form.

To aid the development and exploration of different SFHs when fitting observational data, \prospect{} comes with a useful variety built-in and ready to use. The functional forms included in the package cover a large range of reasonable parameterisations with differing degrees of flexibility and restrictiveness. The user is encouraged to adapt these to suit their own purposes, but in practice they cover a diverse range of physical SFH classes (as we will see in detail later in this paper).

{ Below are a number of useful SFH parameterisations with accompanying figures (\ref{fig:SFH_massfunc_const}--\ref{fig:SFH_massfunc_snorm_trunc}) demonstrating the fitting behaviour when assuming linear priors on all of the specified parameters. The presence of over-densities in the greyscale lines demonstrates potential biases when using a given functional form for fitting purposes, and the coloured lines give a clearer view of the kind of SFH diversity possible when using a specific function.}

In brief, \prospect{} includes the following mass formation functions with the stated default argument values (note all mentions of star formation rates are in units of \msol{} / yr, and variable ages are in Gyrs, but the first argument {\tt age} is in years to be consistent with the stellar libraries). { The full mathematical description of each form is provided in the documentation included with the \prospect{} package.}

\begin{itemize}

\item {\tt massfunc\_const(age, mSFR = 1, magemax = 13.8)} has control parameters of constant star formation rate (mSFR) and the maximum age of star formation (magemax), returning the star formation rate at the specified ages (age). See Figure \ref{fig:SFH_massfunc_const} to see the distribution of random parameter samples, reflecting the natural coverage of the possible SFHs.

\begin{figure}
\begin{center}
\includegraphics[width=9cm]{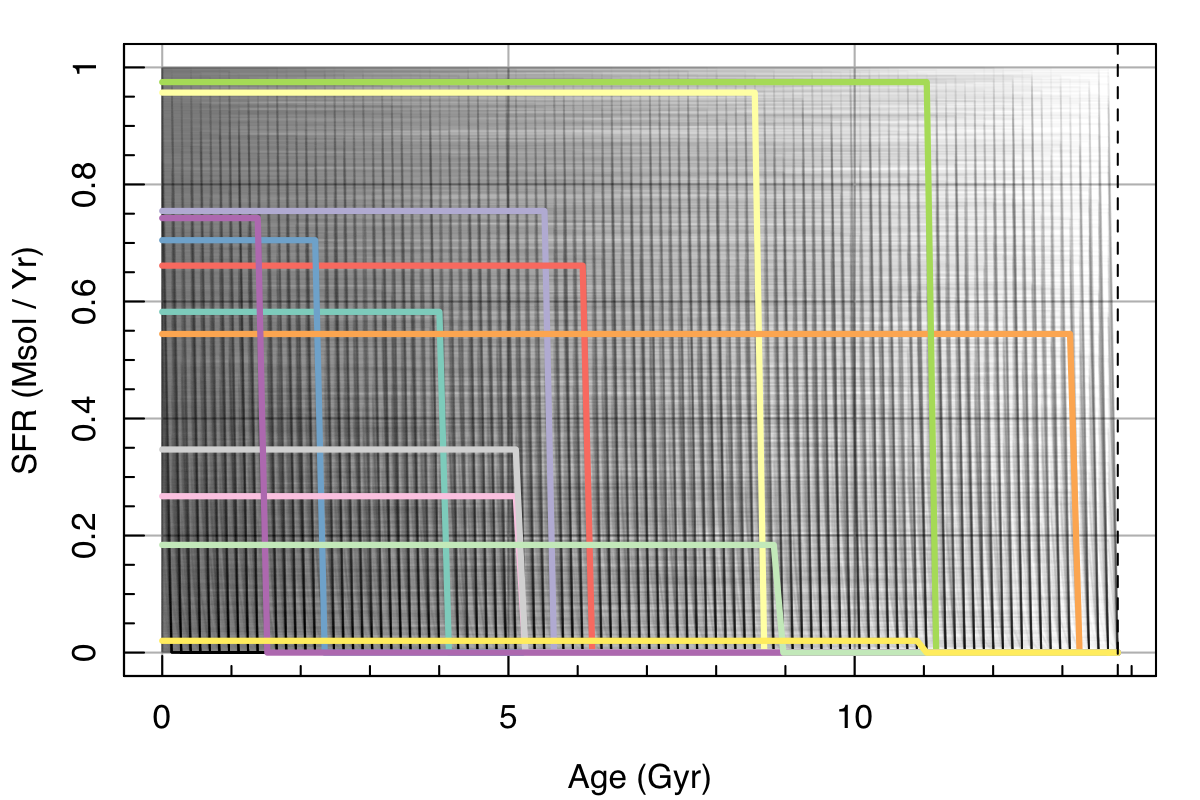}
\caption{{\tt massfunc\_const} SFHs from 10,000 samples of the mSFR [0,1] and magemax [0,13.8] parameters, with 12 reference SFHs in bold colour.}
\label{fig:SFH_massfunc_const}
\end{center}
\end{figure}

\item {\tt massfunc\_p2(age, m1 = 1, m2 = m1, m1age = 0, m2age = magemax, magemax = 13.8)} a linear interpolation model that has control parameters for star formation rate at two nodes (m1 / m2), with ages (m1age / m2age), and the maximum age of star formation (magemax), returning the star formation rate at the specified ages (age). Figure \ref{fig:SFH_massfunc_p2} shows the distribution of random parameter samples, reflecting the natural coverage of the possible SFHs.

\begin{figure}
\begin{center}
\includegraphics[width=9cm]{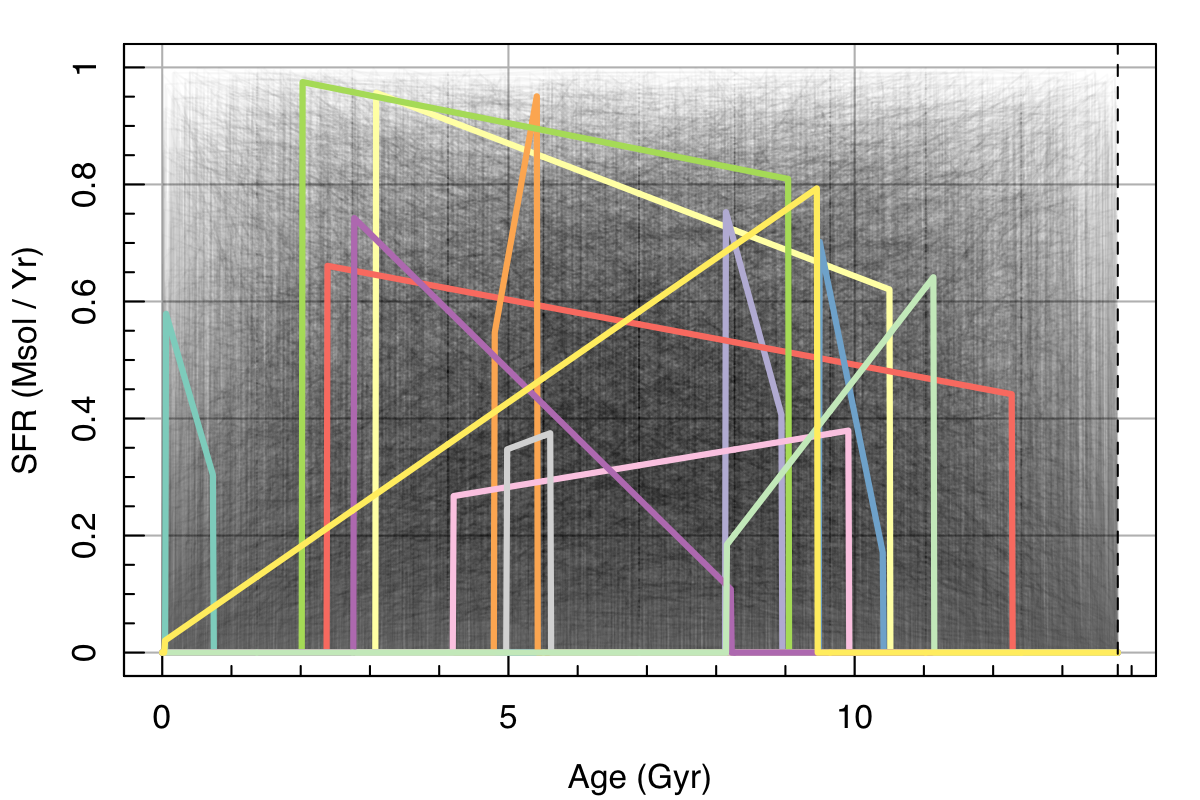}
\caption{{\tt massfunc\_p2} SFHs from 10,000 samples of the m1 / m2 [0,1], and m1age / m2age [0,13.8] parameters, with 12 reference SFHs in bold colour. Note that given the flexibility to adjust all age nodes, a more diverse range of SFHs is in practice possible.}
\label{fig:SFH_massfunc_p2}
\end{center}
\end{figure}

\item {\tt massfunc\_p3(age, m1 = 1, m2 = m1, m3 = m2, m1age = 1e-4, m2age = 7, m3age = 13, magemax = 13.8)} a monotone Hermite spline interpolation model that has control parameters for star formation rate at three nodes (m1 / m2 / m3), with ages (m1age / m2age / m3age), and the maximum age of star formation (magemax), returning the star formation rate at the specified ages (age). Figure \ref{fig:SFH_massfunc_p3} shows the distribution of random parameter samples, reflecting the natural coverage of the possible SFHs.

\begin{figure}
\begin{center}
\includegraphics[width=9cm]{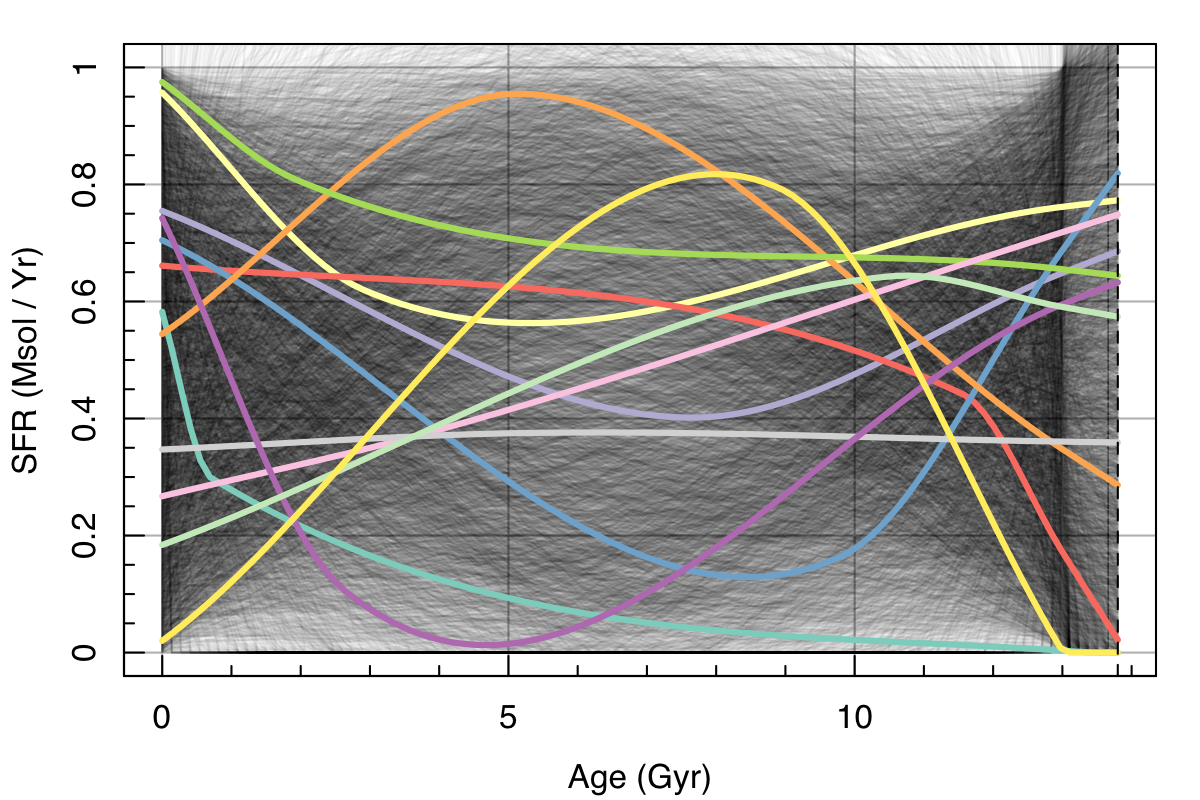}
\caption{{\tt massfunc\_p3} SFHs from 10,000 samples of the m1 / m2 / m3 [0,1] and m2age [$10^{-4}$,13] parameters, with 12 reference SFHs in bold colour. Note that given the flexibility to adjust all age nodes, a more diverse range of SFHs is in practice possible.}
\label{fig:SFH_massfunc_p3}
\end{center}
\end{figure}

\item {\tt massfunc\_p3\_burst(age, mburst = 0, m1 = 1, m2 = m1, m3 = m2, m1age = 1e-4, m2age = 7, m3age = 13, mburstage = 0.1, magemax = 13.8)} as above, with the option of adding a burst of higher star formation rate (mburst), for a certain duration (mburstage).

\item {\tt massfunc\_p4(age, m1 = 1, m2 = m1, m3 = m2, m4 = m3, m1age = 1e-4, m2age = 2, m3age = 9, m4age = 13, magemax = 13.8)} a monotone Hermite spline interpolation model that has control parameters for star formation rate at four nodes (m1 / m2 / m3 / m4), with ages (m1age / m2age / m3age / m4age),  and the maximum age of star formation (magemax), returning the star formation rate at the specified ages (age). Figure \ref{fig:SFH_massfunc_p4} shows the distribution of random parameter samples, reflecting the natural coverage of the possible SFHs.

\begin{figure}
\begin{center}
\includegraphics[width=9cm]{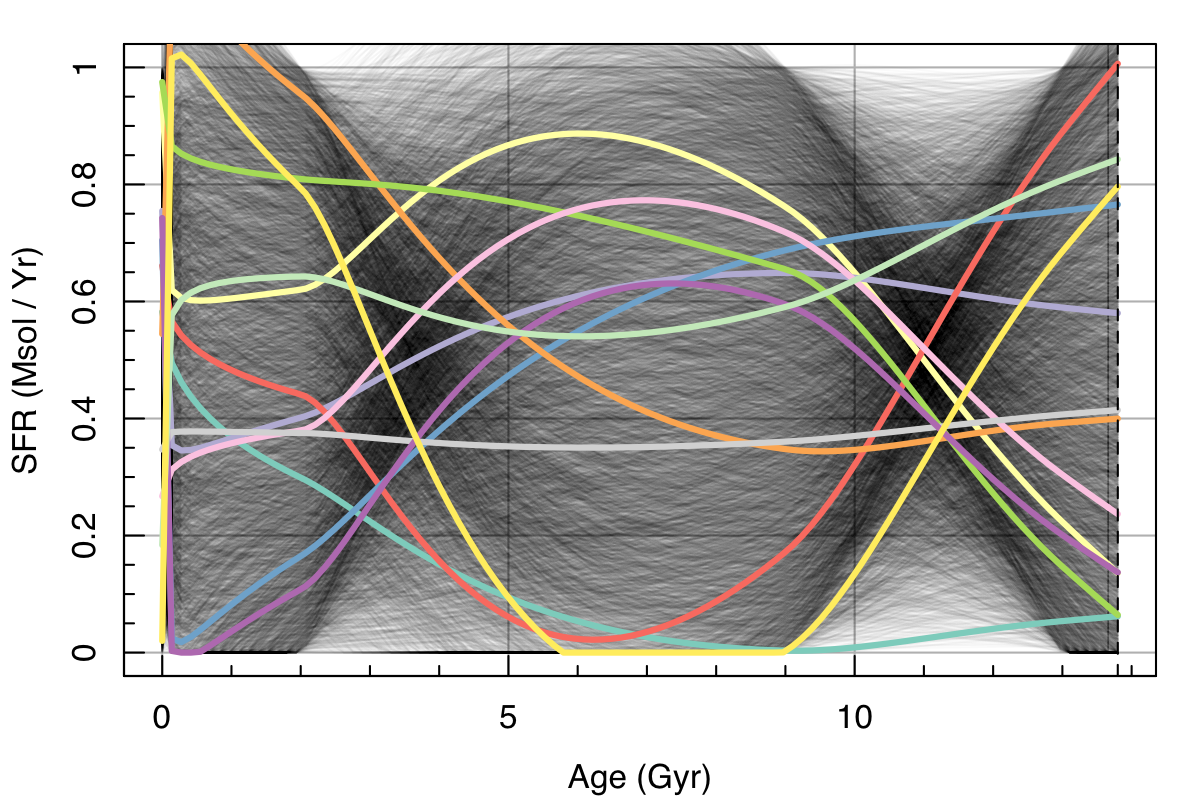}
\caption{{\tt massfunc\_p4} SFHs from 10,000 samples of the m1 / m2 / m3 / m4 [0,1] parameters, with 12 reference SFHs in bold colour. Note that given the flexibility to adjust all age nodes, a more diverse range of SFHs is in practice possible.}
\label{fig:SFH_massfunc_p4}
\end{center}
\end{figure}

\item {\tt massfunc\_p6(age, m1 = 1, m2 = m1, m3 = m2, m4 = m3, m5 = m4, m6 = m5, m1age = 1e-4, m2age = 0.1, m3age = 1, m4age = 5, m5age = 9, m6age = 13, magemax = 13.8)} a monotone Hermite spline interpolation model that has control parameters for star formation rate at four nodes (m1 / m2 / m3 / m4 / m5 / m6), with ages (m1age / m2age / m3age / m4age / m5age / m6age), and the maximum age of star formation (magemax), returning the star formation rate at the specified ages (age). Figure \ref{fig:SFH_massfunc_p6} shows the distribution of random parameter samples, reflecting the natural coverage of the possible SFHs.

\begin{figure}
\begin{center}
\includegraphics[width=9cm]{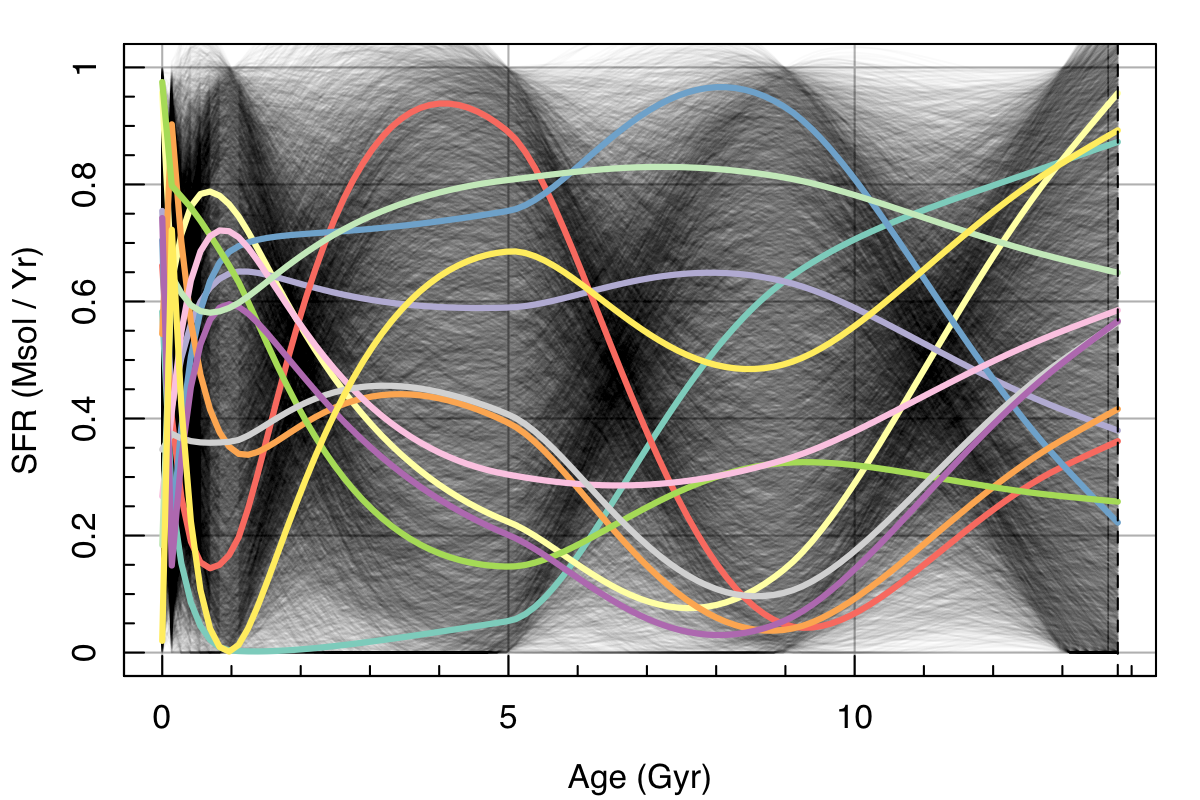}
\caption{{\tt massfunc\_p6} SFHs from 10,000 samples of the m1 / m2 / m3 / m4 / m5 / m6 [0,1] parameters, with 12 reference SFHs in bold colour. Note that given the flexibility to adjust all age nodes, a more diverse range of SFHs is in practice possible.}
\label{fig:SFH_massfunc_p6}
\end{center}
\end{figure}

\item {\tt massfunc\_b5(age, m1 = 1, m2 = m1, m3 = m2, m4 = m3, m5 = m4, m1age = 0, m2age = 0.1, m3age = 1, m4age = 5, m5age = 9, m6age = 13, magemax = 13.8)} a top-hat model that has control parameters for star formation rate at five bins (m1 / m2 / m3 / m4 / m5), with bin age limits (m1age / m2age / m3age / m4age / m5age / m6age), and the maximum age of star formation (magemax), returning the star formation rate at the specified ages (age). Figure \ref{fig:SFH_massfunc_b5} shows the distribution of random parameter samples, reflecting the natural coverage of the possible SFHs.

\begin{figure}
\begin{center}
\includegraphics[width=9cm]{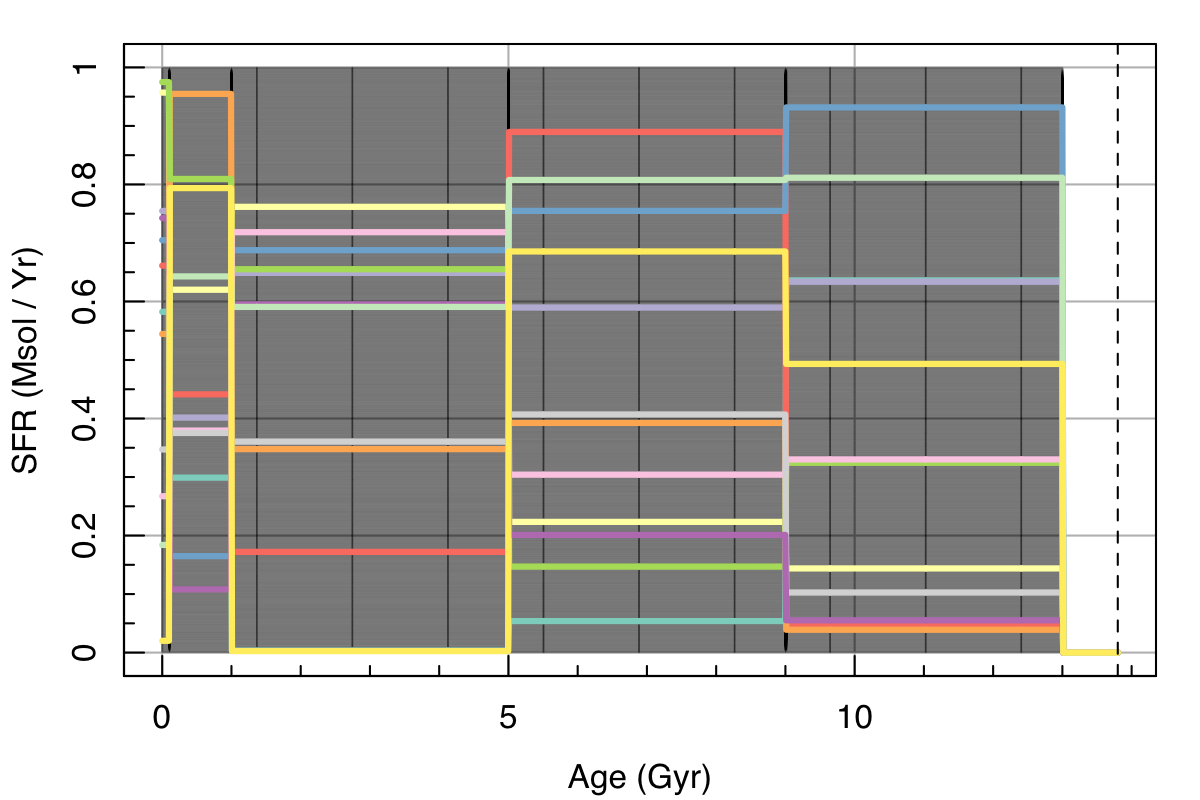}
\caption{{\tt massfunc\_b5} SFHs from 10,000 samples of the m1 / m2 / m3 / m4 / m5 [0,1] parameters, with 12 reference SFHs in bold colour. Note that given the flexibility to adjust all age bins, a more diverse range of SFHs is in practice possible.}
\label{fig:SFH_massfunc_b5}
\end{center}
\end{figure}

\item {\tt massfunc\_exp(age, mSFR = 10, mtau = 1, mpivot = magemax, magemax = 13.8)} an exponentially declining star formation model that has control parameters for the star formation rate (mSFR) at the pivot age, the exponential control parameter $\tau$ (mtau), the pivot age (mpivot) and the maximum age of star formation (magemax), returning the star formation rate at the specified ages (age). Figure \ref{fig:SFH_massfunc_exp} shows the distribution of random parameter samples, reflecting the natural coverage of the possible SFHs.

\begin{figure}
\begin{center}
\includegraphics[width=9cm]{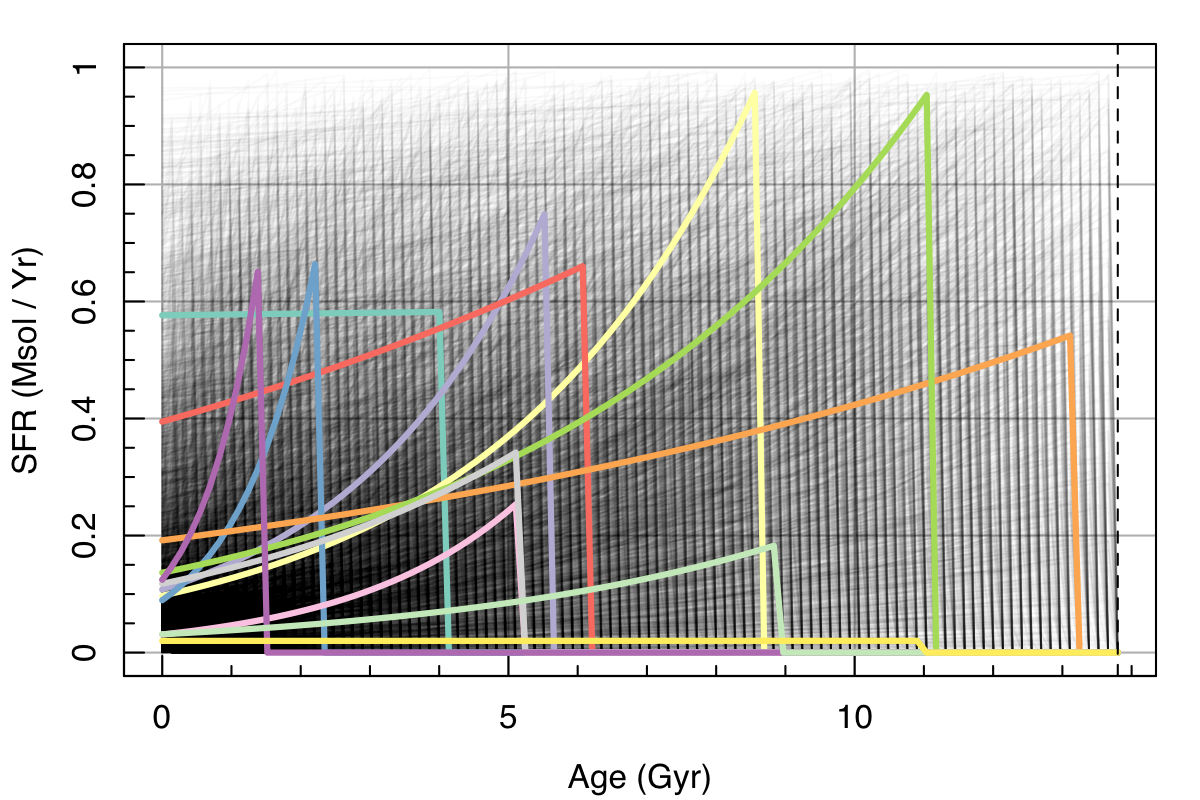}
\caption{{\tt massfunc\_exp} SFHs from 10,000 samples of the mSFR [0,1], mtau [0,3], magemax [0,13.8] parameters, with 12 reference SFHs in bold colour.}
\label{fig:SFH_massfunc_exp}
\end{center}
\end{figure}

\item {\tt massfunc\_exp\_burst(age, mburst = 0, mSFR = 10, mtau = 1, mpivot = magemax, mburstage = 0.1, magemax = 13.8)} as above, with the option of adding a burst of higher star formation rate (mburst) for a certain duration (mburstage). Simple exponentially declining, and exponentially declining with a burst are two of the most popular fiducial models of the SFH used in the modern literature \citep{dacu08, noll09, tayl11, mitc13}.

\item{\tt massfunc\_snorm(age, mSFR = 10, mpeak = 10, mperiod = 1, mskew = 0.5, magemax = 13.8)} a skewed Normal star formation model that has control parameters for the peak star formation rate (mSFR), the age of the peak in star formation (mpeak) the standard deviation of the star formation period (mperiod), the skew of the Normal (mskew, where 0 is perfectly Normal, +ve is skewed to younger ages and -ve is skewed to older ages), and the maximum age of star formation (magemax), returning the star formation rate at the specified ages (age). Figure \ref{fig:SFH_massfunc_snorm} shows the distribution of random parameter samples, reflecting the natural coverage of the possible SFHs. { Since this functional form is novel in the literature we shall specify the mathematical implementation, which is as follows:}

\begin{eqnarray}
SFR(\rm{age}) &=& m_{\rm{SFR}} e^{-\frac{X({\rm age})^2}{2}}, \\
X(\rm{age}) &=& \left[\frac{{\rm age} - m_{\rm{peak}}}{m_{\rm{period}}}\right] \left( {e^{m_{\rm{skew}}}}^{{\rm{asinh}}\left(\frac{{\rm age} - m_{\rm{peak}}}{m_{\rm{period}}}\right)} \right)
\end{eqnarray}

\begin{figure}
\begin{center}
\includegraphics[width=9cm]{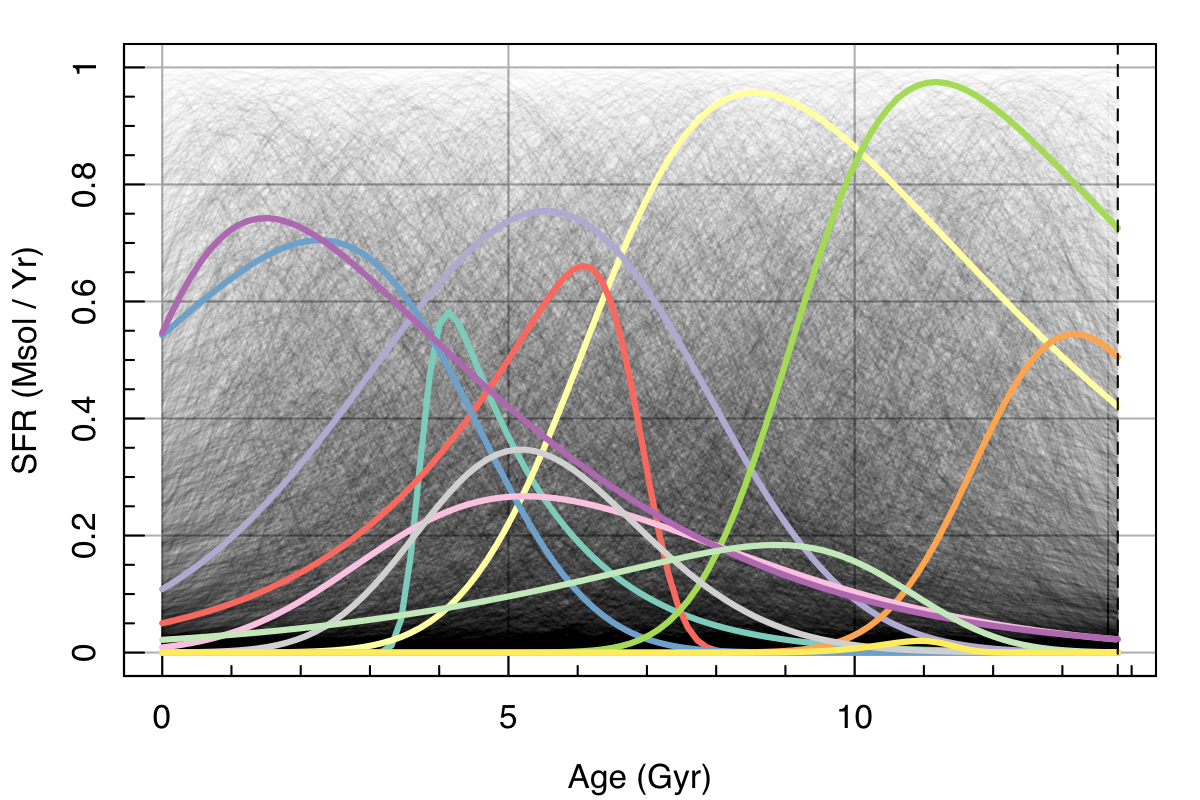}
\caption{{\tt massfunc\_snorm} SFHs from 10,000 samples of the mSFR [0,1], mpeak [0,13.8], mperiod [0.5,3.5], mskew [-0.5,0.5] parameters, with 12 reference SFHs in bold colour.}
\label{fig:SFH_massfunc_snorm}
\end{center}
\end{figure}

\item{\tt massfunc\_snorm\_burst(age, mburst = 0, mSFR = 10, mpeak = 10, mperiod = 1, mskew = 0.5, mburstage = 0.1, magemax = 13.8)} as above, with the option of adding a burst of higher star formation rate (mburst) for a certain duration (mburstage).

\item{\tt massfunc\_snorm\_trunc(age, mSFR = 10, mpeak = 10, mperiod = 1, mskew = 0.5, mtrunc = 2, magemax = 13.8)} a skewed Normal star formation model that has control parameters for the peak star formation rate (mSFR), the age of the peak star formation (mpeak) the standard deviation of the star formation period (mperiod), the skew of the Normal (mskew, where 0 is perfectly Normal, +ve is skewed to younger ages and -ve is skewed to older ages), the maximum age of star formation (magemax), and how sharp the early-time truncation is (mtrunc, where value around 2--3 are fairly strong truncations, and 0 is no truncation), returning the star formation rate at the specified ages (age). Figure \ref{fig:SFH_massfunc_snorm_trunc} shows the distribution of random parameter samples, reflecting the natural coverage of the possible SFHs. This is very similar to Figure \ref{fig:SFH_massfunc_snorm_trunc}, but with clearly sharper growth in star formation rate near the age limit due to the additional mtrunc parameter. SFHs that do not have significant early-time star formation rates are largely unaffected by this new parameter. As such, when SED modelling a real galaxy such a functional form might be preferable since it forces the SFH to grow from a 0 rate rather than starting at the mode, which is unphysical in any reasonable galaxy formation scenario.

\begin{figure}
\begin{center}
\includegraphics[width=9cm]{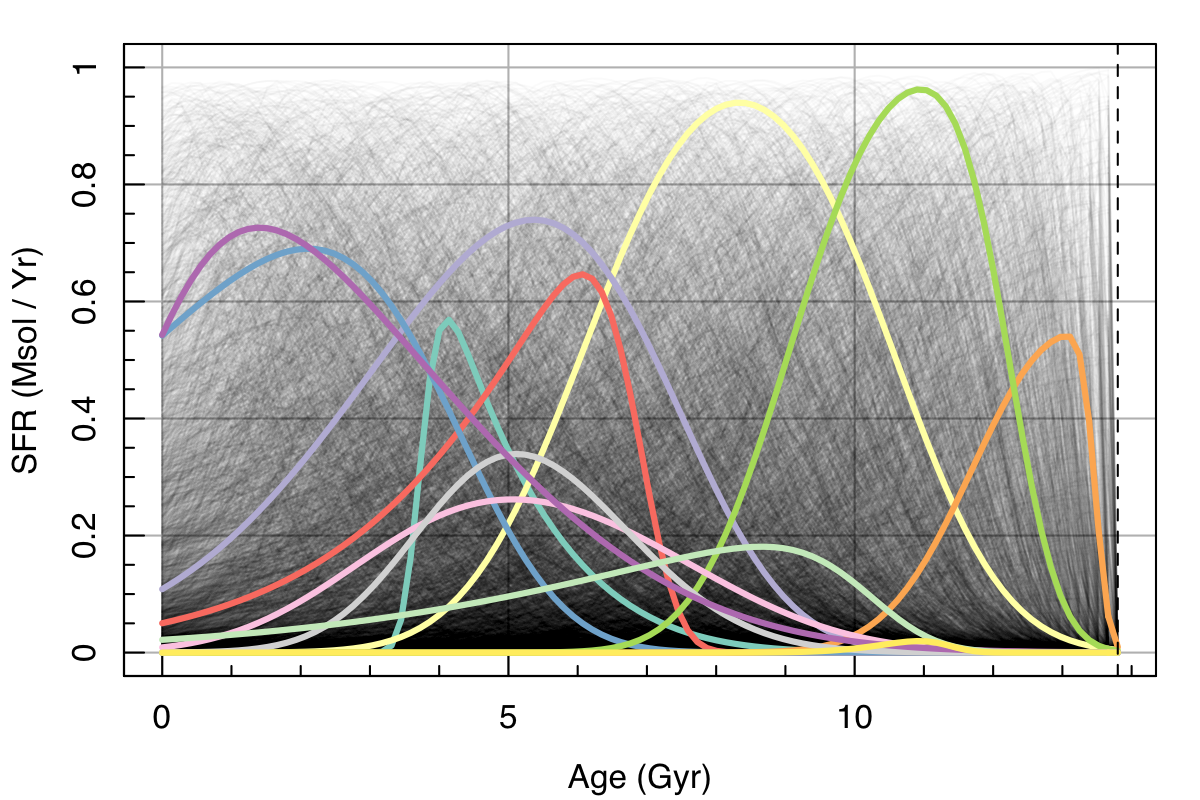}
\caption{{\tt massfunc\_snorm\_trunc} SFHs from 10,000 samples of the mSFR [0,1], mpeak [0,13.8], mperiod [0.5,3.5], mskew [-0.5,0.5], mtrunc[2,3] parameters, with 12 reference SFHs in bold colour.}
\label{fig:SFH_massfunc_snorm_trunc}
\end{center}
\end{figure}

\item{\tt massfunc\_snorm\_burst\_trunc(age, mburst = 0, mSFR = 10, mpeak = 10, mperiod = 1, mskew = 0.5, mburstage = 0.1, mtrunc = 2, magemax = 13.8)} as above, with the option of adding a burst of higher star formation rate (mburst) for a certain duration (mburstage).

\end{itemize}

In principle all of the above parameters can be used as free parameters when fitting a star formation history. In practice, since solutions can become degenerate, it is a good idea to fix some of the parameters (or similarly make use of highly constraining priors), and potentially make use of conditional parameters (which are offered in \prospect).

\subsection{Metallicity Histories}
\label{sec:ZH}

{ All current SED software that we are aware of make simple yet highly impactfulassumptions regarding the evolution of star forming gas metallicity along the lifetime of star formation: either it is fixed to a fiducial value (often solar metallicity), or treated as a variable but constant value. Both of these treatments make erroneous assumptions regarding the evolution of star formation, where it is well understood that for any reasonable model of star formation metallicity increases over time. Given the well understood degeneracy between stellar age and metallicity (higher metallicity young SSPs appear similar to lower metallicity old SSPs, \citealt{wort99}), introducing a more physically motivated model for metallicity evolution is a notable advance of \prospect. This advance should allow for the more accurate extraction of galaxy star formation histories, and it utilises more of the complexity in simulated star formation histories when used in a generative mode.}

\prospect{} allows the user to define (and potentially fit) the star forming gas metallicity history (ZH) of a galaxy in much the same manner that we define and fit the star formation history, with the output value being the fraction of mass in metals ($Z$) rather than the star formation rate. The main user-visible difference is that, in order to avoid variable clashes, the leading letter of the variable becomes a `Z' rather than an `m', e.g.\ we would use `Z1' as variable name rather than `m1'.

{ Internally, \prospect{} implements a functional form of the metallicity evolution by mixing discrete SSPs with varying ages and metallicities (this weighted mixing scheme is discussed earlier in this paper). This approach means typically four model SSP spectra have to be mixed via bi-linear weighting to achieve a desired stellar population age and metallicity.} Whilst this adds some computational and memory overhead (potentially all six metallicities available with the BC03 SSPs have to be mixed), this route offers substantial advantages (discussed later in this work) over simpler schemes of fixing the metallicity to a fiducial value, allowing it to be free but constant at a few discrete values, or allowing it to be free but constant and interpolating between library values \citep[e.g.][are all variants of these simpler schemes]{dacu08, tayl11, mitc13}

In brief, we include the following metallicity functions with the stated default argument values (note variable ages are in Gyrs, but the first argument `age' is in years to be consistent with the stellar libraries in \prospect):

\begin{itemize}

\item {\tt Zfunc\_p2(age, Z1 = 0.02, Z2 = Z1, Z1age = 0, Z2age = Zagemax, Zagemax = 13.8)} a linear interpolation model that has control parameters for star formation rate at two nodes (Z1 / Z2), with ages (Z1age / Z2age), and the maximum age of metal evolution (Zagemax), returning the metallicity at the specified ages (age). See Figure \ref{fig:SFH_Zfunc_p2} to see the distribution of random parameter samples, reflecting the natural coverage of the possible metallicities. With the previous {\tt massfunc\_p2} the SFH was 0 outside of the specified age range, but to be more physically sensible for {\tt Zfunc\_p2} it is the value of Z2 at older times than Z2age and Z1 at younger times than Z1age.

\begin{figure}
\begin{center}
\includegraphics[width=9cm]{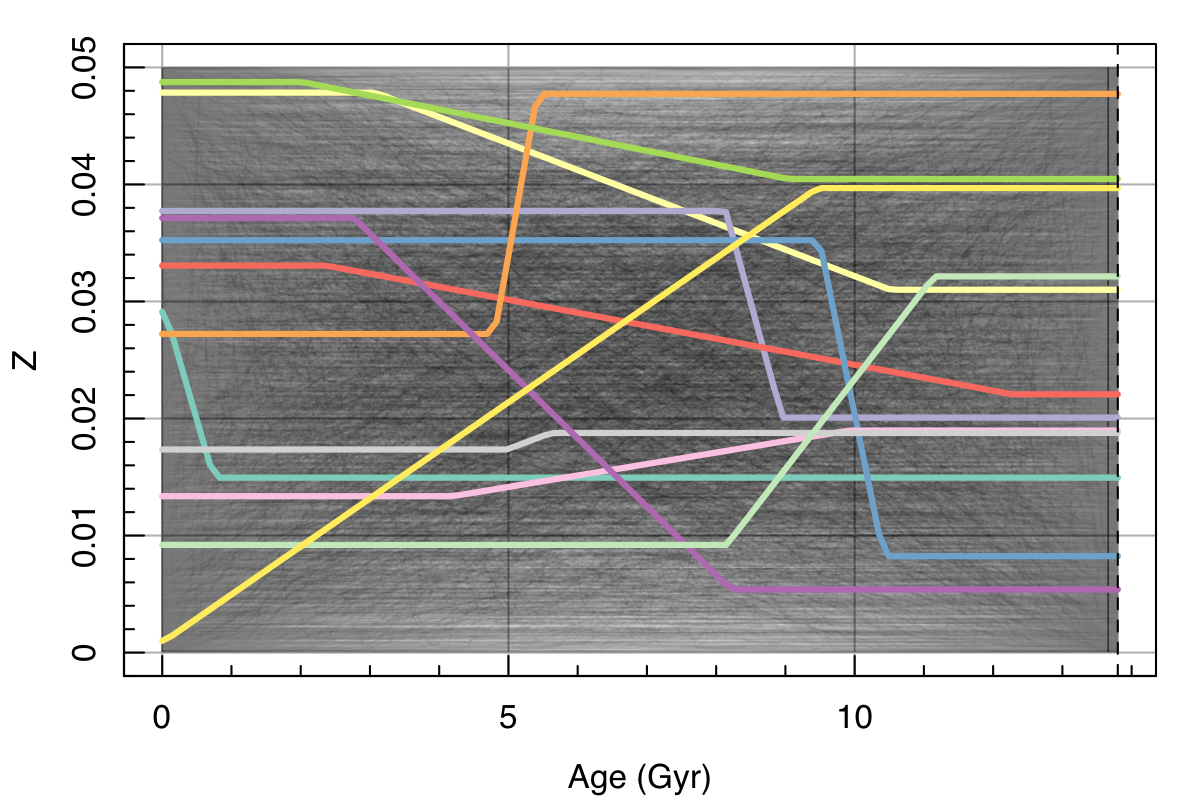}
\caption{{\tt Zfunc\_p2} ZHs from 10,000 samples of the Z1 / Z2 [0,0.05], Z1age / Z2age [0,13.8] parameters, with 12 reference metallicity histories in bold colour.}
\label{fig:SFH_Zfunc_p2}
\end{center}
\end{figure}

\item {\tt Zfunc\_massmap\_lin(age, Zstart=1e-4, Zfinal=0.02, Zagemax=13.8, massfunc, massfunc arguments)} a linear SFH-to-metallicity mapping model as per \citet{driv13} that has control parameters for the starting and finishing metallicity (Zstart / Zfinal), and the maximum age of metal evolution (Zagemax), returning the metallicity at the specified ages (age). The basic idea in this model is that metal enrichment follows 1:1 with mass build up, so when e.g.\ half of a galaxies mass has been assembled half of its chemical enrichment will have also occurred. This model is precisely what would be expected when star formation proceeds in a closed-box but with a constant ejecta metallicity regardless of the metallicity of the gas that formed the stars (so dropping the derived yield). Whilst perfectly closed-box star formation is not supported by detailed chemical abundance observations of galaxies \citep[e.g.\ the g-dwarf problem][]{roch96}, analysis using \shark{} suggests this indeed a reasonable approximation to make in practice (see Section \ref{sec:shark}).

This linear mass mapping metallicity model naturally introduces low initial metallicity for the earliest phases of star formation, and broadly is a consequence of quasi closed-box star formation. In fact, unless there is extreme gas inflow of low metallicity gas it is hard in practice to drastically break this type of metal evolution for realistic galaxy formation. Simulations show that a significant fraction of gas is expected to be recycled, and such pristine infall is likely to be rare \citep{uble14}.

This functional form of ZH is therefore a recommended type to use when attempting to fit a real SED, with the Zfinal parameter (the current gas phase metallicity of the galaxy) kept free when fitting. See Figure \ref{fig:SFH_Zfunc_massmap_lin} to see the distribution of random parameter samples, reflecting the natural coverage of the possible metallicities for the {\tt massfunc\_snorm} SFH model.

\begin{figure}
\begin{center}
\includegraphics[width=9cm]{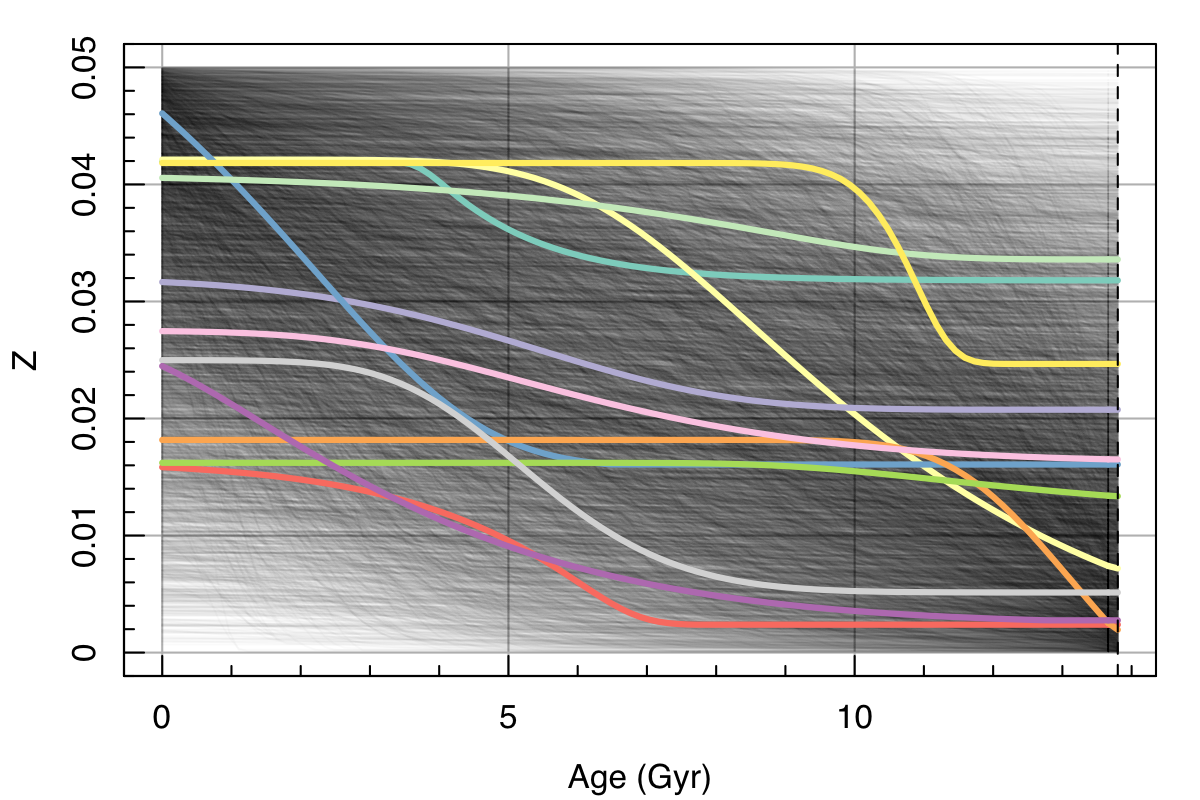}
\caption{{\tt Zfunc\_massmap\_lin} ZHs from 10,000 samples of the Zstart [0,0.05] and Zfinal [0,0.05] parameters, with 12 reference metallicity histories in bold colour. The model SFHs (and associated random sampling of those parameters) used here are the same as in Figure \ref{fig:SFH_massfunc_snorm}, and the colours can be compared directly for an impression for how rapid or slow star formation affects the enrichment timescale.}
\label{fig:SFH_Zfunc_massmap_lin}
\end{center}
\end{figure}

\item {\tt Zfunc\_massmap\_box(age, Zstart=1e-4, Zfinal=0.02, yield=0.03, Zagemax=13.8, massfunc, massfunc arguments)} a closed-box fixed yield metallicity mapping that has control parameters for the starting and finishing metallicity (Zstart / Zfinal) the fixed yield (yield), and the maximum age of metal evolution (Zagemax), returning the metallicity at the specified ages (age). The fixed yield approximation is popular in the literature and is used in various semi analytic models \citep[e.g.][]{lace16, lago18}, and can be specified such that $Z_{\rm final}= Z_{\rm start} - \rho \ln(\mu)$, where $\rho$ is the fixed yield and $\mu$ is the gas fraction. This is another variant of a closed-box enrichment model (as above) with the key difference being we now assume a fixed (rather than evolving, in practice declining) yield. Internally for a given $Z_{\rm final}$ and fixed yield ($\rho$) the current gas fraction is derived using $\mu_{\rm final}=\exp(-(Z_{\rm final} - Z_{\rm start}) / \rho)$. With this computed, the build up of metallicity is then linearly mapped between a gas fraction of 1 (0 stars formed) and this derived value (the current total stellar mass formed). {\tt Zfunc\_massmap\_box} is the other type of metallicity evolution recommended to users for most fitting purposes (along with {\tt Zfunc\_massmap\_lin}).

Figure \ref{fig:SFH_Zfunc_massmap_box} shows the distribution of random parameter samples, reflecting the natural coverage of the possible metallicities for the {\tt massfunc\_snorm} SFH model. It should be clear that the differences between this metallicity and {\tt Zfunc\_massmap\_lin} are relatively small in practice.

\begin{figure}
\begin{center}
\includegraphics[width=9cm]{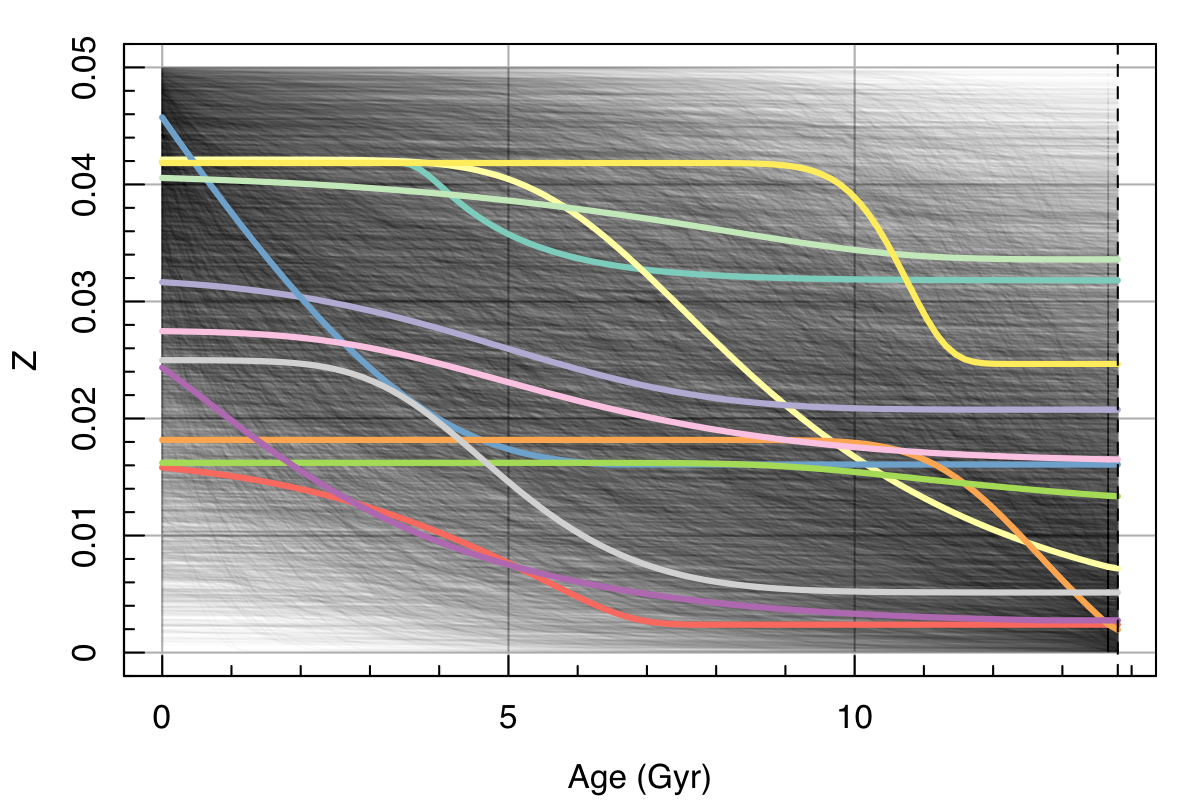}
\caption{{\tt Zfunc\_massmap\_box} ZHs from 10,000 samples of the Zstart [0,0.05], Zfinal [0,0.05] and yield [0.01,0.04] parameters, with 12 reference metallicity histories in bold colour. The model SFHs (and associated random sampling of those parameters) used here are the same as in Figure \ref{fig:SFH_massfunc_snorm}, and the colours can be compared directly for an impression for how rapid or slow star formation affects the enrichment timescale.}
\label{fig:SFH_Zfunc_massmap_box}
\end{center}
\end{figure}


The strict definition of the yield is the ratio of the mass of metals added to the inter stellar medium (ISM) divided by the mass locked up in stars. The fraction of mass locked in stars is usually denoted as $\alpha$, where for a C03-IMF $\sim 20$\% of mass is in stars larger than 10 \msol{} which will enrich the ISM on a rapid timescale. Since the fraction of mass retuned as metals for a typical type-II supernova is $Z\sim0.1$ the typical yield is usually close to $\rho \sim 0.1 (1 - \alpha) / \alpha \sim 0.03$. The approximation of a fixed yield $\rho$ breaks down when gas phase metallicities start to become an appreciable fraction of the metallicity of a supernova event since the yield depends on the mass of metals {\it added} to the ISM, and supernova metallicity is only a weak function of the stellar metallicity. This is the difference in the assumption in Figure \ref{fig:gasfrac_Zgas} between {\tt Zfunc\_massmap\_lin} (dashed, evolving yield) and {\tt Zfunc\_massmap\_box} (solid, constant yield), where at the extreme low gas fraction end we might compute $Z_{\rm gas}$ values that differ by $\sim 30$\%.

\begin{figure}
\begin{center}
\includegraphics[width=9cm]{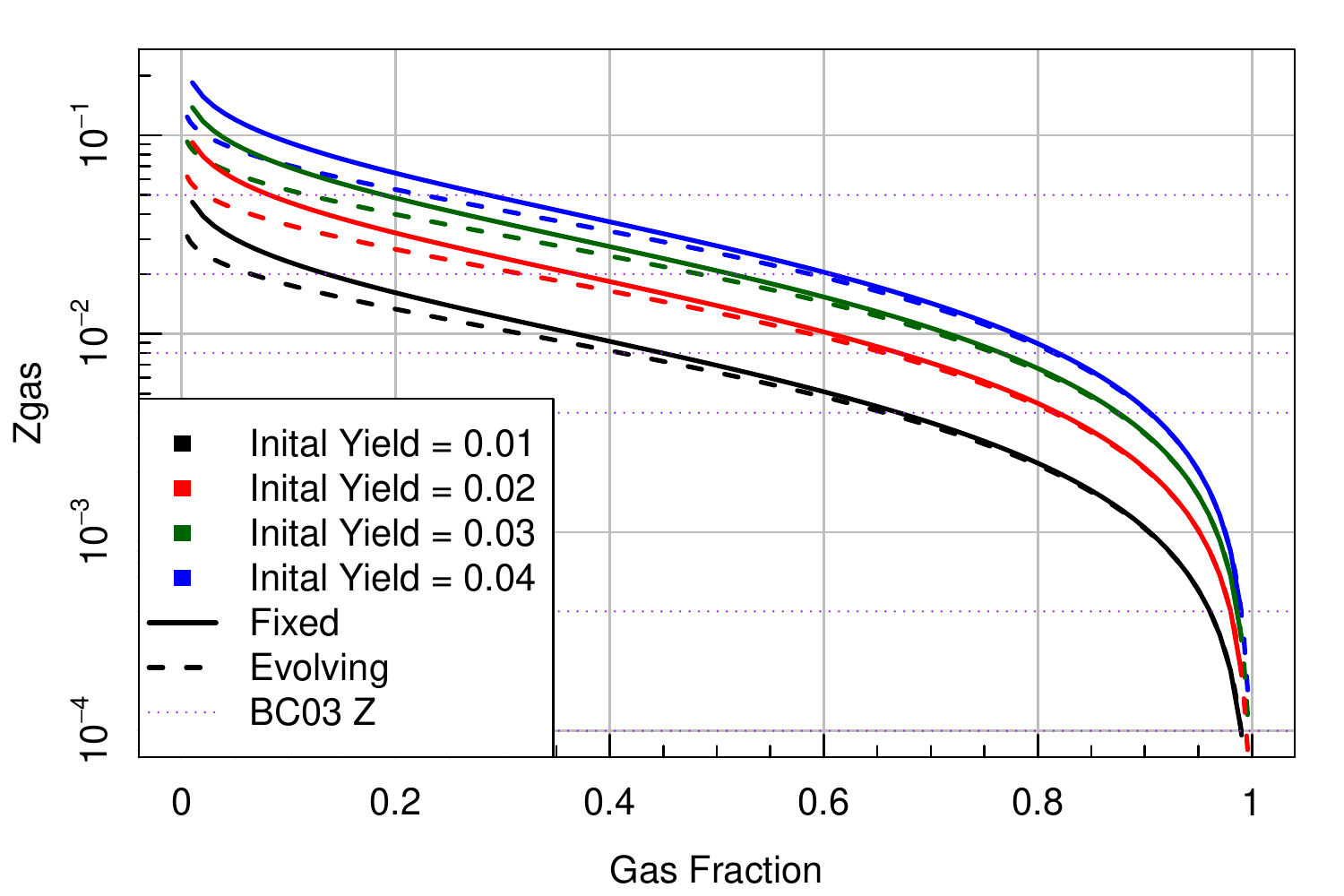}
\caption{Comparison of gas fraction ($\mu$) and the computed $Z_{\rm gas}$ for a range of reasonable yields. For highly enriched low gas fraction systems there is a difference between the predicted $Z_{\rm gas}$, i.e.\ this is the difference in the assumption between {\tt Zfunc\_massmap\_lin} and {\tt Zfunc\_massmap\_box}.}
\label{fig:gasfrac_Zgas}
\end{center}
\end{figure}

\end{itemize}

As with the available SFHs, all of the metallicity evolution parameters can be used in fitting, but in practice many of these should be fixed to avoid degeneracy problems. For instance, if we are using the fixed yield metallicity history specification ({\tt Zfunc\_massmap\_box}) then it is illogical to leave both Zfinal and the yield as free parameters given either one functionally predicts the other exactly.

\subsection{Dust Model}

{ 

As discussed briefly earlier, \prospect{} uses a simple CF00 model for dust attenuation, where the flux observed at a given wavelength ($\lambda$) is modified by the attenuation factor $A$:

$$
A(\lambda) = e^{-\tau (\lambda / \lambda_0)^\nu},
$$

\noindent where $\lambda$ is the wavelength of interest, $\lambda_0$ is the pivot wavelength (5500 {\rm \AA} by default), $\tau$ is the effective optical depth of attenuation, and $\nu$ is the modifying power (-0.7 by default). The three attenuating components within \prospect{} are the ISM dust ($A_{\rm{ISM}}$), which is applied to all light leaving the galaxy; the birth cloud dust ($A_{\rm{BC}}$), which is applied to all light generated by stars younger than 10 Myrs; and the AGN torus dust ($A_{\rm{AGN}}$) which is applied to all light generated by the AGN component.

As per the approximate geometry presented in Figure \ref{fig:ProSpect_schematic}, all light produced by stars older than 10 Myrs is only modified by a single factor $A_{\rm{ISM}}$; light produced by stars younger than 10 Myrs is modified by $A_{\rm{BC}}$ followed by $A_{\rm{ISM}}$; and AGN light is modified by the product of $A_{\rm{AGN}}$ followed by $A_{\rm{ISM}}$.

At each stage of attenuation, an energy balance prescription re-emits the bolometric sum of attenuated light with one of the D14 FIR dust templates. These have a free parameter $\alpha$ that specifies the power law of the radiation field heating the dust, where lower values of $\alpha$ roughly correspond to hotter dust. At each attenuation stage, the re-emitted dust spectrum is added onto the post-attenuation spectrum. This is more accurate than treating the two-phase ISM and birth cloud dust (or ISM and AGN torus dust) as a single combined attenuation process. It also allows for different values of $\alpha$ to be applied to the ISM, birth cloud and AGN torus components (in order of increasing dust temperature and decreasing $\alpha$ for a typical galaxy).

The various D14 templates have a variable dust mass-to-light as a function of $\alpha$, where hotter dust produces much more flux per unit mass. In principle it is possible to use \prospect{} to infer the dust mass present in galaxies, but it is important to note that this property is strictly derived (dust mass is not a parameter that can be fitted directly). For this reason, the FIR dust modelling of \prospect{} is usually treated as a necessary nuisance parameter, and care must be taken when trying to infer dust properties of galaxies. In general, these same caveats should be applied when using any grey body or other dust templates via energy balance modelling.

\subsection{Active Galactic Nuclei}

\prospect{} includes a single broad spectral range AGN taken from \citet{andr18} that has been constructed to appear unattenuated by dust. Within \prospect{} this base AGN template can be attenuated both by the AGN torus itself and the ISM screen dust, and this light is re-emitted using D14 templates at both stages. The AGN torus dust is hot by default, although the effective temperature can be defined by the user.

Figure \ref{fig:AGN_SED} shows a plausible range of AGN templates with different degrees of torus dust attenuation and hot re-emission in the FIR. Due to the energy balance between the templates, the pivot wavelength (where the effect of dust attenuation and re-emission cancels itself out) is in the MIR.

\begin{figure}
\begin{center}
\includegraphics[width=9cm]{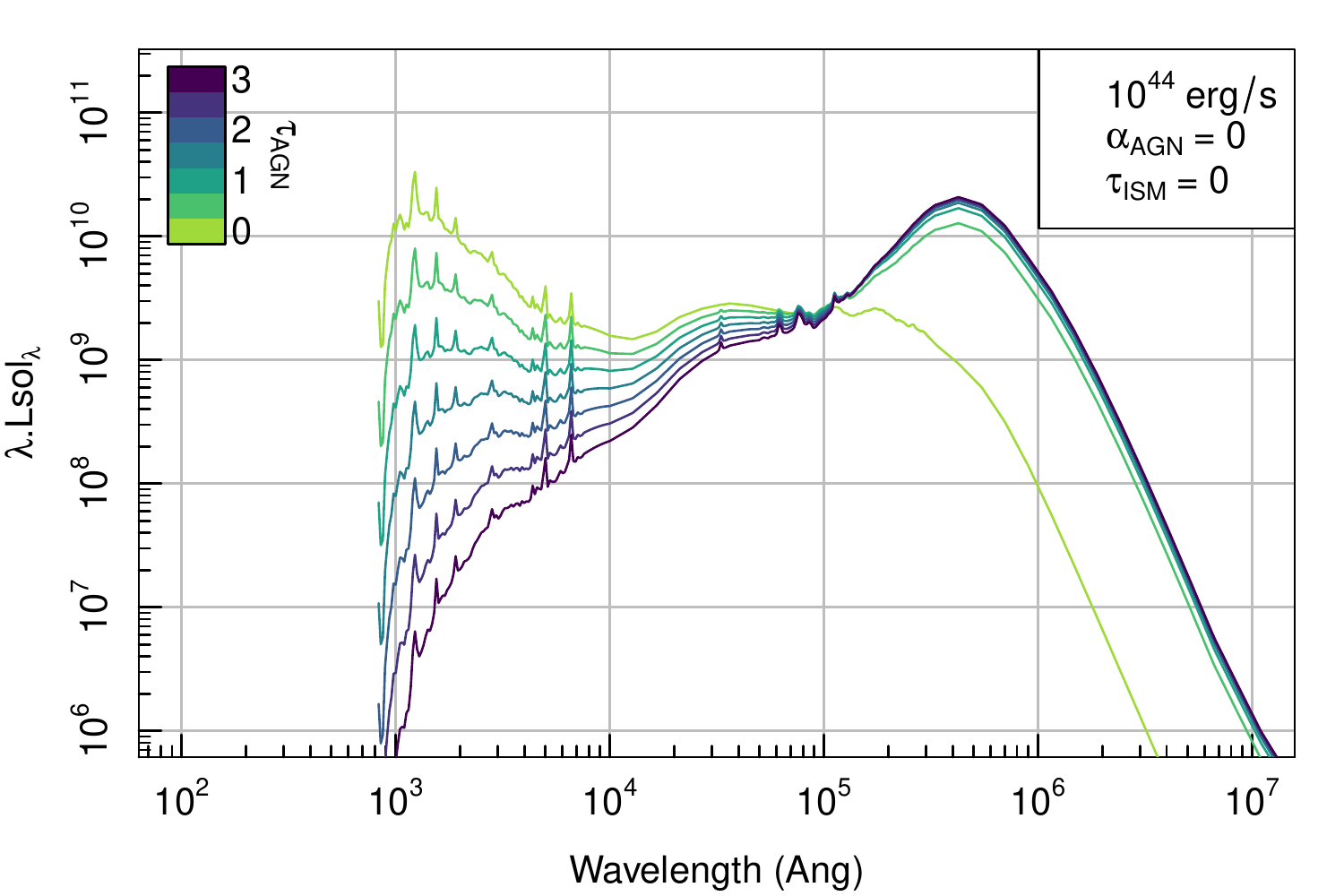}
\caption{The range of possible unattenuated ($\tau_{AGN}=0$) and highly attenuated ($\tau_{AGN}=3$) accessible in \prospect. In all cases the example AGN has a total luminosity of $10^{44}$erg/s and produces hot dust remission with a D14 radiation field parameter of $\alpha_{AGN}=0$ (the hottest available, and default option), and there is no further screen attenuation and re-emission ($\tau_{ISM}=0$). Both the total AGN luminosity and the dust re-emission radiation field can be adjusted or fit as appropriate.}
\label{fig:AGN_SED}
\end{center}
\end{figure}

Currently the AGN in \prospect{} only covers the UV to FIR range. Planned future work is to include physically-motivated models and/or templates to couple the AGN in this regime to emission in the X-ray and radio continuum. The current limitations are a lack of full-spectrum SEDs for AGN \citep[although see][for recent efforts in this regard]{brow19}, and self-consistent theoretical models that fully capture jet formation, torus effects and radio emission \citep[although see][]{turn18}.

\subsection{Nebular Emission Features}

{ As well as stellar light being attenuated by dust and re-emitted in the FIR, stellar spectra are also absorbed by electron excitation in neutral or partially ionised gas, i.e.\ the energy of a photon excites an electron, moving it from one energy level to a higher one, or possibly entirely ionising an electron from an atom. Given gas in the Universe is dominated by hydrogen, hydrogen line transitions dominate this process. For this reason, UV radiation short of the Lyman limit (911.8 {\rm \AA}) is the dominant regime for stellar spectra absorption via hydrogen excitation. Electrons ionised during this absorption process are then free to recombine with ionised atoms, releasing photons with specific energies as they cascade through energy levels (partially excited electrons will also cascade down these same energy levels). This is the origin of the strong narrow nebular emission lines common in star-forming galaxies, these being a source of both UV-bright young stars and intervening gas that acts as a site of ionisation and recombination.}

\prospect{} incorporates a simple energy balance scheme to produce star formation nebular emission features for a range of gas phase metallicities. The key default assumption is that flux short of the Lyman limit is absorbed by an efficiency determined by the UV photon escape fraction (which is 0 by default). The integrated intrinsic stellar flux is then re-emitted using line energies determined by Mapping-III as per the tables provided by \citet[][LKL10 from here]{lkl10}. The full range of optical nebular emission lines available in \prospect{} can be found in Appendix \ref{sec:app_lkl10} Table \ref{tab:lkl10}.

Since here we only focus on star formation emission lines (AGN features are planned for future work), we use the suggested electron density of 100 per cm$^3$ (in future we may expand the range of available electron densities). Whilst it is possible to state the radiation field power law via the free parameter $q$, we by default make use of the metallicity to $q$ mapping provided by \citet{orsi14}, i.e.:

\begin{equation}
q(Z) = q_0 (Z/Z_0)^{\gamma_0},
\end{equation}

\noindent where we used the suggested defaults of $q_0 = 2.8\times 10^7$, $\gamma_0 = -1.3$ and $Z_0 = 0.012$.

The key option for users of \prospect{} is to decide what UV range should be assumed to be ionising the gas. Figure \ref{fig:wave_dom} shows the flux dominance between young stars (less than 10 Myr old) and old stars (older than 10 Myr). It is clear the most important discontinuity occurs short of the Lyman limit (911.8 \rm{\AA}), where intrinsic flux from young type O/B stars completely dominates. This limit is more typically used to determine the ionising flux \citep{orsi14}, and is the default choice in \prospect.

\begin{figure}
\begin{center}
\includegraphics[width=9cm]{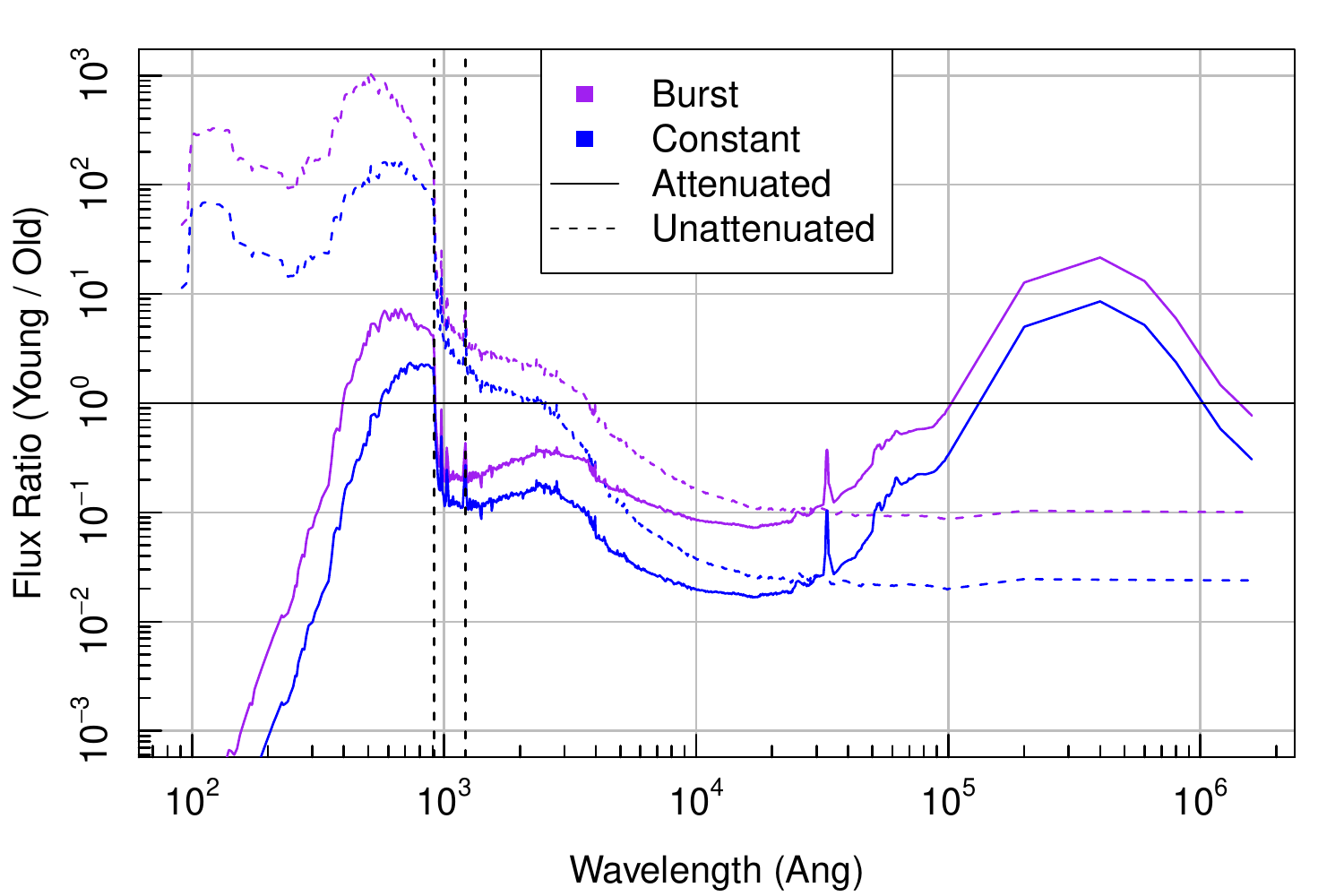}
\caption{Relative domination of the flux contribution between young stars (less than 10 Myr old) and old stars (older than 10 Myr) for either the `burst' or `constant' SFH shown in Figure \ref{fig:SFH_SED}. The two vertical dashed lines highlight the wavelengths of Lyman-$\alpha$ (1215.67 \rm{\AA}) and the Lyman limit (911.8 \rm{\AA}) Even a relatively benign amount of ongoing star formation (constant) will have flux short of the Lyman limit dominated by young stars. For computing ionising flux this is key since efficient Hydrogen ionisation is largely caused by continuum flux short of the Lyman limit, with this ionising flux re-emitted in emission lines, predominantly Hydrogen lines.}
\label{fig:wave_dom}
\end{center}
\end{figure}

With the flux integral determined, the next step is to redistribute this ionising flux across known significant emission features, making use of the standard \prospect{} prescription to interpolate between the gridded value of $Z$ and $q$ available from the tables of LKL10. With the interpolated emission-line fluxes estimated, the features are then attenuated by the same dust prescription as our continuum flux. In practice whilst this creates notable differences in the relative line levels (creating a simulated Balmer decrement) this has very little impact on the energy re-emitted in the FIR (typically a few percent at most for a reasonable SFH).

\begin{figure*}
\begin{center}
\includegraphics[width=17cm]{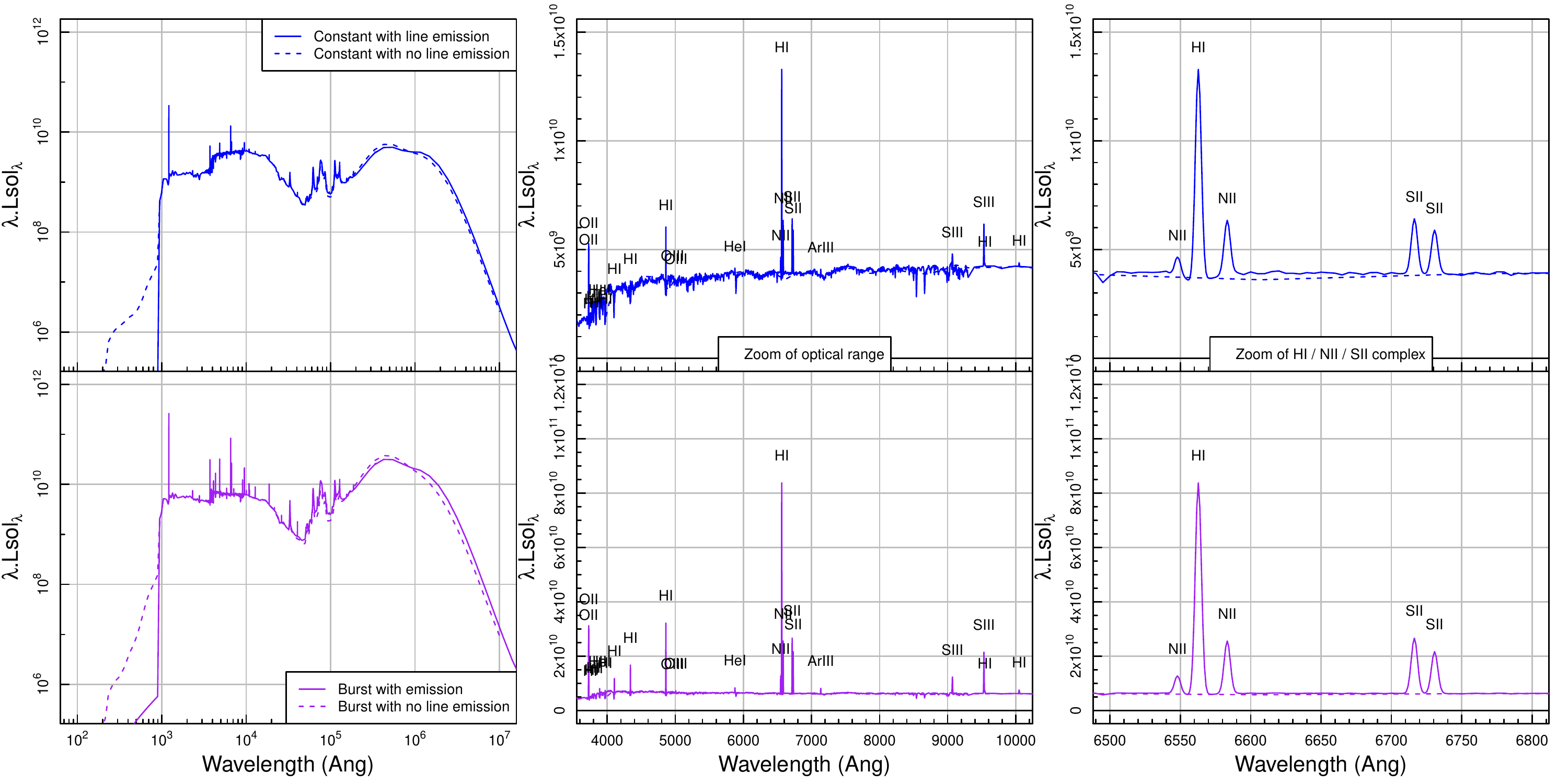}
\caption{Examples of the standard (no emission lines) and emission lines inclusive outputs of prospect{} for a constant (top) and burst (bottom) star formation history galaxy formed with solar metallicity. It is notable in the far left that the emission line free galaxy still has UV flux short of the Lyman limit (911.8 \rm{\AA}), this being entirely absorbed and re-emitted by lines for the emission line variant. All lines are broadened by 100 km/s. Note that here we use the line designations of LKL10.}
\label{fig:emission_lines}
\end{center}
\end{figure*}

The final part of the emission line prescription is to broaden the lines by a desired velocity dispersion assuming a Normal distribution, with the default set to 50 km/s \citep[similar to a typical velocity resolution in low resolution spectroscopic surveys, e.g. GAMA:][]{lisk15}. In principle, more complex mixtures of velocity profiles could be summed to create non-Normal line profiles, but this is left for future work.

Putting these steps together allows us to create realistic emission features that vary sensibly as a function of the amount of ionising flux available (Figure \ref{fig:emission_lines}), and with the gas phase metallicity (Figure \ref{fig:emission_Z}). The computational cost of adding the emission features is notable, increasing a typical run by 50-100\% to around 10 ms (with necessary data pre-loaded). The reason for the increased computational cost is evenly spread between the time spent creating the interpolated emission spectra, the splicing together with the continuum fluxes, and the increased processing expense caused by the higher-resolution spectra (e.g.\ when passing the spectra through target filters etc).

\begin{figure*}
\begin{center}
\includegraphics[width=17cm]{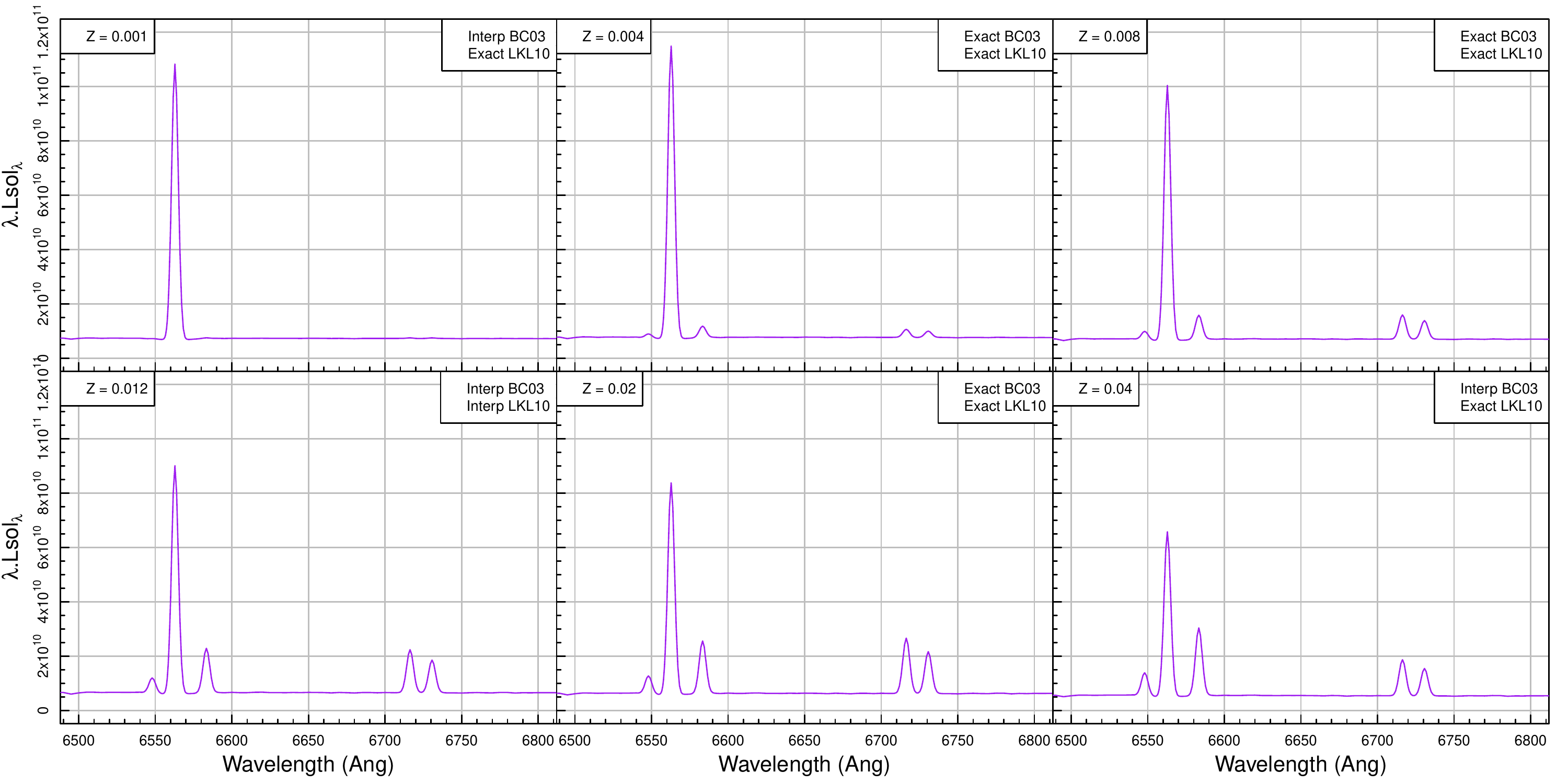}
\caption{How the HII / NII / SII complex varies as a function of metallicity. Due to the discreteness of the available metallicities in BC03 and LKL10, different values need to be interpolated in one of both libraries (as specified in the top right of each panel). Despite this, line ratio and absolute luminosity trends move smoothly over metallicity. All lines are broadened by 100 km/s.}
\label{fig:emission_Z}
\end{center}
\end{figure*}

\prospect{} also includes the ability to scale the emission features via the classic \citet{kenn98} relationship (K98 from here) that scales the strength of the H$\alpha$ line with the star formation rate of stars younger than 10 Myrs. This is included for backwards compatibility with analysis done in this manner, but it cannot ensure proper energy balance, and obviously does not properly adapt the strength of the various features with metallicity (as seen by the varying relative strength of the dominant H$\alpha$ line in Figure \ref{fig:emission_Z}).

The impact of choosing the UV absorption and re-emission route to producing lines versus using the K98 implied SFR to H$\alpha$ relationship is clear in Figure \ref{fig:Ha_scale}, where over the domain of low to solar metallicity the K98 relationship would imply notably lower H$\alpha$ compared to the energy balance method used by \prospect{} by default. This finding is only very weakly sensitive to the star formation rate in question, suggesting that integrating the star formation rate only for stars younger than 10 Myrs is the appropriate temporal range to consider. { In any case, using the simpler K98 prescription versus a full energy balance will produce results that are consistent within a factor of 2, no matter the metallicity. This is also the implied accuracy we can expect when attempting to use emission features to infer the current star formation rate.}

Varying the escape fraction with metallicity would naturally correct the main H$\alpha$ prediction discrepancy (if that is desirable) since \prospect{} is by default redistributing the luminosity of all flux below the Lyman limit across all of the emission lines. Escape fractions near 0.3 for metallicities below solar ($Z<0.02$) would bring the methods into close agreement. Regardless, an H$\alpha$ line prediction ratio of better than a factor of two across such a broad range of metallicity is encouraging given the assumptions and uncertainties implicit when scaling through either route.

\begin{figure}
\begin{center}
\includegraphics[width=9cm]{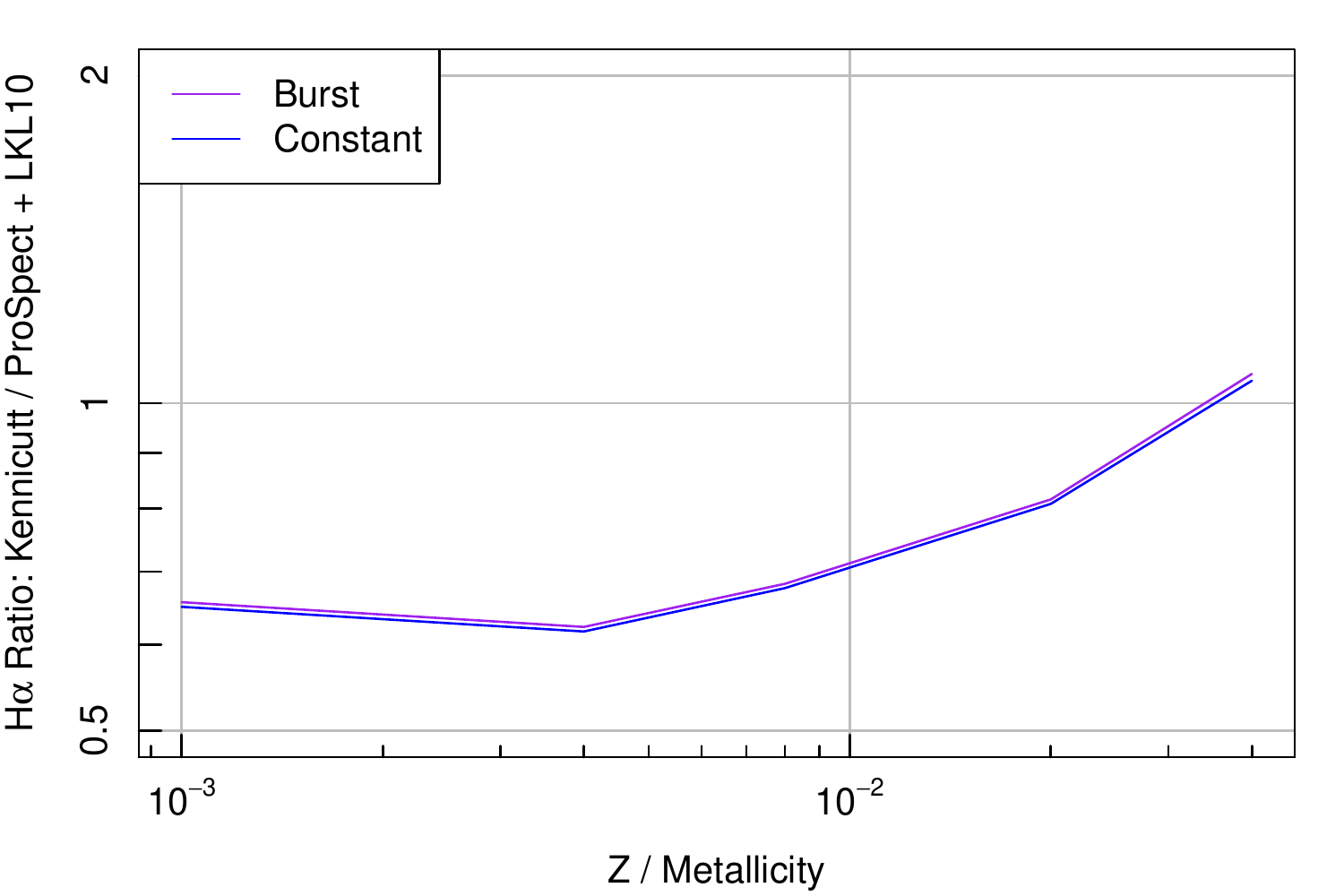}
\caption{Comparison of implied H$\alpha$ flux as a function of constant gas phase metallicity for different SFHs (constant in blue, burst in purple) for the classic K98 relationship and that implied by the energy balance combining \prospect{} with LKL10 assuming an escape fraction of 0. They agree best at the highest metallicities and are only weakly sensitive to the star formation rate, disagreeing by at worst by a factor $\sim2$. The implication of this is that star formation rates derived from the K98 relationship would be overestimated compared to \prospect{} when using the H$\alpha$ feature alone when the metallicity is low.}
\label{fig:Ha_scale}
\end{center}
\end{figure}

 
\section{Usage}

{ As a guide for users, in this Section we discuss some initially simple, and eventually more complex, \prospect{} use cases.} In the following we are leaving all dust and re-emission properties at their default \prospect{} values (as discussed above) and include no AGN contribution. Also, unless otherwise stated, we use solar metallicity libraries ($Z \sim 0.02$).

\subsection{Simple and Interactive}

{ To better understand the utility and plausibility of \prospect{} generated SEDs we will initially investigate a few simple SFHs and compare them to the diversity of galaxy colours observed in the GAMA survey. This will serve to give insight into the diversity of SFHs required to properly capture the range of galaxies present in large surveys of galaxies.}

{ As emphasised in the Methods Section, there are a large number of modes in which \prospect{} can be used. Initially we will generate different SFHs using the {\tt massfunc\_p4} from above (since it is flexible and intuitive to use) and see what impact these have on the output \prospect{} SED for both the unattenuated and attenuated and re-processed light for the BC03 spectral library (which is the default used in \prospect, and unless otherwise specified is the spectral library used in the following parts of this paper).}

\begin{table}
\begin{center}
\begin{tabular}{| l | l | l | l | l |}

Name & m1 & m2 & m3 & m4 \\
\hline
Burst & 12 & 1 & 1 & 1 \\
Constant & 1 & 1 & 1 & 1 \\
Quenched & -20 & 1 & 1 & 1 \\
Dead & -20 & 0 & 1 & 1 \\
Fossil & -20 & -4 & 1 & 1 \\
\end{tabular}
\end{center}
\caption{{\tt massfunc\_p4} parameter values used to create the various SFHs presented in Figure \ref{fig:SFH_SED}.}
\label{tab:params}
\end{table}%

\begin{figure*}
\begin{center}
\includegraphics[width=12cm]{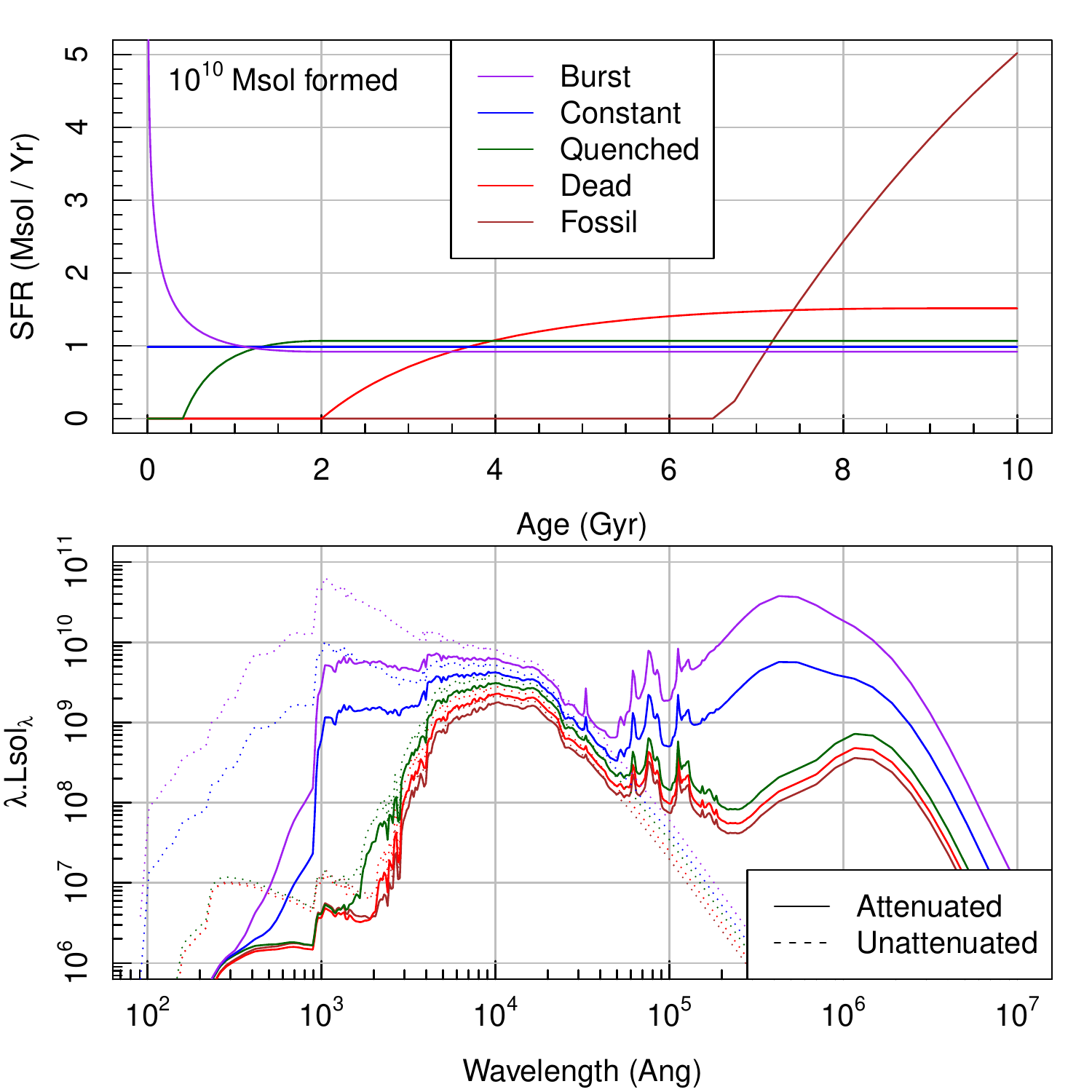}
\caption{Top panel: using the {\tt massfunc\_p4} SFH model we produce five different SFHs named for their conceptual class of star formation. Bottom panel: the different unattenuated and attenuated SEDs produced by these SFH models at $z=0$ are shown as dashed and solid lines respectively.}
\label{fig:SFH_SED}
\end{center}
\end{figure*}

Figure \ref{fig:SFH_SED} shows the SFH models (upper) and output SEDs generated (lower) with modifications made to the various m1 / m2 / m3 / m4 parameters only. { To enable these results to be re-created, we provide the \prospect{} {\tt massfunc\_p4} parameter values used in Table \ref{tab:params}.}. In all cases we are producing exactly the same amount of stellar mass ($10^{10}$\msol). Using just this vanilla mode of \prospect{} we can generate a diverse range of SEDs, from extremely quiescent UV free galaxies with little FIR dust re-emission to vigorously star forming galaxies producing a significant component of hot MIR/FIR dust. It is also easy to see the slow increase in mass-to-light as we move to deader SFH models, with a change of a factor $\times 10$ in the $g$-band flux produced across all of our models. As expected, the mass-to-light variation is smaller in the NIR bands, but still factors of a few. { Appendix \ref{sec:simple_example} explicitly provides the code that produces the quenched galaxy SFH and corresponding flux density, making it easy for the novice user to re-create some of this work.}

With the various libraries and data pre-loaded, each of these full SEDs (with a large number of outputs not presented here) can be generated in around 5 ms on a modern desktop computer, making it easy to interact with the \prospect{} model in realtime when experimenting with SED generation and fitting. To further aid model exploration a GUI interface is included that allows users to directly interact with the main parameters that drive the SED for a simple multi-phase SFH (a restricted version of the {\tt massfunc\_b5} function discussed in detail above). A web interface to this simple GUI is also made available\footnote{\url{http://prospect.icrar.org}}.\newline

\noindent As mentioned in the Methods Section, \prospect{} also includes the EMILES spectral library. This has advantages compared to BC03 in respect to the spectral resolution available, the modernity of the stellar atmospheres used and the metallicity coverage, however, it has notably smaller spectral coverage. This is apparent when comparing instantaneous burst SSPs of different ages at solar metallicity ($Z \sim 0.02$), as shown in Figure \ref{fig:BC03vEMILES}.

\begin{figure}
\begin{center}
\includegraphics[width=9cm]{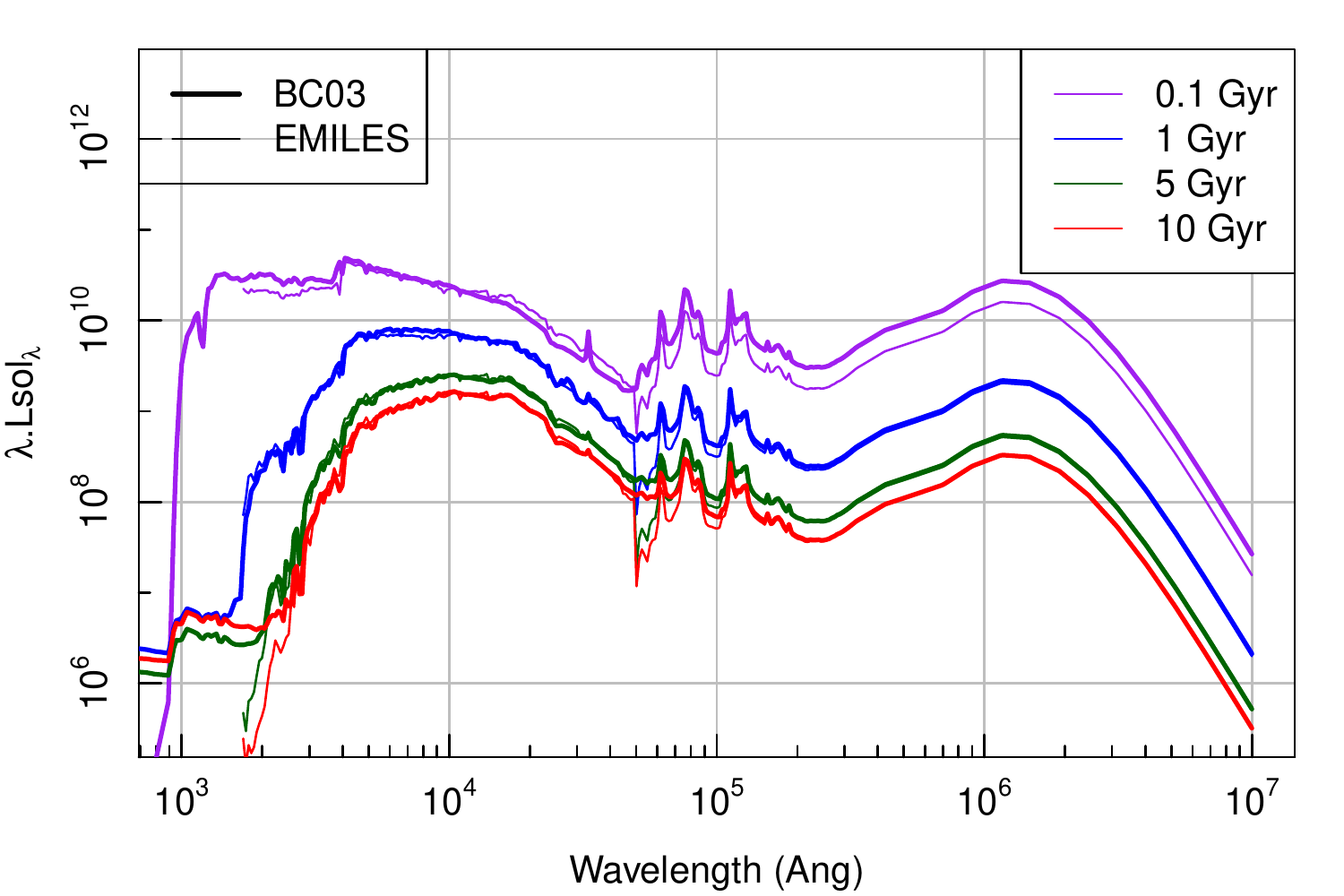}
\caption{Comparison of BC03 and EMILES using four different burst SSP models that all produce $10^{10}$ \msol{} of mass, where all other parameters are at their \prospect{} defaults. It is noticeable that EMILES has a shorter spectral coverage, cutting out at around 2,000 \rm{\AA} in the UV and $5\times10^4$ \rm{\AA} in the NIR (causing the discontinuity seen in the EMILES spectrum at this point).}
\label{fig:BC03vEMILES}
\end{center}
\end{figure}

The two spectral libraries agree quite closely for 1 Gyr stellar populations, with only very small differences in the optical regime. However, there are clear differences in the other age SSPs. The youngest (0.1 Gyr, purple lines) differ throughout the optical and NIR and EMILES clearly has a sudden truncation around $2\times10^2$ \rm{\AA}. This truncation means the integrated dust attenuated light differs markedly, and the amount of re-emitted FIR light changes by around a factor of two (with BC03 producing more flux for the same mass burst).

Less prominently, there are also large differences in the 5 and 10 Gyr SSPs at around $2\times10^2$ \rm{\AA}, with BC03 having a significant UV upturn produced by the inclusion of planetary nebulae in the SSP modelling (these are not included directly in EMILES). This difference has no notable impact on the re-emitted FIR properties however, with the BC03 and EMILES \prospect{} models agreeing very closely beyond $10^5$ \rm{\AA} in the MIR. The end result of this comparison suggests that some care and caveats are required when modelling very recent star formation (usually considered to be anything sub 0.1 Gyr), and when incorporating UV observational data in general. Since later fitting focuses on GAMA data that have observational photometry extending into the FUV \citep{lisk15}, we will concentrate our ongoing discussion on the BC03 spectral library since it better covers this regime.\newline

\noindent With this in mind, we will generate a few simple star formation histories using BC03 for galaxies placed at different redshifts and compare the observed frame $g-i$ to what we find in the GAMA survey \citep[almost complete for galaxies with $r<19.8$;][]{lisk15}. For this application we are leaving all dust and re-emission properties at their default \prospect{} values (as discussed above) and include no AGN contribution. Also, we limit the star formation history so that stars can only form after a current lookback time of 13.8 Gyrs (i.e. they cannot form stars before the Universe began, no matter what redshift they are placed at). In all cases we are producing exactly the same amount of stellar mass ($10^{10}$\msol), but since we are only looking at photometric colours (a relative flux measurement) this is not an important factor.

Figure \ref{fig:redshift_g-i} presents the same five SFHs as above, but now using three different fixed metallicities ($Z=0.0001 / 0.02 / 0.05$). The main bimodality tracks are easily identified, with quenched (or deader) galaxies existing on the GAMA red sequence, and a mixture of SFHs contributing to the visibly broader blue cloud.

\begin{figure}
\begin{center}
\includegraphics[width=9cm]{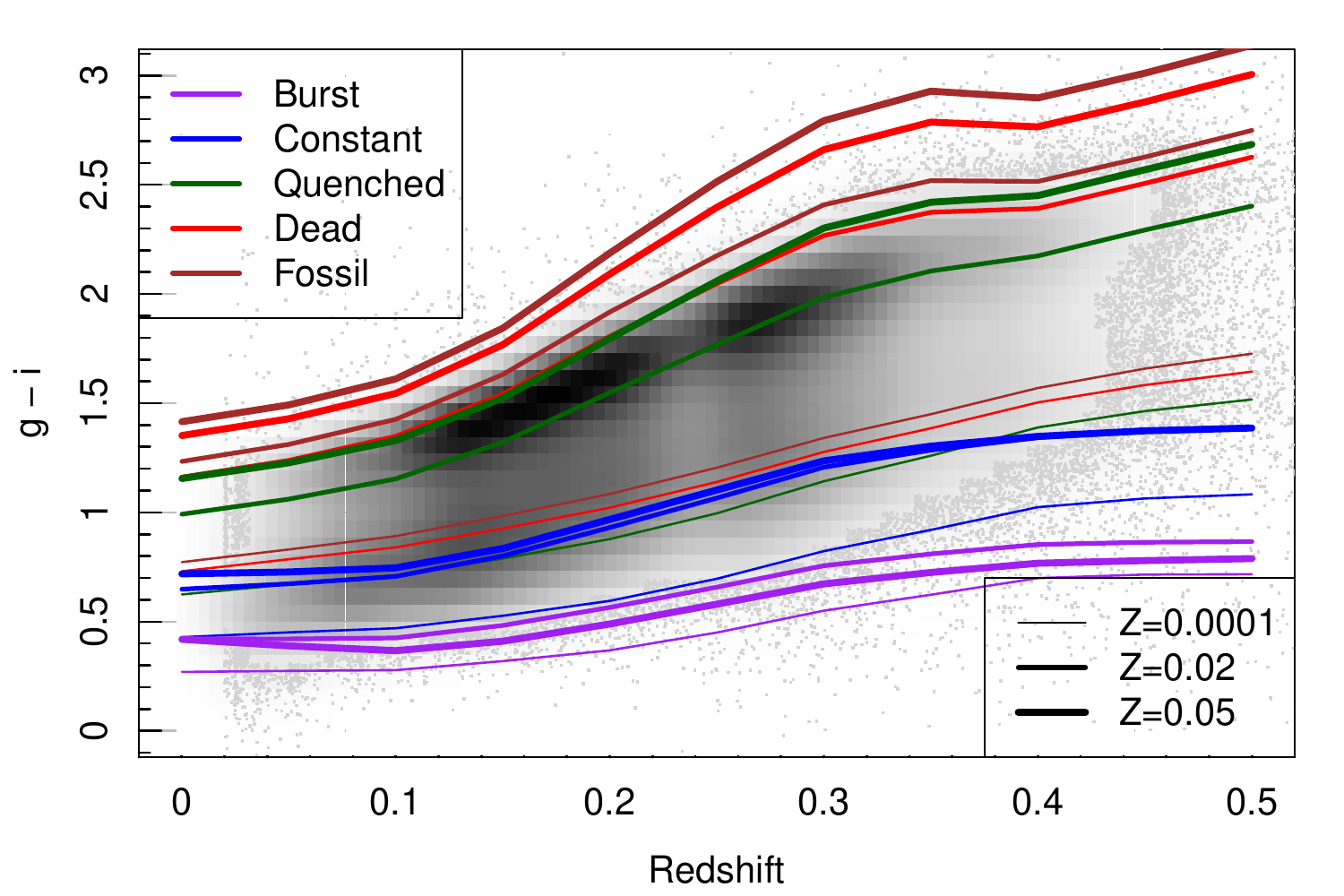}
\caption{Redshift versus $g-i$ observed frame colour for GAMA (grey scale density and points) and a variety of different SFHs (lines). It is clear we can reconstruct the main colour bimodality evolution, with some age and metallicity degeneracies evident. GAMA colours beyond the extrema lines shown here are not physically plausible, and are likely due to be photometry processing errors.}
\label{fig:redshift_g-i}
\end{center}
\end{figure}

{ Interestingly, the general tracks for the blue cloud beyond $z=0.3$ \citep[which in GAMA are dominated by galaxies more massive than $M^*$ above $z\sim0.3$;][]{tayl11} are qualitatively better described by very low metallicity quenched (or less star-forming) galaxies rather than star forming galaxies. However, it is more accurate to say that such a simple diagnostic cannot distinguish between competing galaxy formation scenarios. Below $z=0.3$ (dominated by galaxies less massive than $M^*$) the blue cloud track in GAMA tightens up, and is better described by ongoing (i.e. `constant') star formation models. In the GAMA selection at no point are we significantly populated by bursting star formation, which is consistent with the picture that below $z=1$ highly energetic star formation bursts are somewhat rare.}

The main takeaway from this high level view of SED generation when compared to GAMA is that \prospect{} is capable of generating a plausibly complete suite of colour distributions for moderate redshift galaxies. This suggests it is reasonable to assume \prospect{} might serve as an informative SED fitting tool, at least in application to GAMA data. This will be explored in more detail later in this paper.

\subsection{Application to Semi-Analytic Models of Galaxy Formation}
\label{sec:shark}

It is simple to use \prospect{} on the outputs of semi-analytic models of galaxy formation (SAMs). Typically we might expect a given SAM to produce a SFH and ZH, and both of these can be fed directly into \prospect{} at the resolution they are generated (ideally a few hundred Myr temporal resolution). An example of just such an application is the \shark{} SAM that has been run on the SURFS suite of N-body simulations \citep{elah18,lago18,lago19}.

To aid the production of full SEDs from \shark{} a binding interface (\viperfish) was built that allows for the rapid generation of photometry from the HDF5 outputs generated by \shark{} \citep{lago19}. This binding interface works both on the individual snapshots (where galaxies within a given volume are at the same redshift / age) and lightcones generated by \stingray{} (where every galaxy is placed at a unique redshift / age; Obreschkow et al., in prep.). { Depending on the precise format of the SFH and ZH generated by a SAM, different \viperfish-like interfaces might have to be written. \viperfish{} is freely available online to aid users in writing their own interface\footnote{\url{https://github.com/asgr/Viperfish}}.}

\prospect{} can be run with suggested default dust parameters (which are reasonable local Universe fiducial values), but for more realistic SED generation, especially at high redshift, improvements to the outputs are possible by modifying dust properties in a physically motivated manner. To this end, radiative transfer modelling outputs from the EAGLE simulation were calibrated against the properties available in \prospect{} \citep{tray20}, allowing for more realistic dust attenuation and re-emission on a per galaxy basis. This improved modelling is discussed and applied extensively in \citet{lago19}, where we find that significant improvements to global galaxy photometry properties are achievable through such techniques, with better luminosity functions and colours generated at all redshifts. Ongoing work (Bravo et al., submitted) investigates the quality of galaxy colours as a function of stellar mass in detail. In brief, the agreement is generally excellent, but some there is some shift in the stellar mass $g-i$ colour relationship which requires accounting for (this is discussed later in this paper).

\viperfish{} is a very light interface to \prospect{}, and it is simple for other SAMs to make use of \prospect{} in a similar manner with relative ease. It also scales very well to big simulations since it can parallelise the generation of SEDs trivially, dealing with the book keeping complexities that occur even for embarrassingly parallel problems.


\subsection{Application to Galaxy SED Fitting}

{ \prospect{}, by virtue of its fully generative nature, can be easily utilised for the problem of galaxy SED inference. SED generation is the vital first step when attempting to infer combinations of parameters that best recreate target data. Each set of parameters generated can be compared against our observations, and some goodness-of-fit quantification (usually a likelihood) is then used to determine which parameters best describe our data. There are a number of methods for exploring this parameters space (far too many to comprehensively cover here, but see \citealt{robo17} for discussion of some of the methods available within the \R{} ecosystem), where often researchers try to find the `most likely' solution, and then explore the range of uncertainty around this solution. One popular family of methods for exploring the range of plausible parameters are Markov Chain Monte Carlo (MCMC) samplers. These build up a stochastic picture of plausible parameter combinations, and keep the history of these explorations in posterior sampling chains.}

Any of the parameters discussed thus far can be used as part of this inference process, where the output of interest is always the posterior model samples. Since our \prospect{} model will always be much simpler than a true galaxy, the aim is that any parameters of interest are at least informative and useful, e.g.\ the current stellar mass remaining, star formation rate and metallicity. Other parameters are perhaps better viewed as nuisance parameters to be marginalised over (e.g.\ the interpretation of some of the dust modelling parameters should not be pushed too far). \prospect{} has already been successfully used in this mode for recent work (e.g.\ Seymour et al., in press; Tiley et al., submitted; Allison et al., submitted).

In this sub section we explore the impact of varying the photometric errors on the quality of the MCMC posterior sampling, and the impact of attempting to fit an incorrect model to a given star formation history. Throughout we use the Component-wise Hit And Run Metropolis (CHARM) MCMC algorithm that is included in the {\sc LaplacesDemon} open source inference package available for \R{} through CRAN \citep[as discussed in ][]{robo15,robo17}. { This algorithm is relatively slow, but it is well suited to sampling complex posterior space with correlation between parameters. Such correlation tends to be common between parameters describing SFHs and ZHs, since they tend to be parameterised in a manner that is intuitive rather than guaranteeing orthogonality, so CHARM is the recommended algorithm for the novice user.}

\subsubsection{Impact of Photometric Error}

Even in the regime of knowing a-priori what star formation and metallicity model we should use to fit a given set of photometric data (impossible in reality) there is still the issue of photometric uncertainty (i.e.\ flux error). Clearly if the per-band error is smaller it should be possible to constrain a given model to better accuracy than if the error were significantly higher.

To assess the broad impact of photometric error we re-fit the same intrinsic {\tt snorm\_trunc} model with four different grades of per band error: 0.001 mag (the best photometry you would typically see presented), 0.01 mag (typical of good quality photometry with no systematics or source confusion), 0.1 mag (typical of lightly blended photometry) and 0.5 mag (typical of the faintest sources in a given source extraction). The results of this experiment are shown in Figure \ref{fig:Refit_worm}, where we see a trace plot of all posterior chain samples for each parameter for three of the different photometric errors (the 0.001 mag case is not shown here, because it is visually exactly on top of the input parameters).

\begin{figure*}
\begin{center}
\includegraphics[width=12cm]{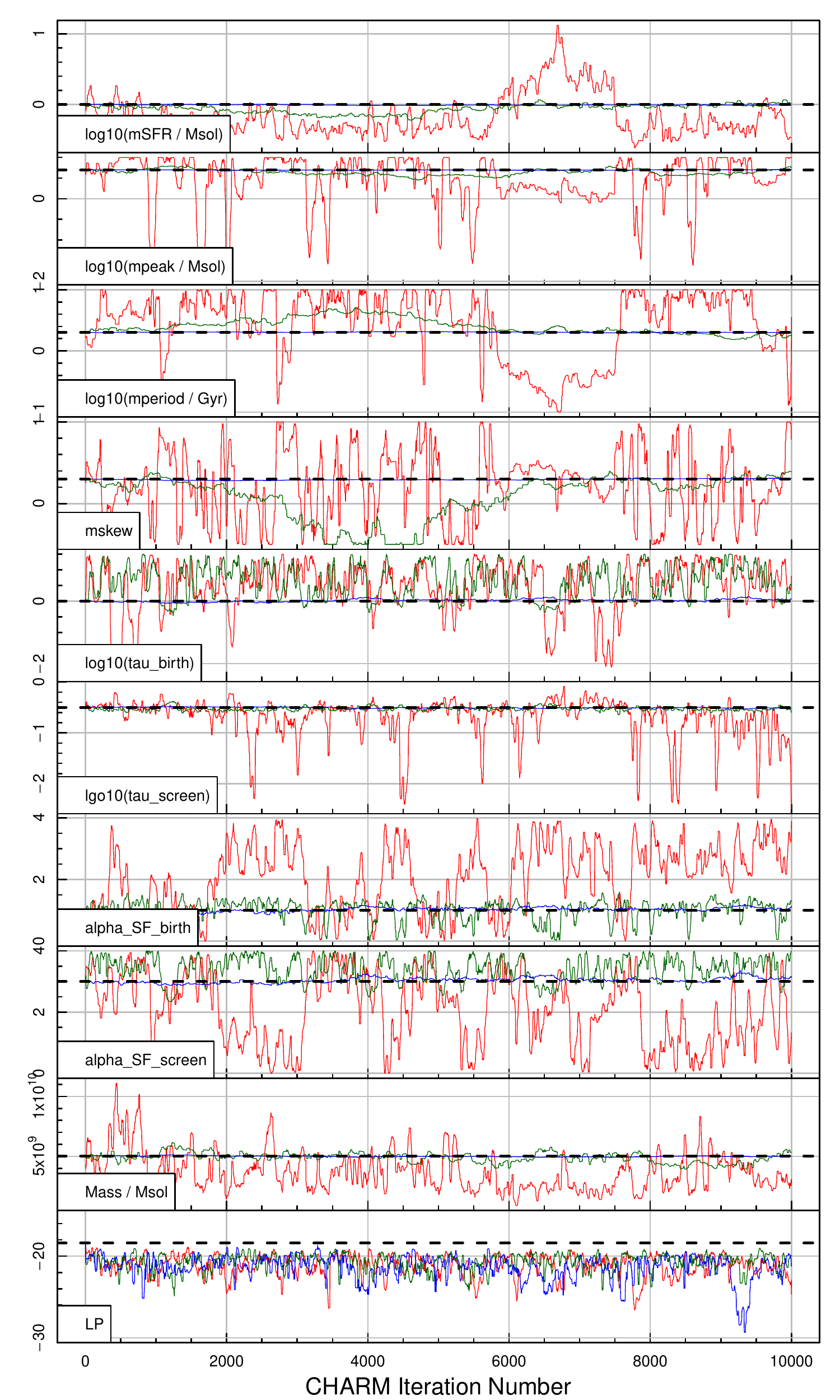}
\caption{An example multi-parameter MCMC model fit where the true parameters are shown as the horizontal dashed black line in each panel. The coloured lines represent different errors applied to the generated photometry: 0.01 mag (blue), 0.1 mag (green) and 0.5 mag (red). Interestingly, model degeneracies mean that there is little to no improvement in the quality of the posterior samples when moving from 0.01 mag to 0.1 mag errors, but a marked degradation in quality for 0.5 mag errors.}
\label{fig:Refit_worm}
\end{center}
\end{figure*}

In general the various posterior samples correctly explore the regime around the input parameters. This is especially true for the four star formation history parameters (the top four panels), where we only see large departures in the posterior samples when the photometric errors are 0.5 mag. The birth cloud parameters see the most departure in their posterior samples, which is largely down to the fact they contribute sub-dominant flux at all wavelengths in the spectrum with the parameter set chosen here (low recent SFR).

It is notable that the implied total stellar mass is in general very well behaved. We computed the standard deviation in the logged stellar mass for each of the posterior distributions, and compared this to the input photometric error. Figure \ref{fig:MCMC_Error} shows the result of this, where for the range of GAMA filters explored here we find $0.4 \sigma[mag] = \sigma[\log_{10} M*]$.

\begin{figure}
\begin{center}
\includegraphics[width=9cm]{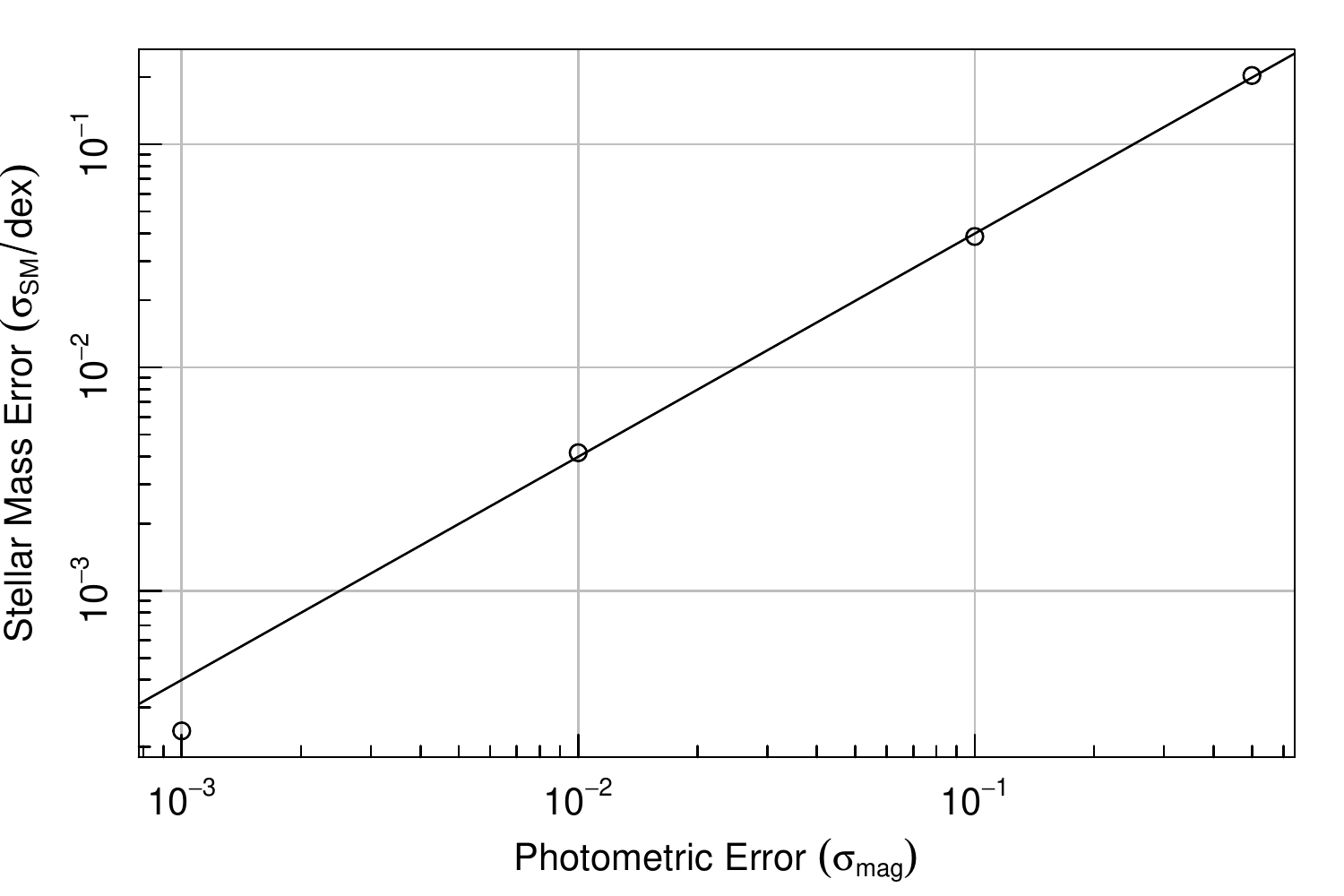}
\caption{Comparison of the input photometric error and the posterior sample implied stellar mass error (in dex). Applying a fixed relationship of $0.4 \sigma[mag] = \sigma[\log_{10} SM]$, we see excellent agreement between the observed stellar mass errors and our expectation.}
\label{fig:MCMC_Error}
\end{center}
\end{figure}

Even in the regime of choosing the correct model, the implication here is that 0.1 mag error or better photometry is required to remove highly erroneous posterior samples for our dust properties, although interestingly the implied star formation history and the final stellar mass are more robust. Assuming no systematic issues in the photometry and the correct model selection, we can assume to measure stellar mass to no better than $0.4 \sigma[mag]$ dex. If posterior sampling implied worse error than this, the assumption can be drawn that we are either not capturing the true photometric error (there are other systematics present not represented in the stated errors) and/or we are mis-specifying the model.

The issue of model mis-specification is a serious one, since it is largely undetectable via our model inversion. Strictly, Bayesian techniques can only inform you of the best parameter choice for a given model, but not whether than model is correct (or `better'). Even popular techniques such as the Bayesian Information Criterion (BIC) and the Akaike Information Criterion (AIC) are only qualitatively useful in this regard, and real data often fail many of the deep assumptions required to apply them meaningfully. In the next sub-Sections we will deliberately mis-specify the model being used for both an idealised model and for one generated from semi-analytic models (with its much noisier and complex star-formation history).

\subsubsection{Fitting Mis-Specified Model}

To test the impact of a slightly mis-specified model, we first create an SED for a {\tt snorm\_trunc} star formation model with a linearly growing metallicity history ({\tt massmap\_lin}). {The four dust parameters ($\tau$ and $\alpha$ for ISM screen and birth cloud dust) are fitted for the purposes of this test to make it more comparable to a real application of \prospect.} The result of fitting the generated photometry with { 0.01} mag errors with the correct model is shown in Figure \ref{fig:MCMC_mtrunc}, where we focus on the implied star formation history (top panel) and metallicity history (bottom panel).

\begin{figure}
\begin{center}
\includegraphics[width=9cm]{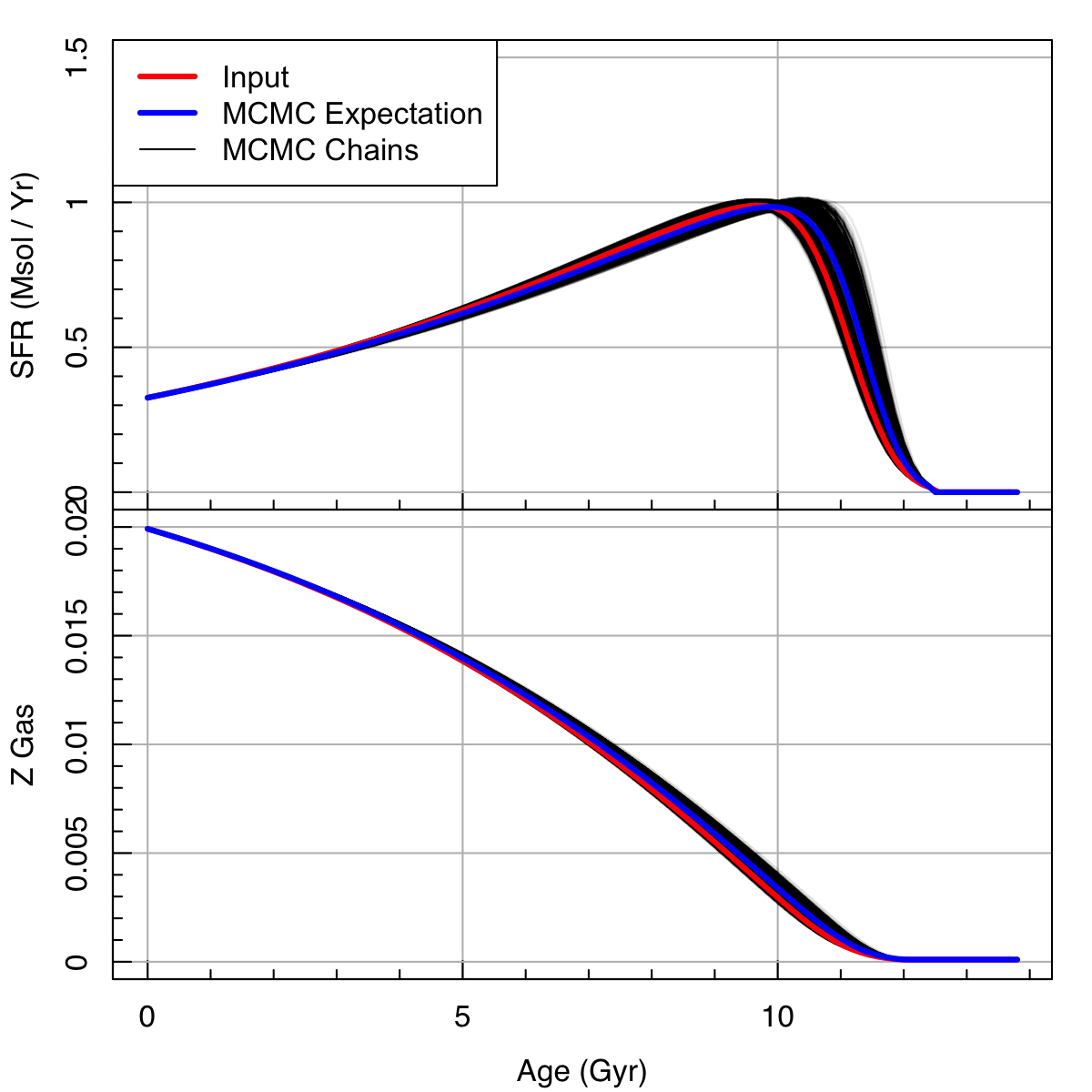}
\caption{Example of a full MCMC exploration of a target {\tt snorm\_trunc} model (faint grey lines, with solid blue line the final expectation) versus the intrinsic {\tt snorm\_trunc} model (solid red line) SFH (top) and Z history (bottom). As should be expected, the posterior samples are highly converged for recent times, but display larger uncertainty for the most ancient periods of star formation. However, the expectation of the samples proves to be a good representation of the true star formation history.}
\label{fig:MCMC_mtrunc}
\end{center}
\end{figure}

To ascertain the minimal impact of mis-specifying the model, we redo this fit, but using the {\tt snorm} star formation model, i.e.\ there is now no guaranteed truncation of the star formation rate at early times (every other aspect of the fitting is the same). The result of fitting this slightly mis-specified model is shown in Figure \ref{fig:MCMC_mtrunc2}. The resultant implied star formation history is much flatter than the input, and the chains are clearly less well converged at early times (there is a lot of variation between samples). We see a systematically less pronounced peak in the star formation rate, and due to the lack of truncation we see much higher star formation rates in the very early Universe.

\begin{figure}
\begin{center}
\includegraphics[width=9cm]{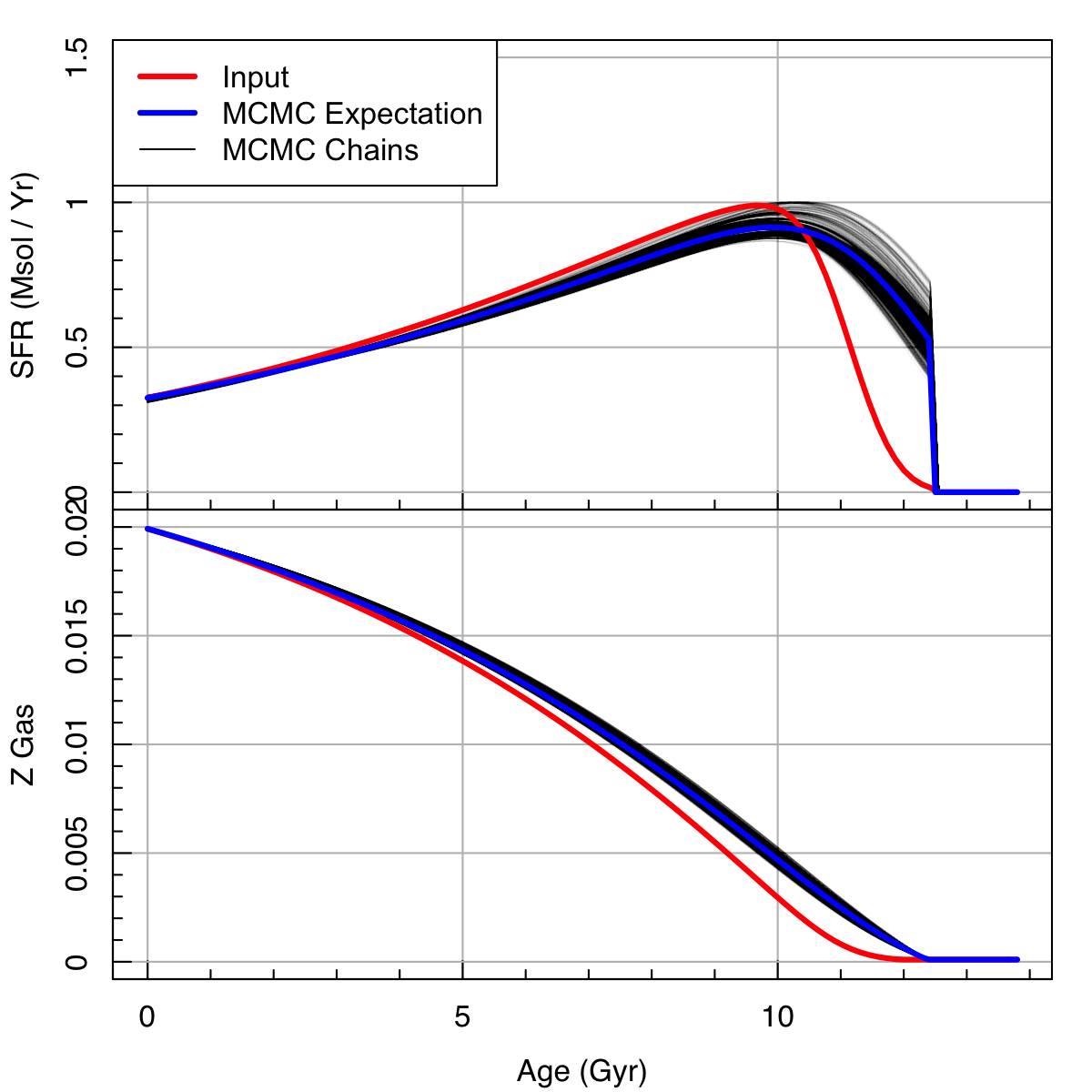}
\caption{Example of a full MCMC exploration of a target  {\tt snorm} model (faint grey lines, with solid blue line the final expectation) versus the intrinsic {\tt snorm\_trunc} model (solid red line) SFH (top) and Z history (bottom). The difference compared to Figure \ref{fig:MCMC_mtrunc} is that here we use the non-truncated form of the {\tt snorm} model. As such the sampling struggles to reproduce the sharp rise in the initial star formation rate, creating a much flatter star formation history than input.}
\label{fig:MCMC_mtrunc2}
\end{center}
\end{figure}

Figure \ref{fig:Model_Masses} shows the inferred stellar mass distribution for the correct model specification shown in Figure \ref{fig:MCMC_mtrunc} (red line) and the incorrect model specification shown in Figure \ref{fig:MCMC_mtrunc2}. Both agree within the stated confidence intervals, which is encouraging for making use of any inferred stellar masses. {In this case the mis-specified model is biased slightly higher and has a 87\% broader distribution in stellar mass samples, with some highly non-Normal features in the distribution (e.g.\ the pronounced tail to higher stellar masses).}

\begin{figure}
\begin{center}
\includegraphics[width=9cm]{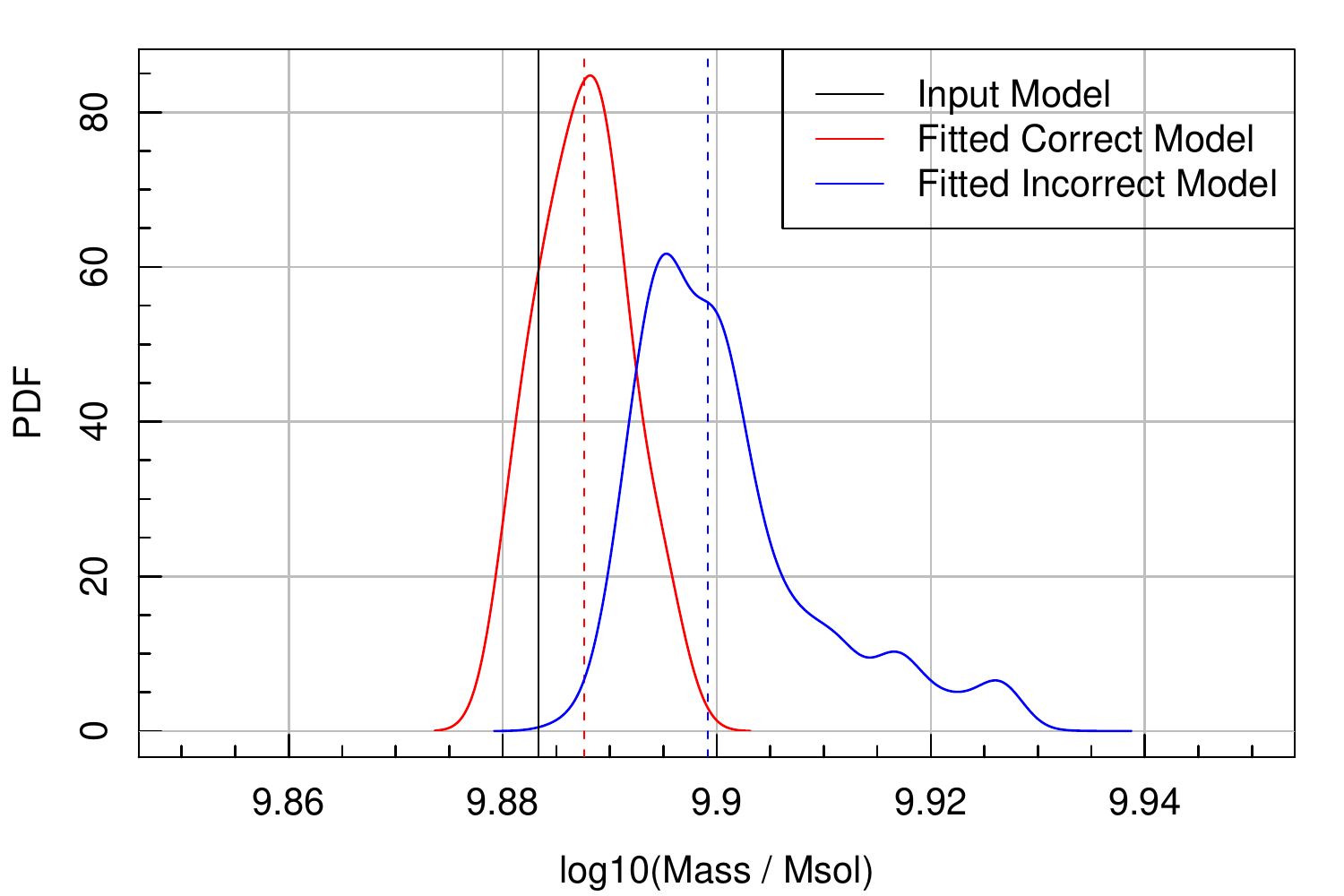}
\caption{Comparison of the input stellar mass formed (vertical black line) to that implied for posterior samples for the correctly specified model (red line) and an incorrectly specified model (blue line).}
\label{fig:Model_Masses}
\end{center}
\end{figure}

In this case we have not attempted to fit for the dust parameters, which in practice can form complex degeneracies with the star formation history parameters. These degeneracies will typically act to increase the spread in any posterior sampling of the stellar mass, but they can also systematically bias the stellar mass, e.g.\ increasing the amount of dust will mimic the effect of forming fewer recent stars (making the SED redder). Since mean stellar age, metallicity evolution and dust all have an impact on the $M/L$ of an SED fit, how these combine can be extremely complex in practice. In other words, mis-specifying the model can produce either systematically small or large stellar masses.

\subsubsection{Fitting Simulated Galaxies from Semi-Analytic Models}

As mentioned above, \prospect{} has already been used to create realistic multi-band SEDs for \shark{}. Given the star formation and metallicity evolution in Shark galaxies is highly complex, much more so than any of the simple parameterisations discussed in Section \ref{sec:SFH} and \ref{sec:ZH}, it is instructive to test how well we can recover galaxy properties using \prospect{} model inversion.

For the purposes of this test we extracted 571 Shark galaxies from the light cone presented in \citet{lago19} with an apparent magnitude limit of $r<19.8$ \citep[which is the limit of the GAMA survey;][]{lisk15}. These galaxies span stellar masses from $10^9$ -- $10^{12}$ \msol, and have physically calibrated dust sampling properties, as outlined in \citet{lago19} and \citet{tray20}. For these tests we applied a 0.1 mag error to all bands, simulating good quality but not exceptional high $S/N$ data.

To these data we fit a {\tt snorm\_trunc} star formation history model with a closed-box metallicity evolution, with free parameters for all of the galaxy dust properties. This makes the fitting process similar to what we would likely apply to real galaxy data, where it would not be reasonable to fix any of the dust properties. For a reference we fit an {\tt exp\_burst} SFH with a constant (but free) metallicity to the same data, mimicking the more typical parameterisation used in other codes \citep{dacu08, tayl11}.

To achieve a reasonable inversion we run our MCMC sampler using CHARM and $10^4$ samples, taking around 20 minutes per galaxy on a single core. Better parameter exploration is certainly possible with more posterior samples, or even possibly a different sampler (e.g.\ Hamiltonian Monte Carlo or No-U-Turns Sampling), however in preliminary tests CHARM appeared to be particularly well suited to the class of problem being tackled here.

\begin{figure*}
\begin{center}
\includegraphics[width=5.5cm]{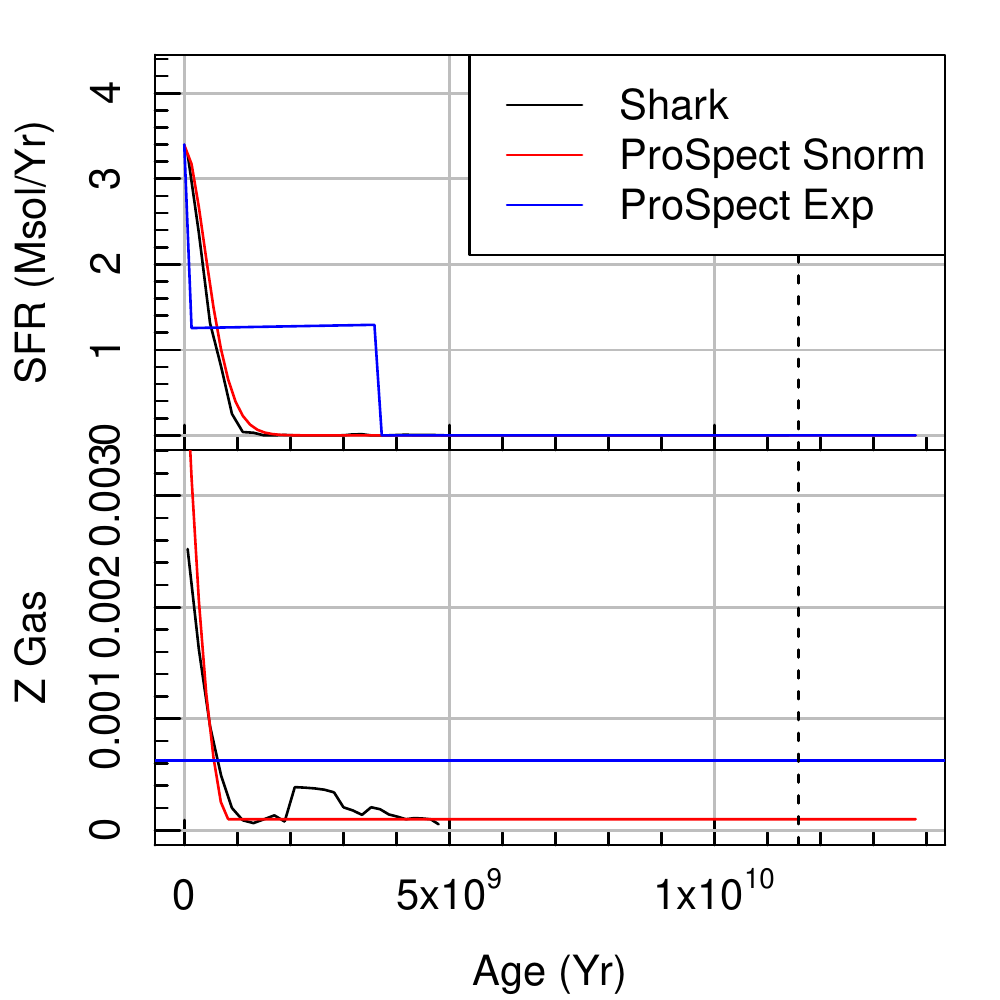}
\includegraphics[width=5.5cm]{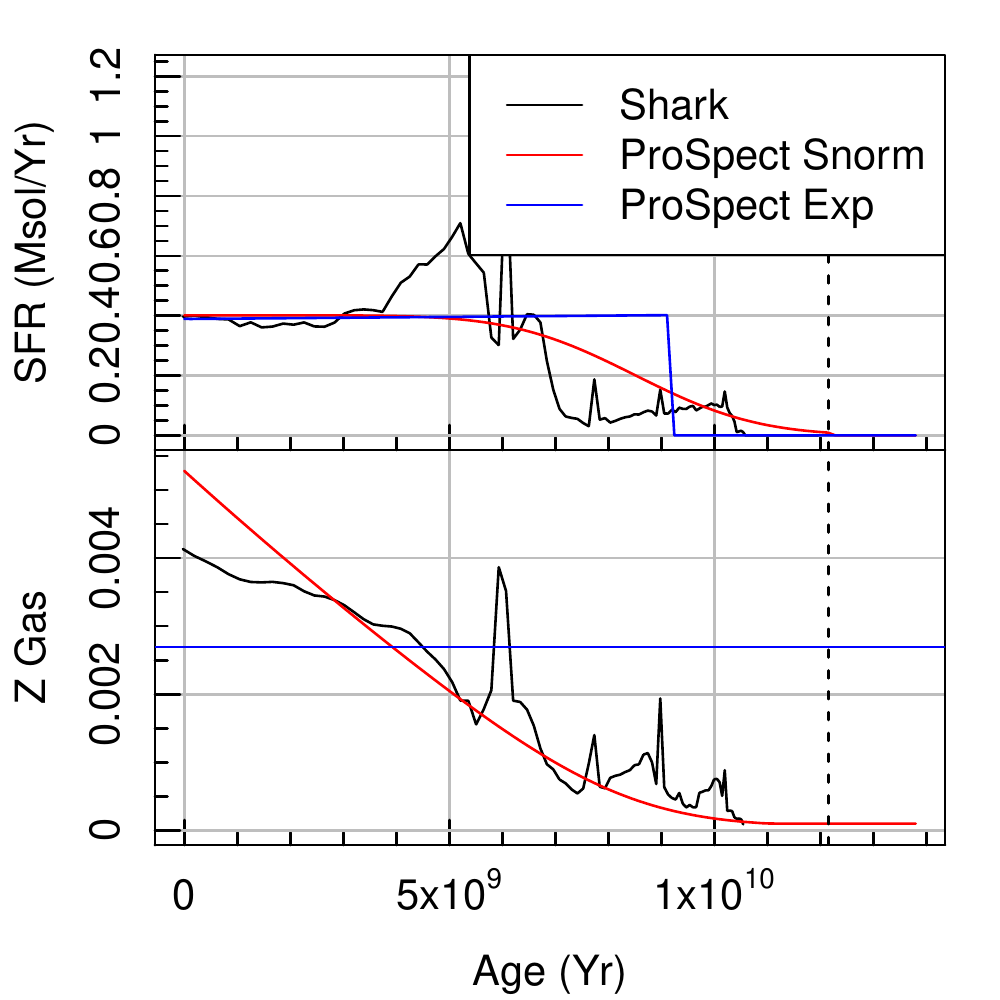}
\includegraphics[width=5.5cm]{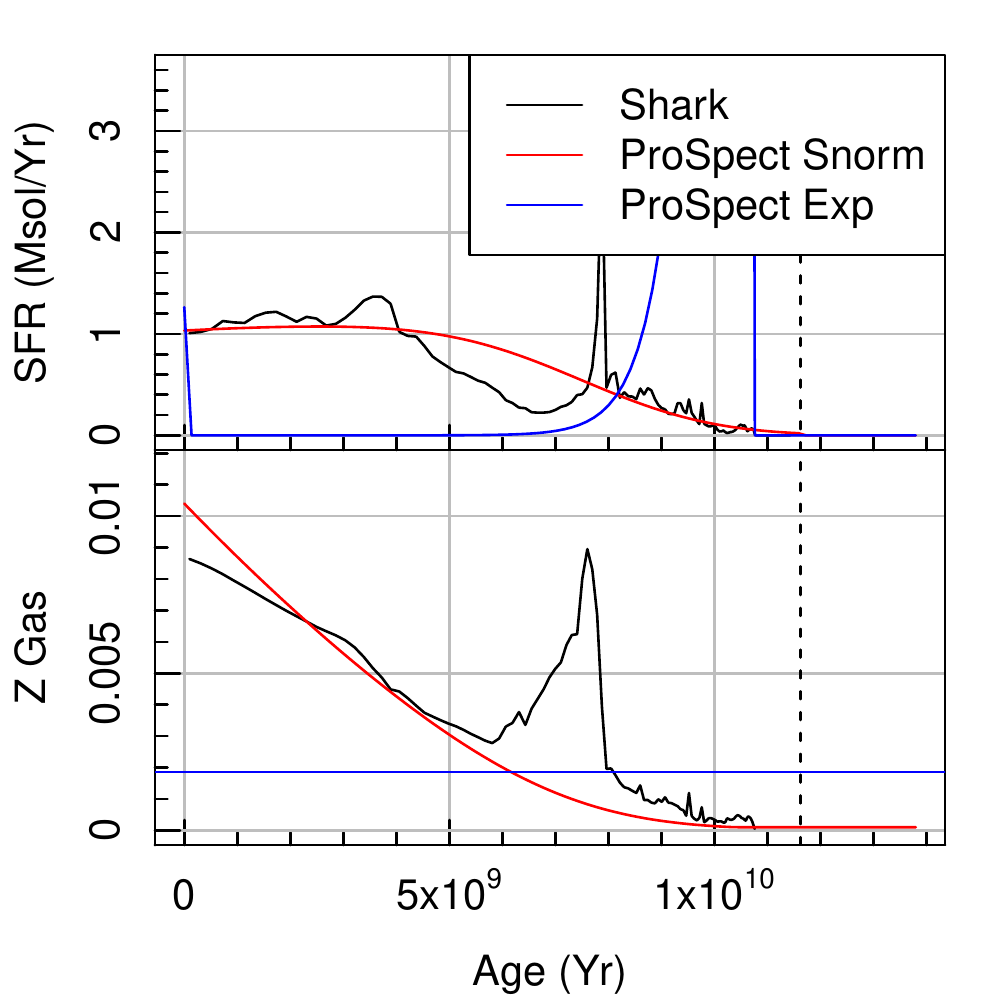}
\\
\includegraphics[width=5.5cm]{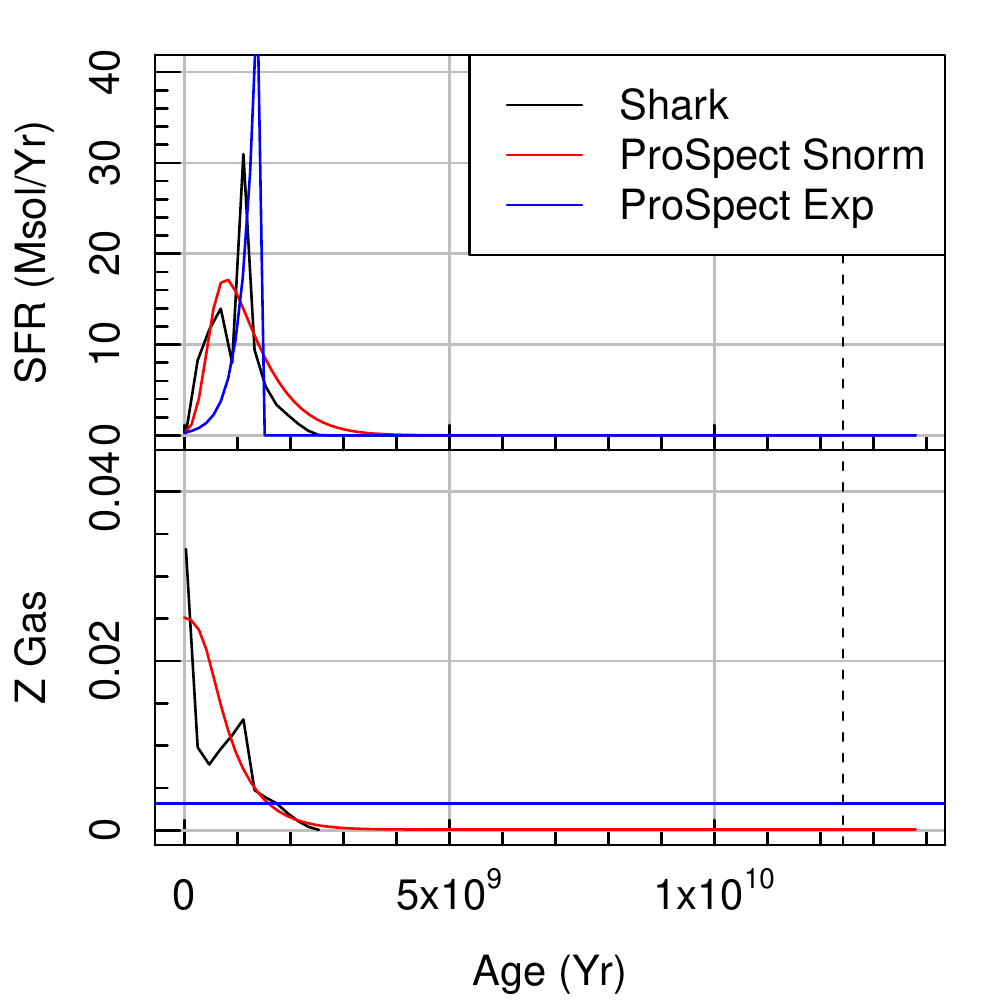}
\includegraphics[width=5.5cm]{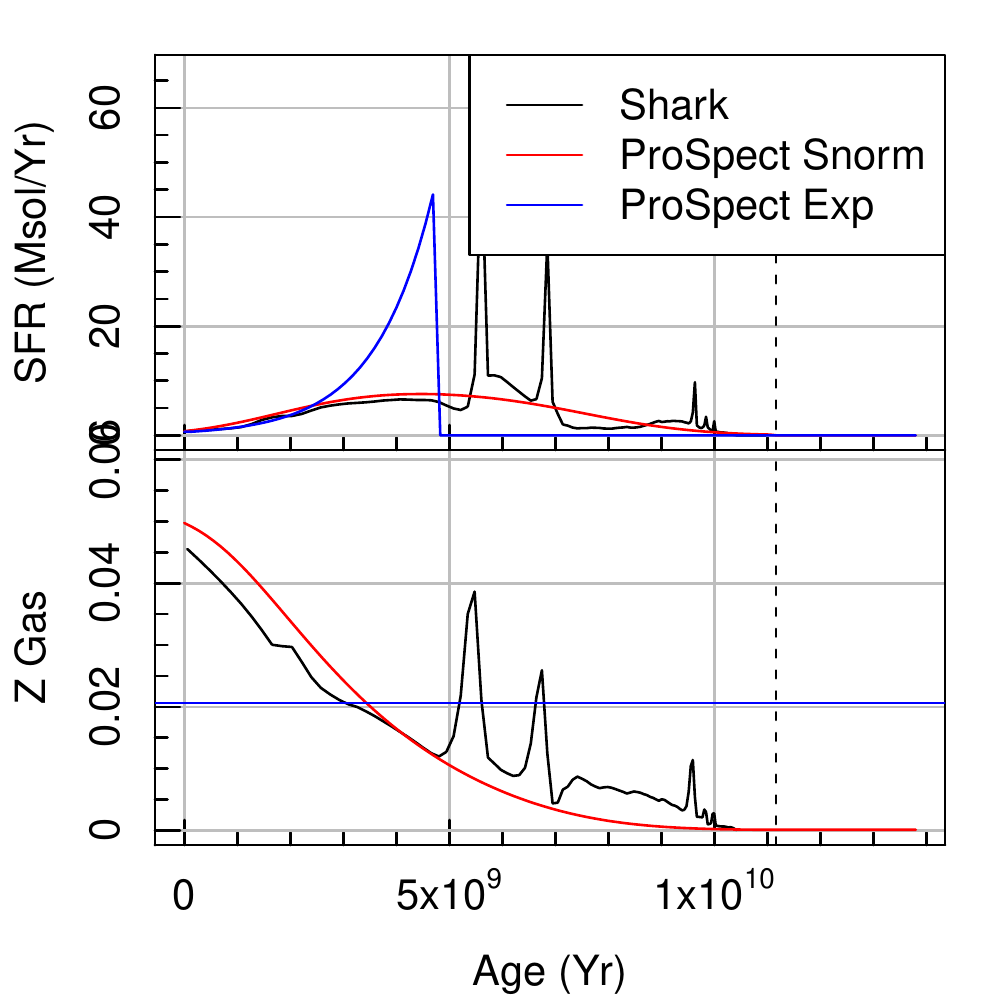}
\includegraphics[width=5.5cm]{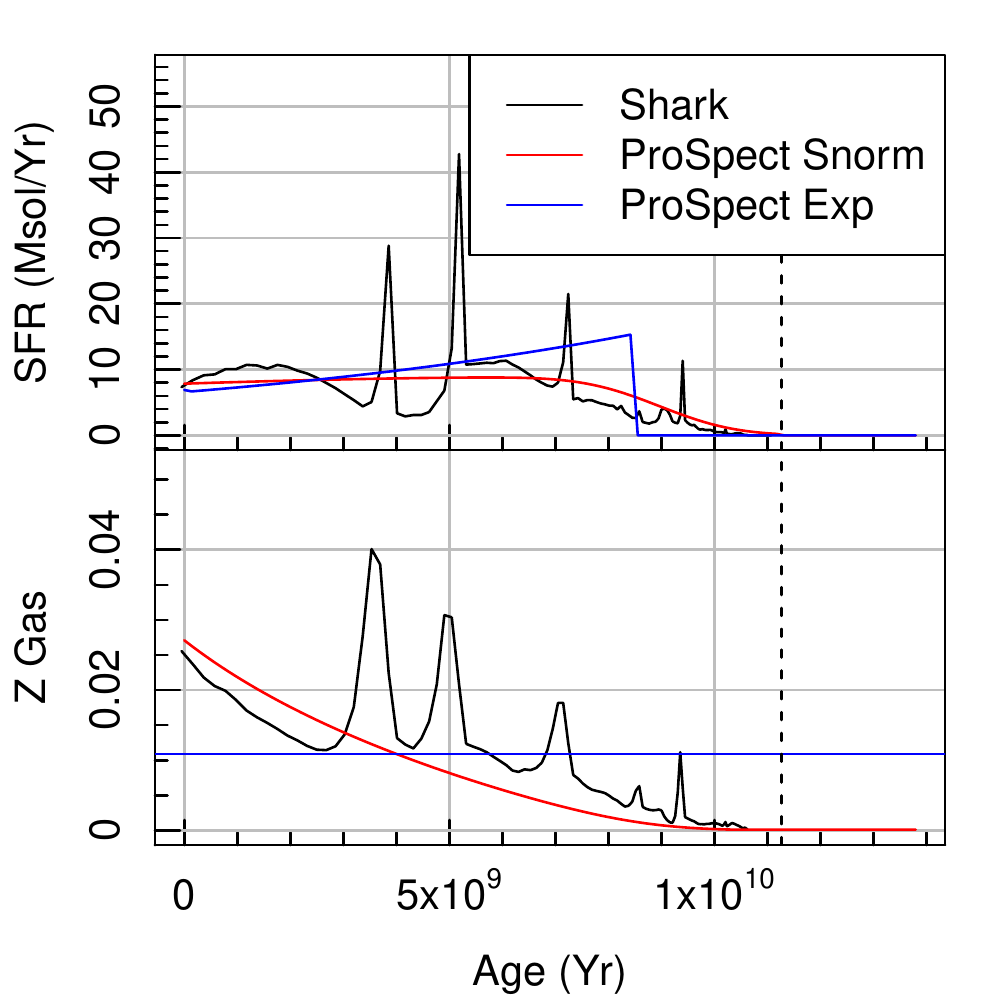}
\\
\includegraphics[width=5.5cm]{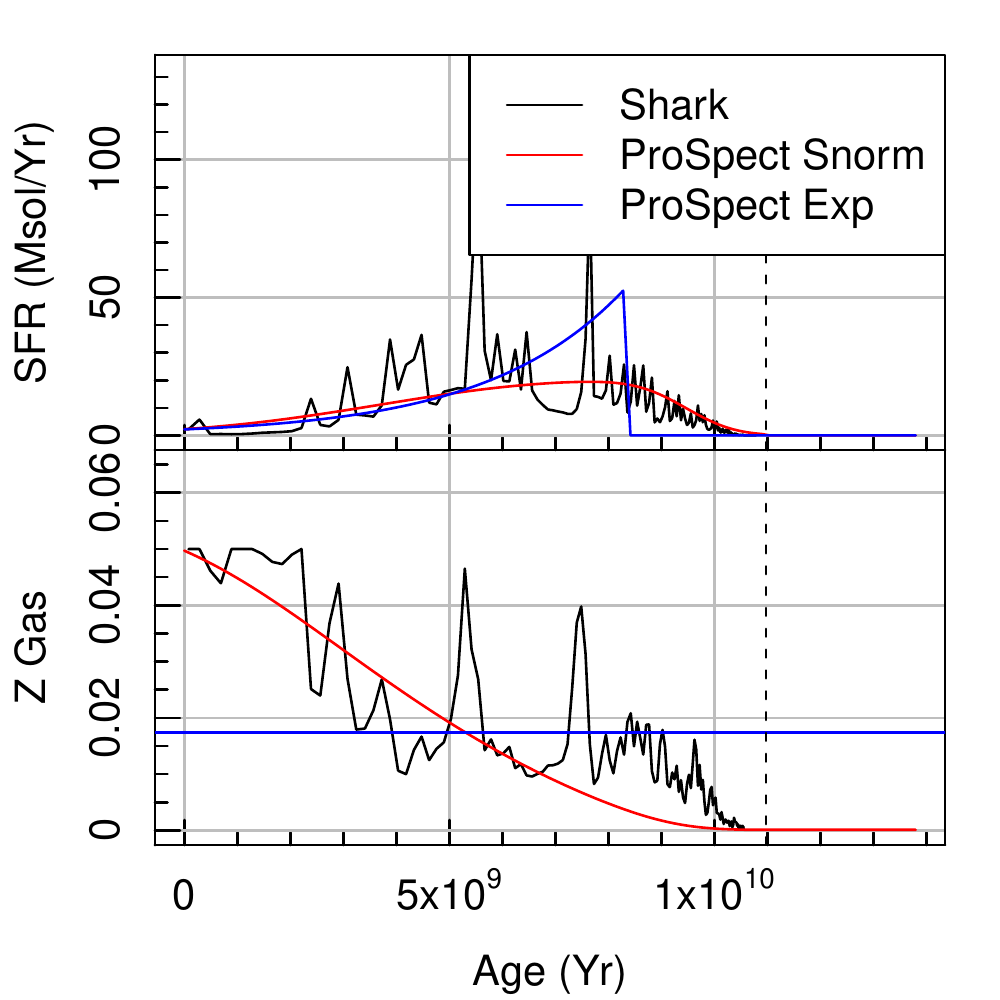}
\includegraphics[width=5.5cm]{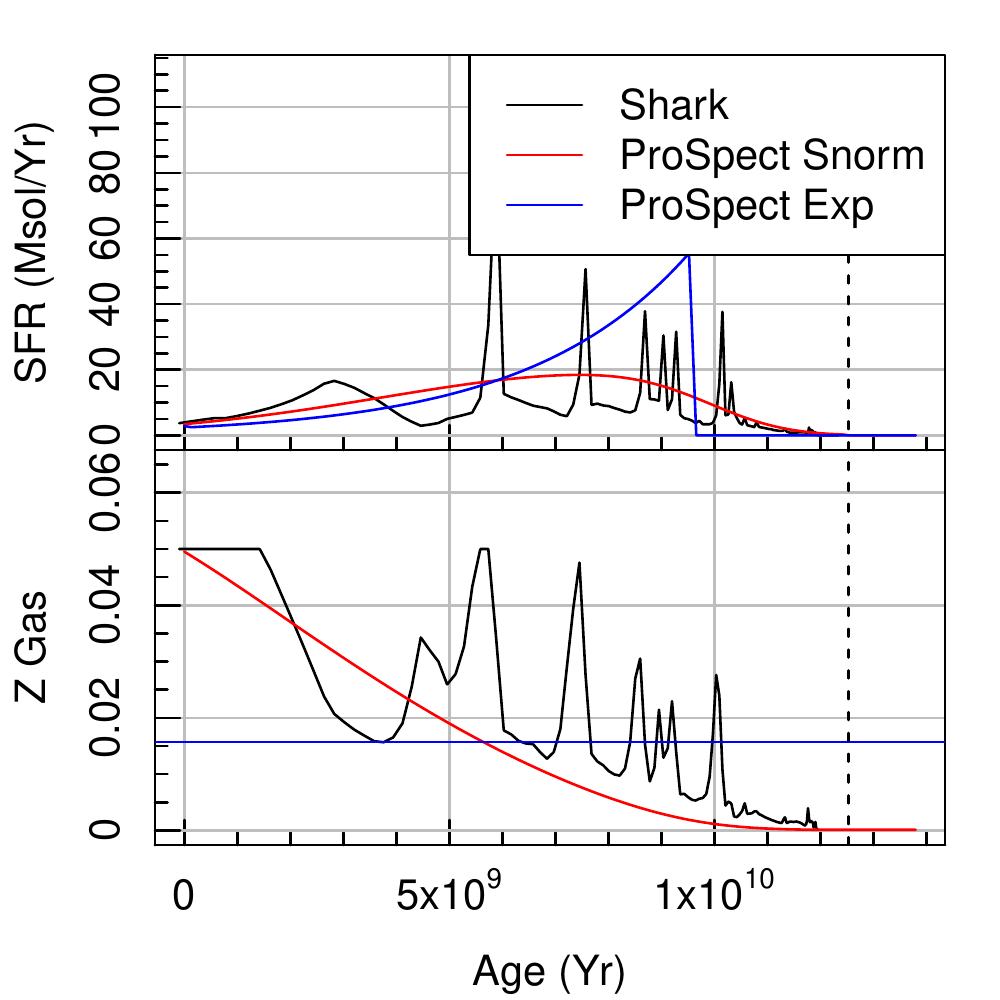}
\includegraphics[width=5.5cm]{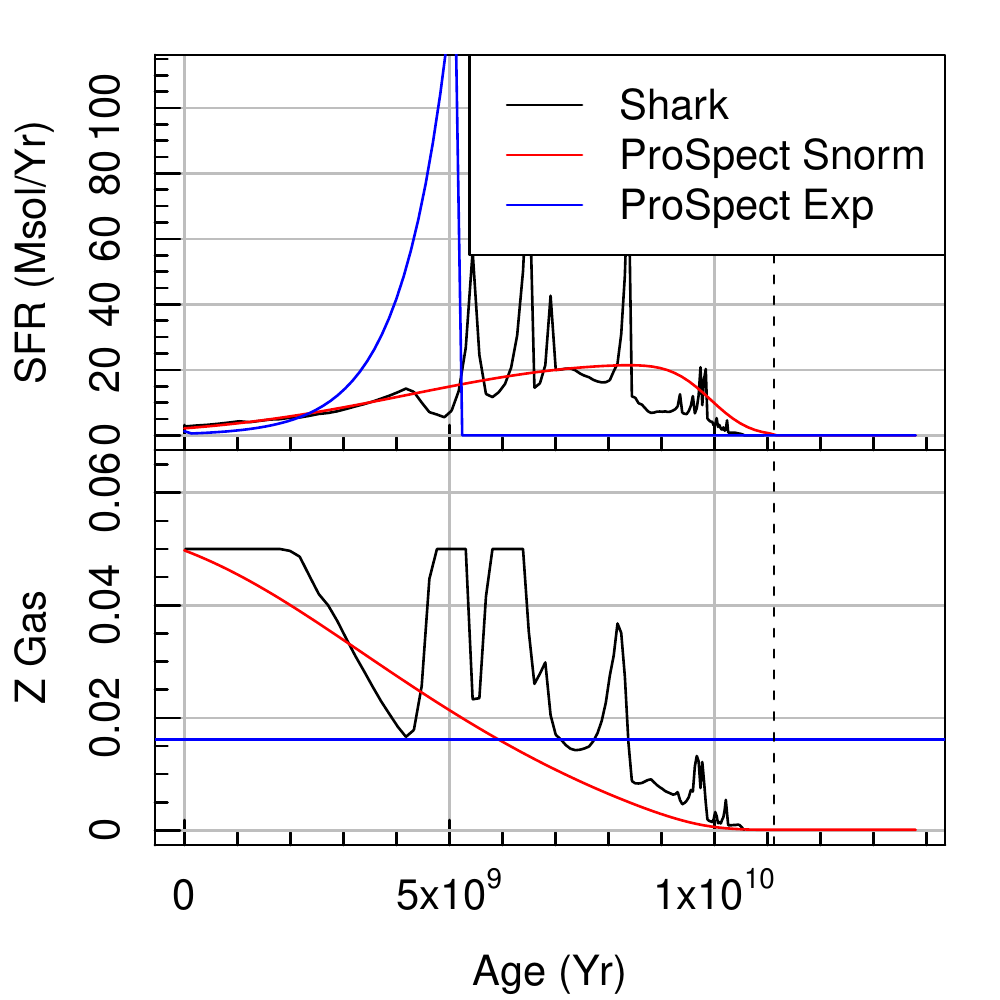}
\\
\caption{Example Shark SFHs and ZHs (black lines), with the inferred best-fit \prospect{} expectations over-plotted (red lines). { Here we have selected a diverse range of declining, constant and rising SFHs that qualitatively span the range observed in \shark.}}
\label{fig:Shark_fits}
\end{center}
\end{figure*}

Some example SFH and ZH inversions, broadly covering the full qualitative diversity seen in these SAMs, are shown in Figure \ref{fig:Shark_fits} (additional fits are shown in Appendix \ref{sec:app_shark} Figure \ref{fig:Shark_fits_extra}). It is clear that whilst it is not possible to capture fine detail in the SFH or ZH, e.g.\ extreme bursts etc, the general smoothed form is certainly recoverable with our combination of {\tt snorm\_trunc} model. This was less clearly expected with the metallicity evolution, where in detail a closed-box model with a free gas fraction does not capture all the complex inflows and outflows that are allowed to occur in \shark{}. In fact, it strongly suggests that a closed-box model does a reasonable job of capturing the longer term trends of metallicity evolution, smoothing over very violent discontinuities in the history.

These fits were also made using the linear CDF mapping form of the ZH. The results of these fits were extremely similar to the closed-box treatment, which is perhaps not surprising given the similarity of the two seen in Figure \ref{fig:gasfrac_Zgas}. For future fitting purposes either of the two metallicity treatments are likely to be reasonable, assuming the Zfinal parameter is left free to be fitted.

Figure \ref{fig:Shark_fits} also shows the result of using the {\tt exp\_burst} model. This is clearly much more restrictive, and adds spurious sharp features not present in the simulated SFHs. The ZH behaves as you might expect, settling on a value that is roughly the mean experienced over the star formation period of the galaxy. For this reason such a fitting approach is much closer to reflecting observed stellar metallicity, since this is what it reflects in principle.

\begin{figure}
\begin{center}
\includegraphics[width=9cm]{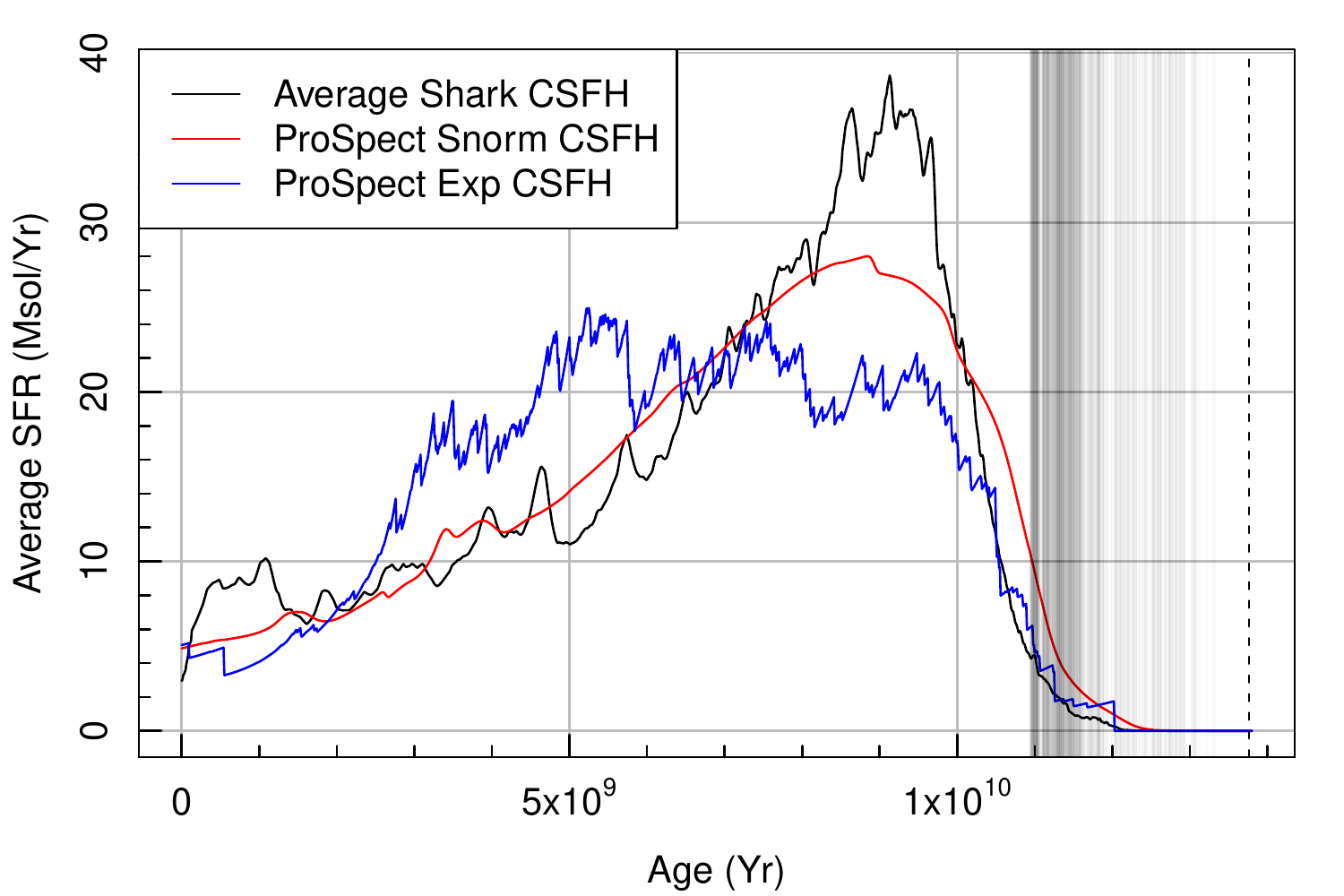}
\caption{Averaged stack of all 571 \shark{} SFHs to produce a pseudo cosmic SFH (black line), comparison \prospect{} {\tt snorm\_trunc} SFH stack (red line), and \prospect{} {\tt exp\_burst} SFH stack (blue line). The faint grey lines on the right indicate the individual maximum ages allowed for the galaxy SFHs, e.g. the age of the Universe at the redshift of simulated observation ($0 < z < 0.22$).}
\label{fig:Shark_CSFH}
\end{center}
\end{figure}

By stacking all of these individual SFHs, we can get an idea of how biased we are when inferring SFHs for a large range of galaxy types. This is shown in Figure \ref{fig:Shark_CSFH}, where we see a shape similar to the classic cosmic SFH \citep[CSFH; see][]{driv18}. In general the average form of the CSFH for a large range of galaxies covering two dex in stellar mass is well recovered in \prospect{} using {\tt snorm\_trunc}. The sharp peak at 9 Gyr is not as strongly defined, but the locus is in the same position. The tail to more ancient star formation does not drop off quite as steeply in \prospect{} either, suggesting our simple parameterisation of the SFH might not perfectly capture the manner of the truncation (or rather build up) at ancient epochs. This could be improved by adjusting the strength of the truncation in \prospect{}, or potentially even fitting for the truncation, but this is beyond the scope of this work. In general it should be noted that it is incredibly difficult to distinguish such ancient stellar populations since their colours are nearly identical, and they differ only in $M/L$. \prospect{} suggests that a more physically motivated manner of SFH and ZH parameterisation are the best route to successfully inferring such epochs.

In Figure \ref{fig:Shark_CSFH} we also see how the stack of {\tt exp\_burst} SFHs compares to the input model and our preferred {\tt snorm\_trunc} SFH. We find it is significantly biased to forming stars at younger ages compared to the known input, and does a systematically worse job of recovering the SFH \citep[differing in conclusion from][]{mitc13}. Given that a key requirement of an SFH parameterisation should be that it can recover the global SFH without undue bias, this Figure strongly demonstrates that we should prefer a {\tt snorm\_trunc} SFH model with a physically coupled (either closed-box or linearly mapped) ZH.

As well as inspecting some individually recovered SFHs and ZHs, we can check how well we recover the various inferred parameters of most interest for our preferred {\tt snorm\_trunc} SFH model: stellar mass and $Z_{\rm gas}$. Figure \ref{fig:Shark_global} is a summary of this recovery, showing how the stellar mass and $Z_{\rm gas}$ found through \prospect{} {\tt snorm\_trunc} inversion compared to the intrinsic properties known from \shark{}. The medians of these distributions are both very close to 1, showing excellent recovery. In the case of stellar mass, we see a 0.1 dex 1$\sigma$ spread in the stellar mass, which is comparable to many other stated literature levels \citep[e.g.][]{bell01, tayl11}. For reference the {\tt exp\_burst} SFH model displays nearer to 0.15 dex 1$\sigma$ spread in the stellar mass recovered, again giving preference to using a {\tt snorm\_trunc} SFH model. 

\begin{figure}
\begin{center}
\includegraphics[width=9cm]{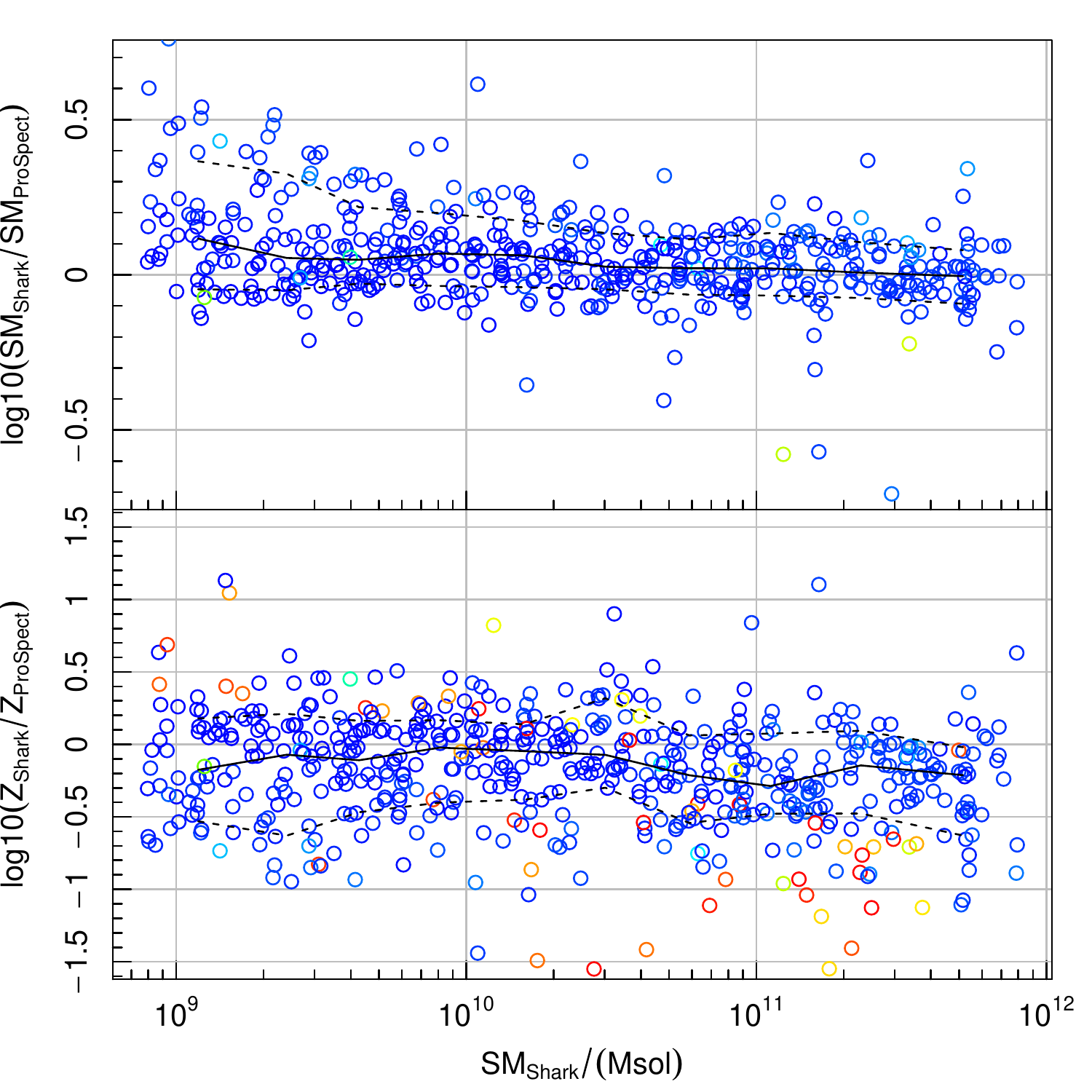}
\caption{Recovery of global properties as a function of the \shark{} stellar mass using \prospect{} with a {\tt snorm\_trunc} SFH model. The colouring reflects the log-likelihood of the fit, where bluer colours are better fits and redder colours are poorer fits in \prospect. The horizontal line shows the median recovery, and the dashed at the 1$\sigma$ limits.}
\label{fig:Shark_global}
\end{center}
\end{figure}

{ Broadly similar uncertainties were seen even when reducing the photometric error to 0.01 mag, suggesting this error should be interpreted as modelling error. As such, this 0.1 dex error in stellar mass is really an absolute best-case scenario when using a similar combination of broad band FUV--FIR filters as used here (which is much more comprehensive spectral coverage than available in typical observations).}

{In practice a real galaxy will have a more complicated dust attenuation and re-emission behaviour, and will not be perfectly modelled by a CF00 and D14 dust model. Even our multi-component model is highly simplified, and is attempting to apply a single best effort global behaviour to a phenomenon that is highly local. On top of this we are applying a single stellar population library (BC03 in this case) with a fixed IMF (C03-IMF), which only partially samples the full range of plausible star formation metallicity. Any stellar population library is also having to bolt together stellar isochrones (or similar) with stellar atmospheres, where the latter in particular are not well understood over all parameter space (observationally or theoretically).

Some of these issues can be broadly thought of as global systematics, e.g.\ if the IMF is universal but is not a C03-IMF then the effect will be to modify all stellar masses and mean stellar ages. If it is more bottom heavy (more low mass high mass-to-light stars) then stellar masses will be consistently higher than returned by \prospect, but if it is top heavy (more high mass low mass-to-light stars) then the stellar masses will be consistently lower. In practice the effect of IMF is likely to be complicated than this picture since it not definitively known to what degree IMF may vary over cosmic time as a function of metallicity, star formation rate, galaxy stellar mass and morphological type. These differential effects will cause both systematic and pseudo random errors in the implied properties of galaxies.

Putting these parts together, the suggested uncertainty of 0.1 dex should be considered a very optimistic lower limit on the stellar mass, since just imperfect modelling of the non-smooth SFH and ZH produce this component of the error budget. This is consistent with the view presented in the recent literature \citep[e.g.][]{tayl11}, and it is hard to reason how the error in stellar mass in any typical galaxy sample could be less than this given the diversity in formation scenarios that we now predict.}

\section{Minimal Photometric Stellar Masses}
\label{sec:photoSM}

Using light cones generated with \shark{} as part of the analysis of \citet{lago19}, we can use the combination of a physically motivated SAM and \prospect{} to create a new set of best-effort two or three band stellar mass predictors using optical \citep[Sloan filters,][]{fuku96} and near infrared colours \citep[VISTA filters,][]{hewe06}. The main motivation for such a set of predictors would be to produce efficient but reasonably accurate stellar masses in the regime where full SED fitting with \prospect{} is not possible, i.e.\ when we have very few available bands of observation.

In the analysis of \citet{lago19} and Bravo et al. (submitted) we find \shark{} and \prospect{} combined does an excellent job of recovering known luminosity and colour distributions out to $z\sim1.5$, so for this reason we limit our new stellar mass calibrations to work within this regime. One caveat is Bravo et al. (submitted) find a $\sim 0.3$ dex shift in the stellar mass colour distribution in \shark{} compared to GAMA, where \shark{} galaxies are too massive for their $g-i$ colour distribution. To account for this colour shift we apply a 0.3 dex adjustment to all \shark{} stellar masses. For clarity, all stellar mass predictors discussed below are producing {\it remaining} stellar mass (not formed).

\subsection{\shark{} Derived Observed Frame Apparent Magnitude Stellar Masses}

For high utility, we first create a set of stellar mass predictors using only observed frame quantities. We limit the analysis to $g-i$ and $g-r$ colours, and two colour predictors with the short filter fixed to $g$ and the long filter set to $r$ or longer. The advantage of such a calibration is that potentially complex $k$ (filter transform), $e$ (evolution) and $d$ (dust) corrections can be ignored by the user since these are all incorporated into the representative \shark{} model generated. These predictors will not be as good as the fully $k$ corrected rest frame absolute magnitude predictors discussed in the next Section, but in many cases they are still usefully accurate within the specified redshift range.

We parameterise the functional form of our predictor as follows:

\begin{eqnarray}
\label{eqn:SMapp}
B_{\textrm{ab}} = B_{\textrm{ap}} - \mu - 5\log_{10}h_{67.8} - 2.5\log_{10} (1+z) \\
\log_{10}(M / {\rm M} \odot) = \alpha B_{\textrm{ab}} + \beta C + \gamma z + \delta \pm \sigma,
\end{eqnarray}

{ \noindent where $B_{\textrm{ap}}$ refers to a target observed frame apparent magnitude; $B_{\textrm{ab}}$ is the pseudo absolute magnitude where we have corrected for luminosity distance and band pass stretching, $C$ refers to a target observed frame colour; $\mu$ is the usual distance modulus calculated assuming $H_0 = 67.8 \rm{km/s} / \rm{Mpc}$, where the user can then apply the desired correction for their target $H_0$ cosmology (where $h_{67.8} = H_0 / (67.8 \rm{km/s} / \rm{Mpc})$); and $\alpha$, $\beta$, $\gamma$, $\delta$ and $\sigma$ are all terms to be fit using \hyperfit. Note that the $\mu - 5\log_{10}h_{67.8}$ term can be replaced with the distance modulus using the desired $H_0$ explicitly, the formula above just highlights the $h$ dependency and makes it explicit that these stellar mass calibrations assume an $H_0 = 67.8 \rm{km/s} / \rm{Mpc}$ Universe. The use of the $\gamma z$ term, as opposed to any other redshift dependency, was arrived at via fitting a large range of parameterisations (including $\gamma \log_{10}(z)$, $\gamma \log_{10}(1+z)$ and $\gamma \log_{10}(1+z^2)$). This form produced the most linear stellar mass mappings with the smallest scatter, so was selected for our approximate relationships.}

When attempting each fit combination, the upper redshift limit is modified to ensure the expected scatter on the predictor remains under 0.3 dex (considered a reasonable maximum level of desirable stellar mass uncertainty). In some cases we are able to return a good predictor across the entire $0<z<1.5$ range we are exploring in this work, but others are limited to $0<z<1.0$ and $0<z<0.5$. Care must be taken if extending any of the predictors beyond these specified ranges. { To do the fitting we randomly extract $10^4$ galaxies in 0.5 dex bins between $10^8$ and $10^{12}$ \msol{} using the cone presented in \citet{lago19} with a Y$<23$ photometric and $z<1.5$ redshift selection, ensuring we achieve excellent stellar masses throughout a broad range of redshift {\it and} stellar mass. Figure \ref{fig:shark_masslims} presents the result of applying this selection to our \shark{} sample of galaxies, with the red line showing a limit of reliability for any approximate stellar masses generated (i.e.\ below this line there might be large biases in the generated stellar masses).}

\begin{figure}
\begin{center}
\includegraphics[width=9cm]{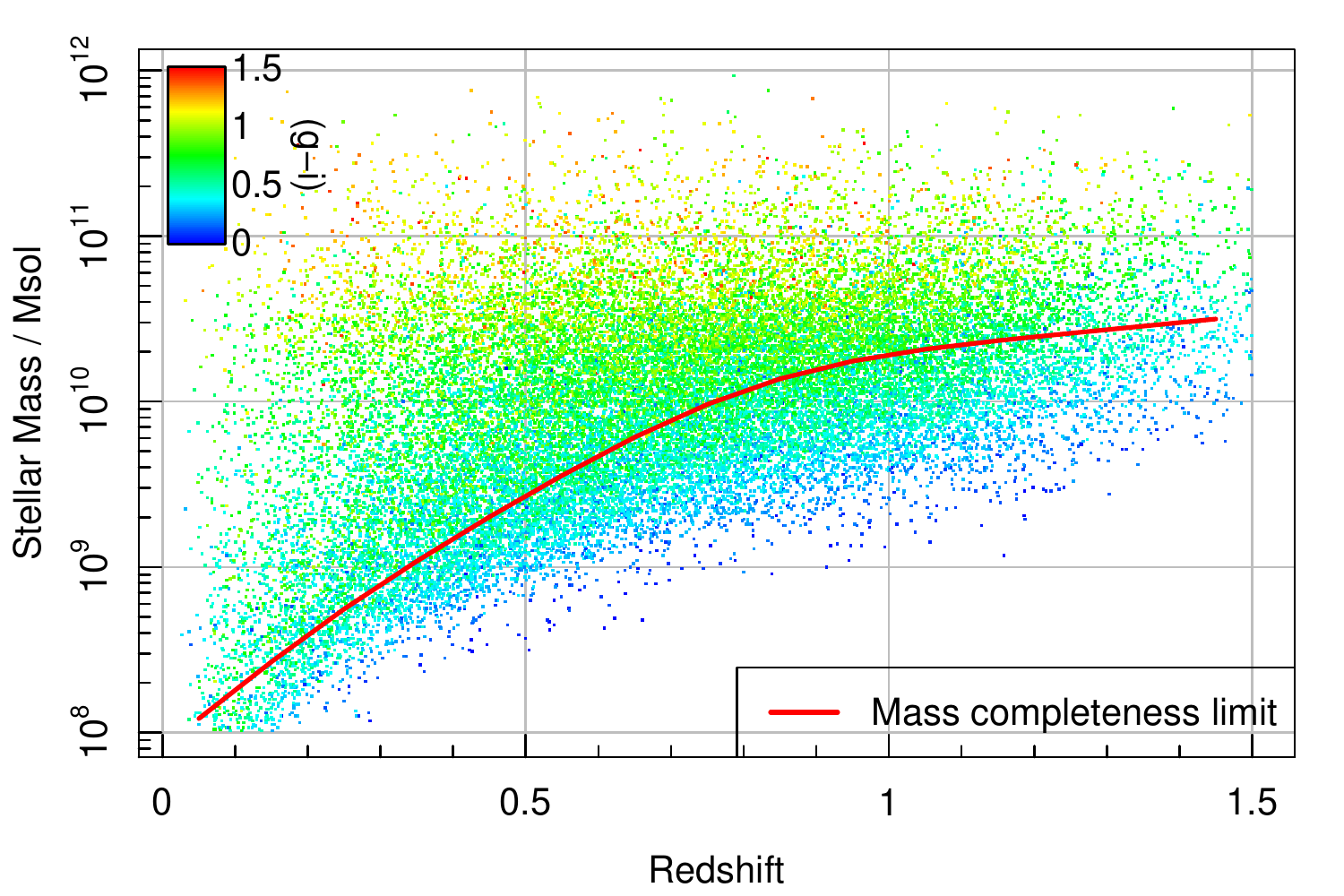}
\caption{\shark{} redshift versus stellar mass relationship using the selection applied for calculating approximate photometric stellar masses. Care should be taking when generating approximate stellar masses below the red line (showing the mass completeness limit) since these will be outside the range of mass complete data, and there may be biases not captured by the linear fitting process.}
\label{fig:shark_masslims}
\end{center}
\end{figure}

\begin{table}
\centering
\begin{tabular}{llrrrrrl}
  \hline
$B$ & $C$ & $\alpha$ & $\beta$ & $\gamma$ & $\delta$ & $\sigma$ & $z$ max \\ 
  \hline
  g & (g-i) & -0.443 & 1.032 & -1.514 & 0.547 & 0.275 & 1.0 \\ 
  r & (g-i) & -0.423 & 0.859 & -2.098 & 0.932 & 0.245 & 1.0 \\ 
  i & (g-i) & -0.443 & 0.626 & -1.893 & 0.604 & 0.238 & 1.0 \\ 
  z & (g-i) & -0.459 & 0.446 & -1.617 & 0.319 & 0.231 & 1.0 \\ 
  Y & (g-i) & -0.506 & 0.215 & -1.230 & -0.520 & 0.282 & 1.0 \\ 
  J & (g-i) & -0.483 & 0.155 & -1.143 & -0.099 & 0.250 & 1.5 \\ 
  H & (g-i) & -0.461 & 0.098 & -1.086 & 0.329 & 0.217 & 1.5 \\ 
  Ks & (g-i) & -0.444 & 0.074 & -1.101 & 0.683 & 0.199 & 1.5 \\ 
  W1 & (g-i) & -0.450 & 0.014 & -1.564 & 1.010 & 0.207 & 1.5 \\ 
  W2 & (g-i) & -0.436 & 0.062 & -1.573 & 1.418 & 0.200 & 1.5 \\ 
  S1 & (g-i) & -0.450 & 0.019 & -1.594 & 1.051 & 0.202 & 1.5 \\ 
  S2 & (g-i) & -0.437 & 0.061 & -1.576 & 1.379 & 0.197 & 1.5 \\
   \hline
\end{tabular}
\caption{Formula terms for \shark-derived observed frame apparent magnitude photometry using $g-i$ colours. See Equation \ref{eqn:SMapp}.}
\label{tab:app_gi}
\end{table}

\begin{table}
\centering
\begin{tabular}{llrrrrrl}
  \hline
$B$ & $C$ & $\alpha$ & $\beta$ & $\gamma$ & $\delta$ & $\sigma$ & $z$ max \\ 
  \hline
  g & (g-r) & -0.404 & 1.672 & -3.458 & 1.491 & 0.166 & 0.5 \\ 
  r & (g-r) & -0.405 & 1.277 & -3.526 & 1.487 & 0.169 & 0.5 \\ 
  i & (g-r) & -0.405 & 1.061 & -3.199 & 1.444 & 0.152 & 0.5 \\ 
  z & (g-r) & -0.405 & 0.918 & -2.973 & 1.432 & 0.139 & 0.5 \\ 
  Y & (g-r) & -0.401 & 0.852 & -2.885 & 1.485 & 0.138 & 0.5 \\ 
  J & (g-r) & -0.567 & -0.004 & -1.618 & -1.401 & 0.321 & 1.5 \\ 
  H & (g-r) & -0.501 & 0.007 & -1.289 & -0.274 & 0.244 & 1.5 \\ 
  Ks & (g-r) & -0.483 & -0.030 & -1.326 & 0.109 & 0.227 & 1.5 \\ 
  W1 & (g-r) & -0.471 & -0.069 & -1.712 & 0.728 & 0.220 & 1.5 \\ 
  W2 & (g-r) & -0.466 & -0.031 & -1.769 & 1.017 & 0.220 & 1.5 \\ 
  S1 & (g-r) & -0.468 & -0.052 & -1.723 & 0.800 & 0.213 & 1.5 \\ 
  S2 & (g-r) & -0.466 & -0.028 & -1.759 & 0.997 & 0.215 & 1.5 \\
   \hline
\end{tabular}
\caption{Formula terms for \shark-derived observed frame apparent magnitude photometry using $g-r$ colours. See Equation \ref{eqn:SMapp}.}
\label{tab:app_gr}
\end{table}

\begin{table}
\centering
\begin{tabular}{llrrrrrl}
  \hline
$B$ & $C$ & $\alpha$ & $\beta$ & $\gamma$ & $\delta$ & $\sigma$ & $z$ max \\ 
  \hline
  z & (g-z) & -0.459 & 0.310 & -1.093 & 0.234 & 0.269 & 1.5 \\ 
  Y & (g-Y) & -0.440 & 0.261 & -1.064 & 0.574 & 0.212 & 1.5 \\ 
  J & (g-J) & -0.423 & 0.194 & -1.007 & 0.899 & 0.194 & 1.5 \\ 
  H & (g-H) & -0.404 & 0.143 & -0.960 & 1.275 & 0.173 & 1.5 \\ 
  Ks & (g-Ks) & -0.393 & 0.117 & -1.005 & 1.546 & 0.164 & 1.5 \\ 
  W1 & (g-W1) & -0.393 & 0.093 & -1.454 & 1.976 & 0.175 & 1.5 \\ 
  W2 & (g-W2) & -0.387 & 0.107 & -1.553 & 2.329 & 0.174 & 1.5 \\ 
  S1 & (g-S1) & -0.395 & 0.093 & -1.500 & 1.993 & 0.173 & 1.5 \\ 
  S2 & (g-S2) & -0.389 & 0.105 & -1.560 & 2.263 & 0.172 & 1.5 \\ 
   \hline
\end{tabular}
\caption{Formula terms for \shark-derived observed frame apparent magnitude photometry using two bands. Note $i$ \& $(g-i)$ and $r$ \& $(g-r)$ fits are presented in tables \ref{tab:app_gi} and \ref{tab:app_gr} respectively. See Equation \ref{eqn:SMapp}.}
\label{tab:app_2b}
\end{table}

Tables \ref{tab:app_gi}, \ref{tab:app_gr} and \ref{tab:app_2b} present the formula coefficients for the optimal observed frame apparent magnitude $g-i$, $g-r$, and two band fits respectively. The best $g-i$ fit (i.e.\ producing the least stellar mass predicted scatter, $\sigma$ in Table \ref{tab:app_gi}) uses the Ks bands for $B$, although the W1, W2 and S1 photometry is almost as successful at recovering approximate stellar mass. In these cases we achieve close to 0.16 dex scatter in stellar mass, which compares to 0.1 dex when using the full \prospect{} fitting machinery. The best full redshift range $g-r$ fit (Table \ref{tab:app_gr}) uses S1 for the $B$ reference photometry, with Ks, W1, W2 and S2 close behind. We also find excellent recovery using $z$ and Y, but this is over a substantially more restrictive redshift range of utility ($0<z<0.5$).

The optimal two band stellar mass recovery is perhaps of most practical utility since it requires much less data to be collected. The only reasonable purely optical recovery uses $z$ for $B$, but this has the worst stellar mass scatter by some margin (0.269 dex). We find substantial improvement when switching the $B$ choice to one of the NIR filters, in particular Ks, which produces only 0.164 dex of scatter over the entire redshift range. It is not too much worse to use W1 for $B$, where the scatter increases to 0.175 dex. Given the all sky coverage available for WISE, this suggests that reasonable stellar masses can be generated with the addition of optical $g$ band data, with no additional need for sophisticated $k$, $e$ or $d$ corrections from $0<z<1.5$.

\subsection{\shark{} Derived Rest Frame Absolute Magnitude Stellar Masses}

For optimal stellar masses, we next create a set of stellar mass predictors using rest frame quantities. We limit the analysis to $g-i$ and $g-r$ colours, and two colour predictors with the short filter fixed to $g$ and the long filter set to $z$ or longer. Since we are now using rest frame absolute magnitudes, the user must be careful to apply the proper $k$ corrections to their data, however $e$ and $d$ corrections are still unnecessary and are captured by our \shark{} generative modelling and \hyperfit{} fitting.

We parameterise the functional form of our predictor as follows:

\begin{equation}
\label{eqn:SMabs}
\log_{10}(M / \rm{M}\odot) = \alpha B + \beta C + \gamma z + \delta \pm \sigma
\end{equation}

\noindent where $B$ refers to a target rest frame absolute magnitude, $C$ refers to a target absolute frame colour, and $\alpha$, $\beta$, $\gamma$, $\delta$ and $\sigma$ are all terms to be fit using \hyperfit.

\begin{table}
\centering
\begin{tabular}{llrrrrrl}
  \hline
B & C & $\alpha$ & $\beta$ & $\gamma$ & $\delta$ & $\sigma$ & $z$ max \\ 
  \hline
  g & (g-i) & -0.388 & 1.267 & -0.110 & 1.395 & 0.116 & 1.5 \\ 
  r & (g-i) & -0.385 & 1.006 & -0.097 & 1.443 & 0.118 & 1.5 \\ 
  i & (g-i) & -0.385 & 0.889 & -0.097 & 1.447 & 0.117 & 1.5 \\ 
  z & (g-i) & -0.379 & 0.793 & -0.095 & 1.561 & 0.113 & 1.5 \\ 
  Y & (g-i) & -0.373 & 0.718 & -0.099 & 1.688 & 0.109 & 1.5 \\ 
  J & (g-i) & -0.369 & 0.654 & -0.109 & 1.797 & 0.107 & 1.5 \\ 
  H & (g-i) & -0.363 & 0.592 & -0.119 & 1.908 & 0.104 & 1.5 \\ 
  Ks & (g-i) & -0.358 & 0.578 & -0.129 & 2.123 & 0.104 & 1.5 \\ 
  W1 & (g-i) & -0.345 & 0.679 & -0.128 & 2.542 & 0.121 & 1.5 \\ 
  W2 & (g-i) & -0.317 & 0.838 & -0.056 & 3.085 & 0.153 & 1.5 \\ 
  S1 & (g-i) & -0.340 & 0.694 & -0.120 & 2.635 & 0.125 & 1.5 \\ 
  S2 & (g-i) & -0.329 & 0.752 & -0.098 & 2.924 & 0.137 & 1.5 \\ 
   \hline
\end{tabular}
\caption{Formula terms for \shark-derived rest frame absolute magnitude photometry using $g-i$ colours. See Equation \ref{eqn:SMabs}.}
\label{tab:abs_gi}
\end{table}

\begin{table}
\centering
\begin{tabular}{llrrrrrl}
  \hline
B & C & $\alpha$ & $\beta$ & $\gamma$ & $\delta$ & $\sigma$ & $z$ max \\ 
  \hline
  g & (g-r) & -0.388 & 1.851 & -0.095 & 1.423 & 0.126 & 1.5 \\ 
  r & (g-r) & -0.382 & 1.503 & -0.071 & 1.519 & 0.127 & 1.5 \\ 
  i & (g-r) & -0.380 & 1.344 & -0.067 & 1.542 & 0.125 & 1.5 \\ 
  z & (g-r) & -0.373 & 1.211 & -0.058 & 1.665 & 0.120 & 1.5 \\ 
  Y & (g-r) & -0.367 & 1.110 & -0.069 & 1.787 & 0.115 & 1.5 \\ 
  J & (g-r) & -0.363 & 1.023 & -0.075 & 1.903 & 0.112 & 1.5 \\ 
  H & (g-r) & -0.357 & 0.938 & -0.084 & 2.016 & 0.109 & 1.5 \\ 
  Ks & (g-r) & -0.351 & 0.920 & -0.092 & 2.230 & 0.108 & 1.5 \\ 
  W1 & (g-r) & -0.336 & 1.088 & -0.082 & 2.662 & 0.124 & 1.5 \\ 
  W2 & (g-r) & -0.306 & 1.361 & 0.003 & 3.231 & 0.159 & 1.5 \\ 
  S1 & (g-r) & -0.331 & 1.118 & -0.071 & 2.763 & 0.128 & 1.5 \\ 
  S2 & (g-r) & -0.319 & 1.221 & -0.042 & 3.063 & 0.142 & 1.5 \\ 
   \hline
\end{tabular}
\caption{Formula terms for \shark-derived rest frame absolute magnitude photometry using $g-r$ colours. See Equation \ref{eqn:SMabs}.}
\label{tab:abs_gr}
\end{table}

\begin{table}
\centering
\begin{tabular}{llrrrrrl}
  \hline
B & C & $\alpha$ & $\beta$ & $\gamma$ & $\delta$ & $\sigma$ & $z$ max \\ 
  \hline
  z & (g-z) & -0.372 & 0.624 & -0.103 & 1.700 & 0.108 & 1.5 \\ 
  Y & (g-Y) & -0.360 & 0.489 & -0.111 & 1.961 & 0.102 & 1.5 \\ 
  J & (g-J) & -0.356 & 0.393 & -0.135 & 2.107 & 0.101 & 1.5 \\ 
  H & (g-H) & -0.350 & 0.316 & -0.155 & 2.232 & 0.101 & 1.5 \\ 
  Ks & (g-Ks) & -0.346 & 0.286 & -0.188 & 2.520 & 0.106 & 1.5 \\ 
  W1 & (g-W1) & -0.376 & 0.204 & -0.414 & 2.451 & 0.158 & 1.5 \\ 
  W2 & (g-W2) & -0.460 & -0.005 & -0.905 & 1.316 & 0.242 & 1.5 \\ 
  S1 & (g-S1) & -0.385 & 0.171 & -0.468 & 2.347 & 0.168 & 1.5 \\ 
  S2 & (g-S2) & -0.417 & 0.086 & -0.660 & 2.034 & 0.200 & 1.5 \\ 
   \hline
\end{tabular}
\caption{Formula terms for \shark-derived rest frame absolute magnitude photometry using two bands. Note $i$ \& $(g-i)$ and $r$ \& $(g-r)$ fits are presented in tables \ref{tab:abs_gi} and \ref{tab:abs_gr} respectively. See Equation \ref{eqn:SMabs}.}
\label{tab:abs_2b}
\end{table}

Tables \ref{tab:abs_gi}, \ref{tab:abs_gr} and \ref{tab:abs_2b} present the formula coefficients for the optimal rest frame absolute magnitude $g-i$, $g-r$, and two band fits respectively. As should be expected given the additional effort required to estimate true rest frame absolute magnitudes (effectively part of what \prospect{} does internally when fitting SEDs), the quality of the stellar mass recovery is improved throughout. The best $g-i$ recovery (Table \ref{tab:abs_gi}) uses Ks or H for $B$, and here we find close to 0.1 dex scatter. Many of the other bands produce similarly good stellar masses, which should be expected since there is no strong $k$ correction effect undermining the conversion of apparent to absolute magnitudes (this effectively limits which bands will produce reasonable observed frame stellar masses). For $g-r$ based stellar masses (Table \ref{tab:abs_gr}, the preference is also to use the Ks band photometry for $B$.

The optimal two band stellar mass recovery (Table \ref{tab:abs_2b}) has changed from the observed frame fits. We now find that Y, J, H and Ks are the preferable $B$ selection, all producing similar to 0.1 dex of scatter. The $z$ band also produces excellent results, as do the $i$ and $r$ bands (available in Tables \ref{tab:abs_gi} and \ref{tab:abs_gr}).

\section{Application To GAMA}

\prospect{} has already been used to infer physical properties for small samples of galaxies (e.g.\ Seymour et al., in press; Tiley et al., submitted). Our first application to a large sample uses data from the GAMA survey. This has recently been reprocessed with \profound{} to produce improved 20 band photometry covering the UV to the FIR (Bellstedt et al, submitted), so it is an ideal data set for \prospect. GAMA also offers an excellent comparison set for our new stellar mass estimates, since we already have high quality estimates from \citet[][version 20]{tayl11} and \magphys{} \citep[][version 6]{dacu08, driv16}.

A complication to the above is that both the \citet{tayl11} and \magphys-derived stellar masses were generated using GAMA's \lambdar-based photometry \citep{wrig16}, and we are moving to \profound{} photometry for our final data release. To properly capture the source of potential differences we ran \prospect{} on a $z<0.06$ sample of GAMA using both \lambdar{} and \profound{} photometry. This means we can assess the degree to which any stellar mass variation is due to the different source photometry versus different stellar mass estimation software.

\subsection{Impact of Using \profound{} and \lambdar{} Photometry with \prospect{}}

The first test was to compare running identical setups of \prospect{} on \profound{} and \lambdar{} processed photometry using all GAMA galaxies with $z<0.06$ (8,712 in total). For this analysis the input pixel data processed by \profound{} and \lambdar{} were nearly identical, but there have been small updates to the input data since \citet{wrig16} which will be discussed in detail in Bellstedt et al. (submitted).

The comparison of these stellar mass estimates is shown in Figure \ref{fig:gama_photo}. We find no significant bias in the stellar mass estimates as a function of stellar mass, and a decrease in scatter from $\sim 0.2$ dex to $\sim 0.1$ dex as we move to brighter photometry and more massive stellar masses. The scatter of 0.1 dex puts a sensible lower limit on the stellar mass error even when assuming \prospect{} is using the correct model. This is reflected in the posterior samples of the stellar mass formed during the CHARM MCMC process, where we find a median stellar mass uncertainty of 0.12 dex.

\begin{figure}
\begin{center}
\includegraphics[width=9cm]{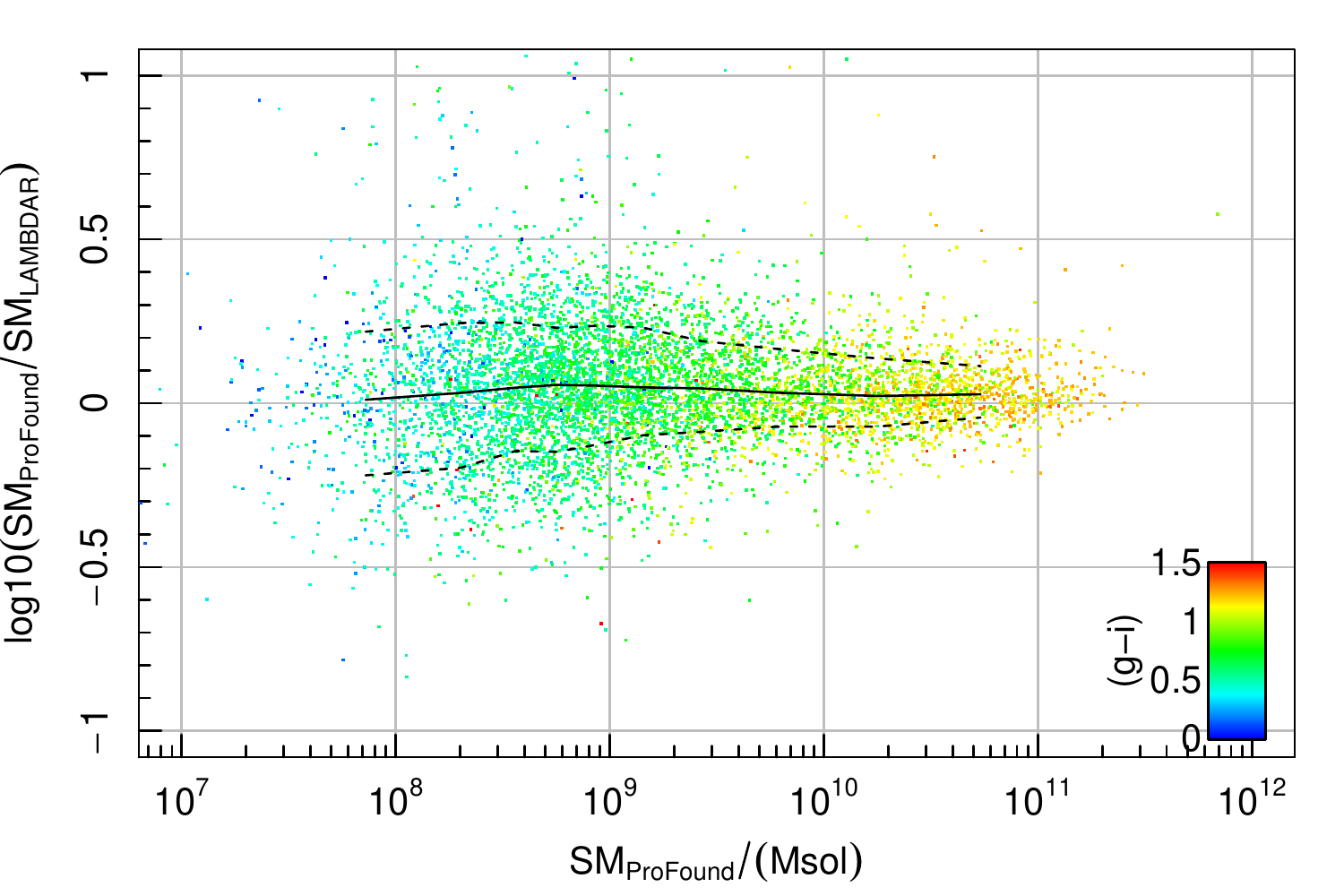}
\caption{In this Figure we compare the impact of running identical code on our different photometric data products: \profound{} and \lambdar. There is no serious bias seen as a function of the stellar mass, and the median scatter is 0.17 dex. The scatter grows as we move to lower stellar mass systems, where the photometry has larger errors and will vary more between \profound{} and \lambdar. There are no strong gradients in $g-i$ colour, beyond more massive galaxies being typically redder.}
\label{fig:gama_photo}
\end{center}
\end{figure}

\subsection{Running Different Software on \lambdar{} Photometry}

The next comparison we made was comparing the stellar mass estimates when we run \prospect{} on the exact same \lambdar{} photometry as previously published work \citep{tayl11, driv16}. This is shown in Figure \ref{fig:gama_code}.

We see a difference between \prospect{} and both previously published efforts, where \prospect{} finds consistently more massive galaxies. All data sets are consistent within the scatter and stated stellar mass errors however. There is the least difference, and smallest scatter, compared to \citet{tayl11}, which is interesting because \prospect{} is conceptually more similar to \magphys{} in regards to how stellar masses are inferred.

\begin{figure}
\begin{center}
\includegraphics[width=9cm]{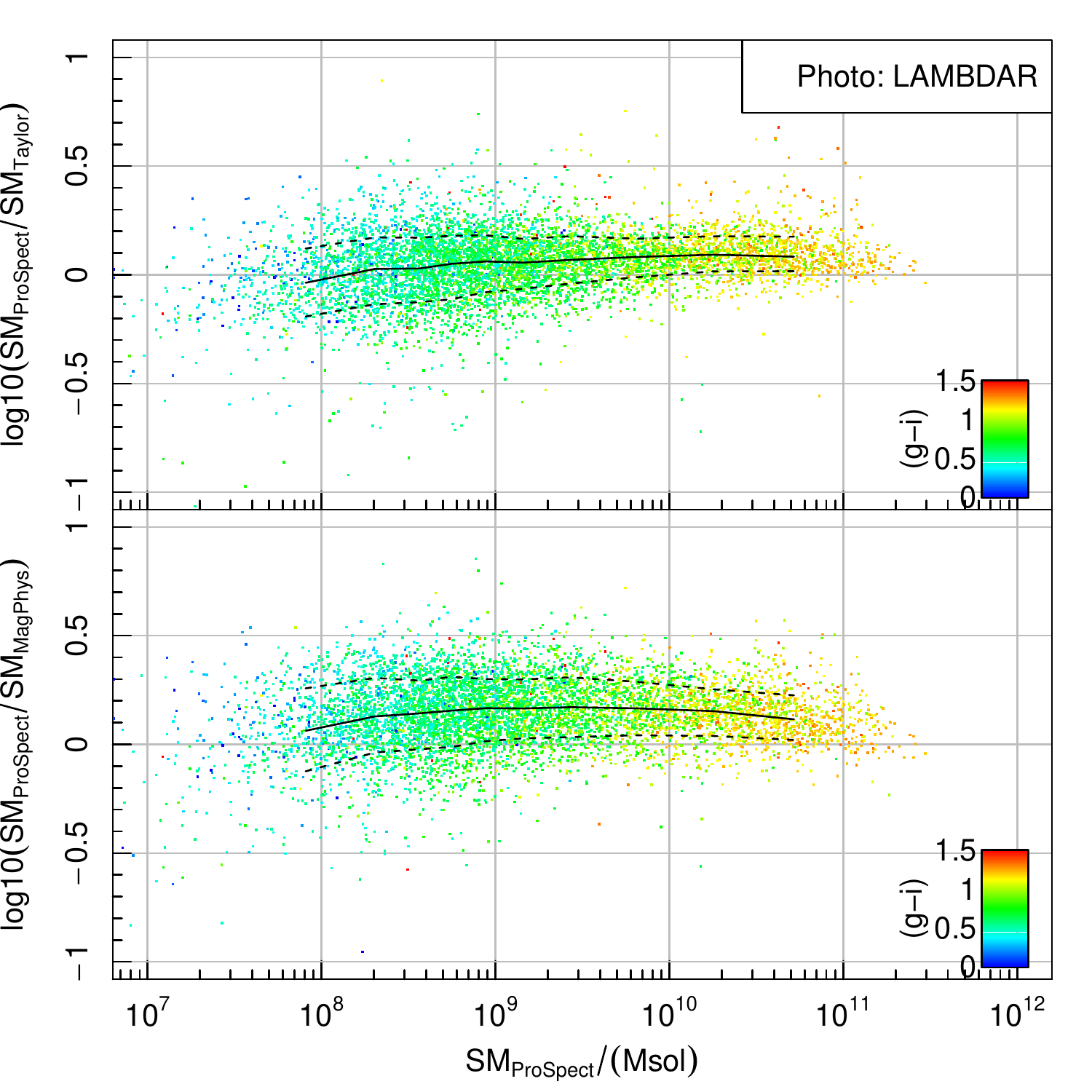}
\caption{In this Figure we compare the impact of running different code on the same photometric data product (\lambdar). There are systematic differences for both comparison sets \citep[\magphys{} and Taylor;][respectively]{dacu08, tayl11}. The median offset to \magphys{} is 0.15 dex with 0.14 dex scatter, and the median offset to Taylor is 0.06 dex with 0.13 dex scatter. This means the different codes are broadly consistent within their expected scatter, but \prospect{} returns systematically more massive galaxies when using the exact same input data. There are no strong gradients in $g-i$ colour, beyond more massive galaxies being typically redder.}
\label{fig:gama_code}
\end{center}
\end{figure}

The obvious explanation for the difference is that \prospect{} is tending to form more older stars to produce the same amount of light, an effect we noticed when switching SFH models with \shark{} in Figure \ref{fig:Shark_CSFH}. Since \prospect{} has more flexibility with its SFH and ZH modelling, we would advocate that \prospect{} is recovering better stellar mass estimates {\it on average}. A full investigation of the implied SFH and ZH recovered in GAMA with our new \prospect{} inversions is left for Bellstedt et al. (in prep.).

\subsection{Final Stellar Mass Comparison}

Since the final stellar masses for GAMA will be based on \prospect{} fits run on \profound{} photometry, it is instructive to make a final comparison of this against the most recent efforts run on \lambdar{}. For this test we were able to use more computing resources to increase the sample size to all GAMA galaxies with $z<0.1$ The results of this comparison is shown in Figure \ref{fig:gama_mix}.

\begin{figure}
\begin{center}
\includegraphics[width=9cm]{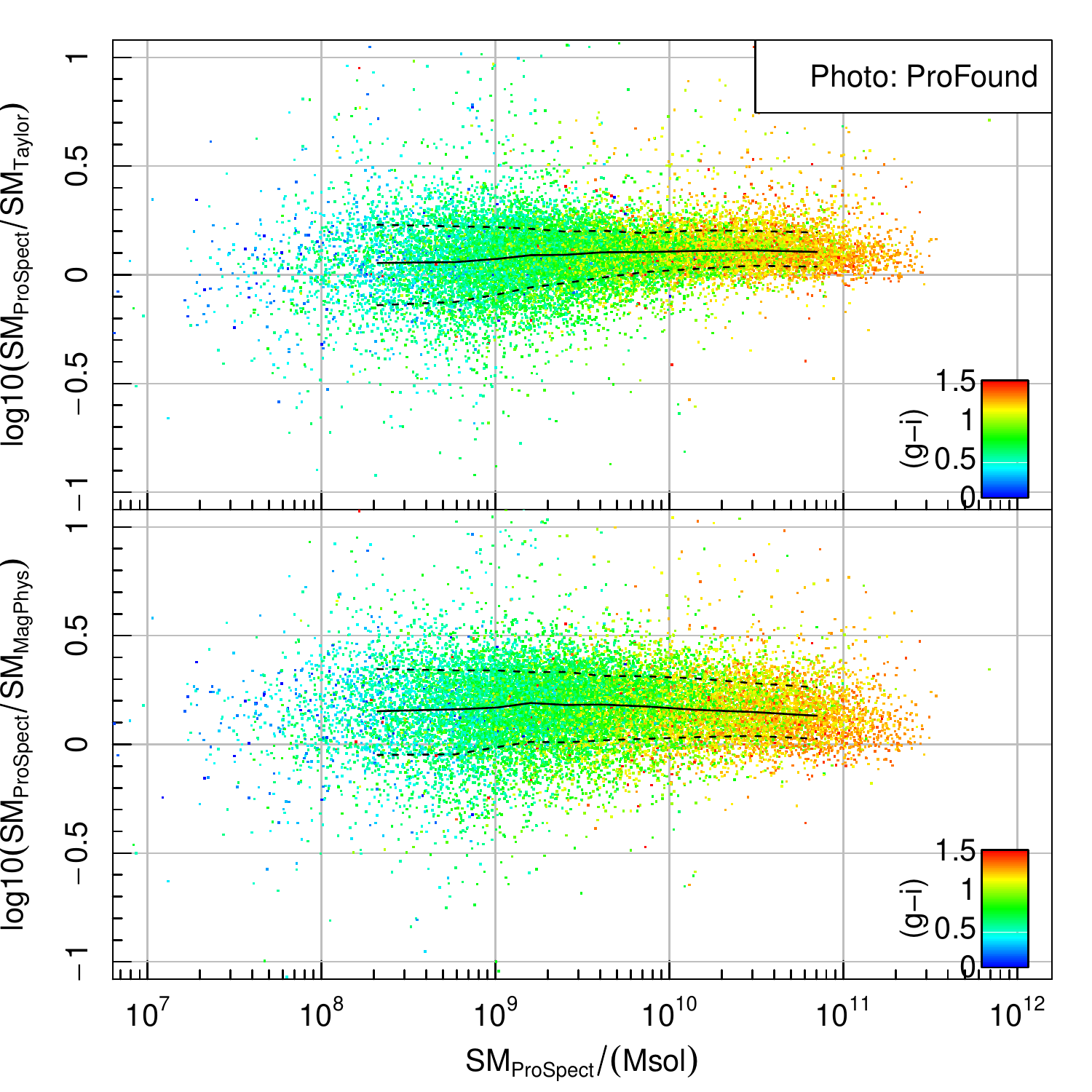}
\caption{In this Figure we compare the impact of running \prospect{} on \profound{} photometry, and \citet{tayl11} and \magphys{} run on \lambdar{} photometry. This is likely to produce the largest differences since we are switching both photometry and stellar mass inversion approaches. The median offset to \magphys{} is 0.17 dex with 0.16 dex scatter, and the median offset to Taylor is 0.1 dex with 0.11 dex scatter. There are no strong gradients in $g-i$ colour, beyond more massive galaxies being typically redder.}
\label{fig:gama_mix}
\end{center}
\end{figure}

The median biases are slightly larger for Figure \ref{fig:gama_mix} than we saw for Figure \ref{fig:gama_code}, and the scatter has also increased for \magphys{} and is almost the same for \citet{tayl11}. The stellar masses returned by \prospect{} are still consistently larger than the two current sets of GAMA stellar masses. Broadly speaking, the stellar masses extracted even when switching both photometry and approach are still in excellent agreement.

\subsection{GAMA Derived Observed Frame Apparent Magnitude Stellar Masses}

Given we now have \prospect-derived stellar masses, it is trivial to also derive essentially entirely observational approximate stellar mass functions in a similar manner to Section \ref{sec:photoSM}. The caveats to this are that we are more limited in redshift coverage (this initial GAMA \prospect{} sample only includes galaxies out to $z=0.1$) and magnitude range ($r<19.8$).

\begin{table}
\centering
\begin{tabular}{llrrrrrl}
  \hline
B & C & $\alpha$ & $\beta$ & $\delta$ & $\sigma$ & $z$ max \\ 
  \hline
  g & (g-i) & -0.388 & 1.332 & 1.196 & 0.000 &  & 0.1 \\ 
  r & (g-i) & -0.382 & 1.073 & 1.310 & 0.000 &  & 0.1 \\ 
  i & (g-i) & -0.368 & 1.037 & 1.513 & 0.000 &  & 0.1 \\ 
  Z & (g-i) & -0.377 & 0.879 & 1.400 & 0.000 &  & 0.1 \\ 
  Y & (g-i) & -0.372 & 0.837 & 1.487 & 0.000 &  & 0.1 \\ 
  J & (g-i) & -0.364 & 0.775 & 1.672 & 0.000 &  & 0.1 \\ 
  H & (g-i) & -0.359 & 0.744 & 1.740 & 0.000 &  & 0.1 \\ 
  Ks & (g-i) & -0.346 & 0.776 & 2.002 & 0.015 &  & 0.1 \\ 
  W1 & (g-i) & -0.316 & 0.985 & 2.676 & 0.055 &  & 0.1 \\ 
  W2 & (g-i) & -0.266 & 1.319 & 3.525 & 0.116 &  & 0.1 \\ 
   \hline
\end{tabular}
\caption{Formula terms for GAMA-derived observed frame apparent magnitude photometry using $g-i$ colours. See Equation \ref{eqn:SMapp}.}
\label{tab:app_gama}
\end{table}

Using a similar methodology to above, we derive new observed frame apparent magnitude based corrections using Equation \ref{eqn:SMapp}. { Because of the low redshift range of the GAMA data used here, we remove the redshift dependent term $\gamma$ from our functional parameterisation presented in Equation \ref{eqn:SMapp} (or, in effect, we set $\gamma = 0$).} Since we are now using observational data, we also pass into \hyperfit{} the fully propagated expected errors for all of the observables (including the inferred uncertainty in the \prospect-derived mass). As such the scatter term ($\sigma$) is effectively the additional stellar mass uncertainty we are adding on top of the \prospect{} uncertainty, which is typically around 0.1 dex for the GAMA sample.

The key parameters from this fitting process are listed in Table \ref{tab:app_gama}. These numbers show small differences in prediction from the earlier \shark-derived relationships depending on the precise stellar mass and redshift of interest. There are likely to be complex reasons for these implied parameter variations. Amongst these are the fact that the GAMA data have observational error that is not present in our \shark{} approximation, the \prospect{} model is in detail imperfect and could be differing systematically from the true SFH and ZH model (biasing these approximations in non-linear ways), and the \shark{} model is also imperfect in representing the true complexity of galaxy formation.

Figure \ref{fig:gama_approx} compares \shark-derived and GAMA-derived approximate photometric predictions for observed frame apparent magnitudes, where in both cases we use the Ks \& $g-i$ fits. The agreement is very good, suggesting that our \shark{} approximations work well in practice. Interestingly the trends seen in $g-i$ colour are in opposite directions, where \shark{} calibrations tend to be too massive for bluer colours and GAMA calibrations tend to be too massive for redder colours. The  $\beta$ term that controls the behaviour with $g-i$ colour is much larger for the GAMA calibration (0.776 versus 0.074 for Ks \& $g-i$ approximate stellar masses). This suggests that in the narrow redshift range of GAMA strongly modifying the mass-to-light (which is mostly what the $g-i$ term controls) is beneficial, but over the broader redshift range used for the \shark{} calibrations it is not appropriate. This seems reasonable, since observed frame $g-i$ probes very different parts of the SED as we move out to higher redshift, with very different consequences for the implied mass-to-light.

\begin{figure}
\begin{center}
\includegraphics[width=9cm]{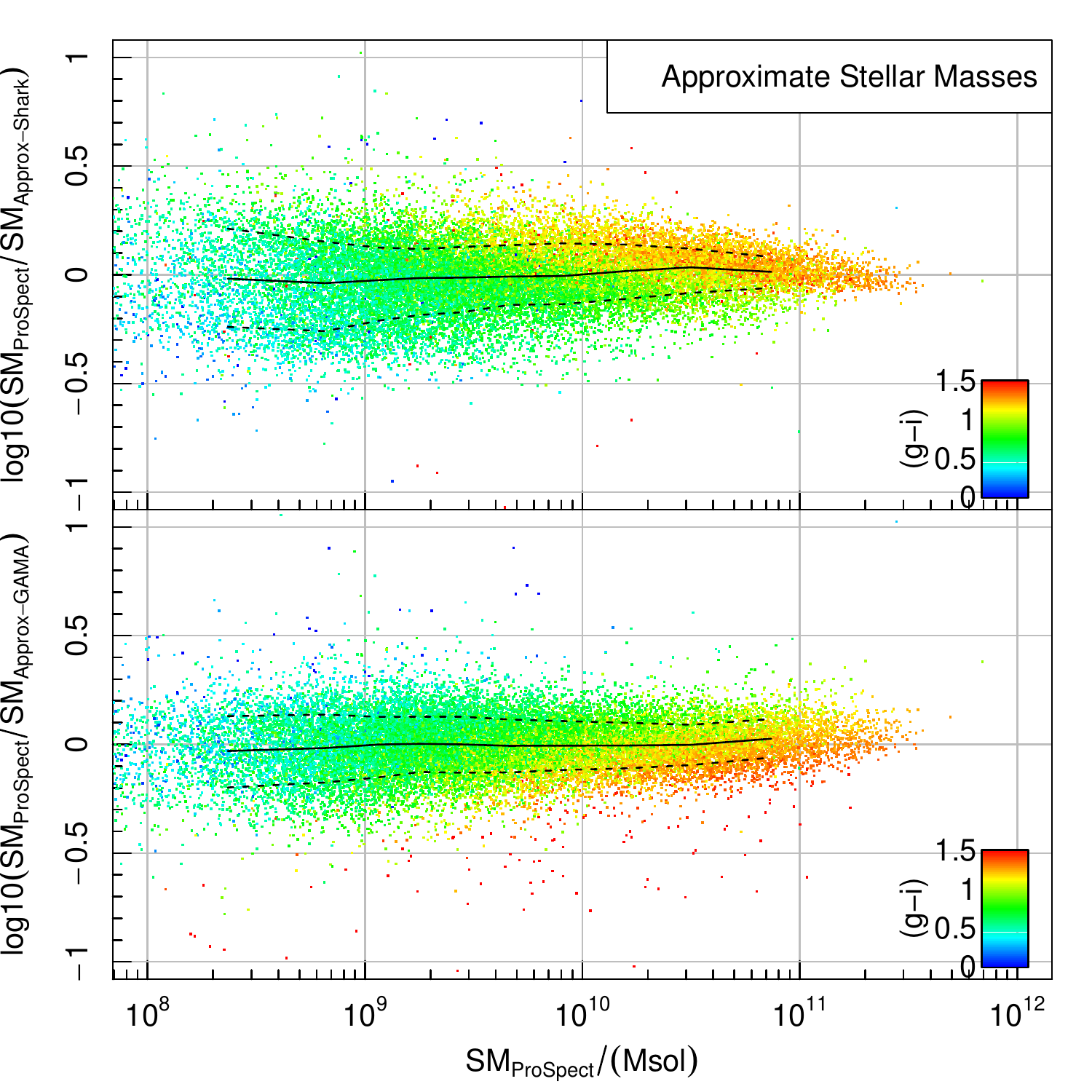}
\caption{In this Figure we compare the \shark-derived (top) and GAMA-derived (bottom) Ks \& $g-i$ approximate stellar masses against their fully fitted GAMA \prospect{} counterparts. In general they agree fairly well in the quality of the prediction throughout the stellar mass range that GAMA is sensitive to. These approximations should be used with caution below $10^8$\msol. There are $g-i$ gradients present for both approximate methods but in opposite direction, where \shark{} calibrations tend to be too massive for bluer colours and GAMA calibrations tend to be too massive for redder colours.}
\label{fig:gama_approx}
\end{center}
\end{figure}

Whether an end user should prefer the combination of a SAM (\shark) and \prospect{} photometry to derive these stellar mass calibrations (with {\it certain} stellar masses), or observational data (GAMA) with \prospect{} used to infer the {\it true} stellar masses is open for debate. Over large redshift ranges ($z>0.5$) and for faint sources ($r>19.8$) the \shark-derived calibrations should certainly be preferred. But at lower redshifts ($z<0.1$) and for brighter sources ($r<19.8$) the GAMA-derived calibration is likely to be preferable. Using both methods will capture some degree of the imperfect nature of the process of deriving stellar masses. Indeed, using both calibration routes with all available photometric approximations should capture the uncertainty of a given stellar mass estimate for a single source. If the known flux errors are also folded in using a Monte Carlo method, then plausible uncertainties should be obtainable.

\section{Conclusions}

In this paper we have presented \prospect{}, a new spectral energy distribution generation and inversion code. In brief it has the following characteristics:

\begin{itemize}

\item contains the full BC03 \citep{bc03} and EMILES \citep{emiles} spectral libraries, organised in a consistent and memory efficient manner;

\item produces intrinsic SEDs for a range of built-in star formation histories and metallicity histories. It is also straight-forward to add user-defined functions as long as certain requirements on the functional inputs and outputs are met;

\item has an energy balance model for dust, using a \citet{cf00} model for the attenuation and the \citet{dale14} template set for re-emission in the far infrared;

\item can produce a simple AGN model that incorporates dust self attenuation due to the presence of a local hot torus; 

\item can produce internally consistent (via energy balance) emission features that vary with metallicity.

\end{itemize}

When working in a purely generative mode (the original design goal of the project), \prospect{} is well suited to processing the outputs of simulations. It has already been applied to the semi-analytic galaxy modelling software \shark{} \citep{lago19}, and has also been applied to the generation of photometry from hydrodynamical simulations (Harbourne et al., in press). When applying \prospect{} to \shark{} we used a simple treatment for dust in galaxies by sampling from the CF00 parameter distributions recovered in \citet{tray20}. The result of this work was the generation of high quality luminosity functions spanning the ultra-violet through to the far-infrared, suggesting that in broad terms \shark{} and \prospect{} are creating galaxy star formation and metallicity histories that are compatible with the real Universe \citep[see][for details]{lago19}.

\prospect{} can also be used as a Bayesian generative modelling tool, allowing for the inversion of all dust and star formation history related parameters (usually around a dozen). Any sampler available to the \R{} eco-system can be used to do this inversion (potentially hundreds), with this work making extensive use of the {\tt CMA} genetic algorithm package and the {\tt LaplacesDemon} optimisation and MCMC package (specifically, the CHARM MCMC algorithm).

Given the reasonable re-creation of galaxy luminosity functions, it was instructive to use \prospect{} in an inversion parameter inference mode to test how well the complex star formation and metallicity history of \shark{} galaxies can be recovered. In general, whilst the fine discontinuous detail of a given star formation history could not be perfectly recovered, the temporally smoothed trends and shapes can be well recovered. The main limitation to this process will likely be the quality of the sampler chosen (where we used CHARM) and the number of effective samples made.

An important result of this fitting work was noting that the {\tt snorm\_trunc} and closed-box metallicity evolution do a reasonable job of extracting the SFH and ZH in an average sense. Whilst they are not perfect, they broadly capture the wide range of galaxy formation seen in \shark{}, which is a good indication that they might also be informative at providing parameter inference for real observations (see Bellstedt et al., in prep.). In particular, we note that the metallicity evolution is much nearer to closed-box than the simple fixed, constant or linearly evolving models that have often been used in previous literature work. We also find that using a simple exponentially declining SFH is highly biased compared to the more physically plausible {\tt snorm\_trunc} model used here. We strongly advocate that any similar SED inversion codes should also encode such closed-box metallicity histories in order to better overcome potentially serious biases produced in the inferred SFHs due to highly erroneous ZHs.

{ Finally, we present the first \prospect{} fits to the final set of GAMA photometry. These produce small systematic differences in stellar masses (on average larger) compared to previous GAMA results using \citet{tayl11} and \magphys. This appears to be due to the new more flexible and physically motivated form of the star formation history being used, allowing star formation to occur at systematically more ancient epochs. The corresponding increase in $M/L$ for older stellar populations naturally gives rise to more massive galaxies since more mass is required to explain the observed amount of galaxy flux.}

{ Given the tests conducted with \shark{} galaxies, we suggest that the new stellar masses and related parameters are more robust using \prospect{} (although the stellar masses are comparable within errors to previous efforts), and these will form one of the core outputs of the final GAMA data release (Robotham et al., in prep.). A key outcome for the future exploitation of \prospect{} is that we will be able to unpick plausible star formation and metallicity histories on a per galaxy basis, which is a significant advance on current software where these properties are not modelled in a physically motivated manner. The star formation histories extracted from GAMA galaxies using \prospect{} will be discussed in extensive detail in upcoming work (Bellstedt et al., in prep.).}

\section*{Acknowledgements}

{ All authors have been supported by the Government of Western Australia funding of ICRAR. SB and SPD acknowledge support by the Australian Research Council's funding scheme DP180103740. CL has received funding from the ARC Centre of Excellence for All Sky Astrophysics in 3 Dimensions (ASTRO 3D), through project number CE170100013. JET and MB are supported by the Australian Government Research Training Program (RTP) Scholarship. This work was supported by resources provided by the Pawsey Supercomputing Centre with funding from the Australian Government and the Government of Western Australia.

GAMA is a joint European-Australasian project based around a spectroscopic campaign using the Anglo-Australian Telescope. The GAMA input catalogue is based on data taken from the Sloan Digital Sky Survey and the UKIRT Infrared Deep Sky Survey. Complementary imaging of the GAMA regions is being obtained by a number of independent survey programmes including GALEX MIS, VST KiDS, VISTA VIKING, WISE, Herschel-ATLAS, GMRT and ASKAP providing UV to radio coverage. GAMA is funded by the STFC (UK), the ARC (Australia), the AAO, and the participating institutions. The GAMA website is http://www.gama-survey.org/ .

Based on observations made with ESO Telescopes at the La Silla Paranal Observatory under programme ID 179.A-2004. Based on observations made with ESO Telescopes at the La Silla Paranal Observatory under programme ID 177.A-3016.

Much of the work presented here was made possible by the free and open \R{} software environment \citep{rcor16}. All figures in this paper were made using the \R{} {\sc magicaxis} package \citep{robo16}.

We would like to thank the referee Edward N. Taylor for his helpful, detailed and expansive comments regarding this work. Many changes were made to improve the clarity of the work, and the context of the results, due to this feedback.}




%
%


\appendix

\section{Simple Example}
\label{sec:simple_example}

{ Here we will briefly show a simple example of running \prospect{} in its generative mode using the settings for a quenched SFH presented in Table \ref{tab:params} with a fixed solar metallicity. The results of running this code are shown in Figures \ref{fig:simple_sfh} and \ref{fig:simple_flux}.}

\begin{verbatim}
library(ProSpect)

GAMA_filters =
  c('FUV_GALEX', 'NUV_GALEX', 'u_SDSS', 'g_SDSS',
  'r_SDSS', 'i_SDSS',  'Z_VISTA', 'Y_VISTA', 'J_VISTA',
  'H_VISTA', 'K_VISTA', 'W1_WISE' , 'W2_WISE',
  'W3_WISE', 'W4_WISE', 'P100_Herschel',
  'P160_Herschel', 'S250_Herschel' , 'S350_Herschel',
  'S500_Herschel')

piv_GAMA = 
  pivwave[pivwave$filter %in% GAMA_filters,"pivwave"]

quench = ProSpectSED(
  massfunc = massfunc_p4, m1=-20, m2=1, m3=1, m4=1,
  agemax = 1e10,
  forcemass = 1e10,
  filters = GAMA_filters
)

magplot(quench$Stars$agevec, quench$Stars$SFR,
  xlim=c(0,1e10), type='l', grid=TRUE, xlab='Age (Yrs)',
  ylab='SFR (Msol / Yr)')

magplot(quench$FinalFlux, log='xy', type='l', grid=TRUE,
  xlim=c(1e3,1e7), ylim=c(1e-8,1e-2),
  xlab='Wavelength (Ang)', ylab='Flux Density (Jky)')
points(piv_GAMA, quench$Photom, col='red')
\end{verbatim}

\begin{figure}
\begin{center}
\includegraphics[width=9cm]{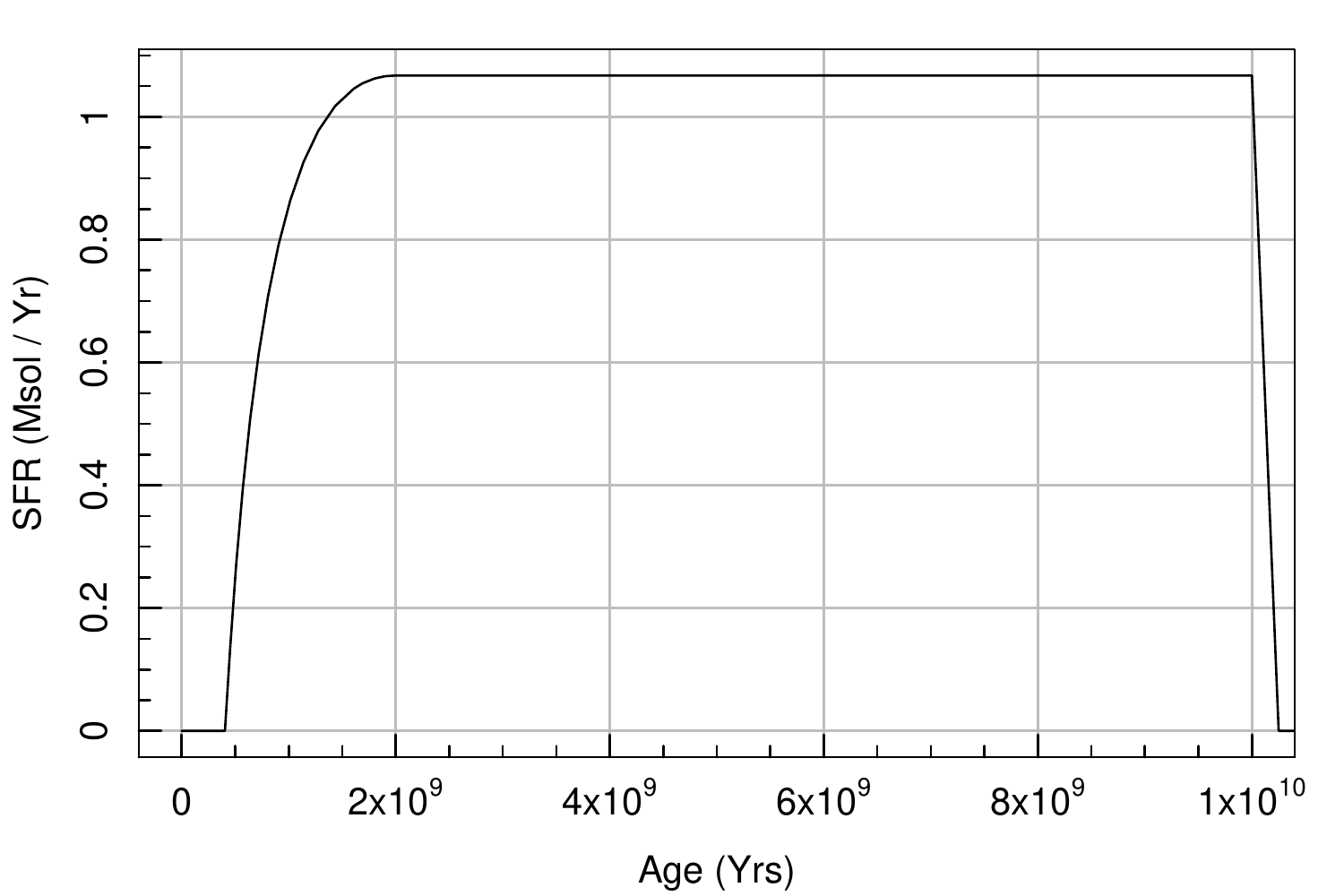}
\caption{Result of the SFH generated by running the simple example code shown in Section \ref{sec:simple_example}.}
\label{fig:simple_sfh}
\end{center}
\end{figure}

\begin{figure}
\begin{center}
\includegraphics[width=9cm]{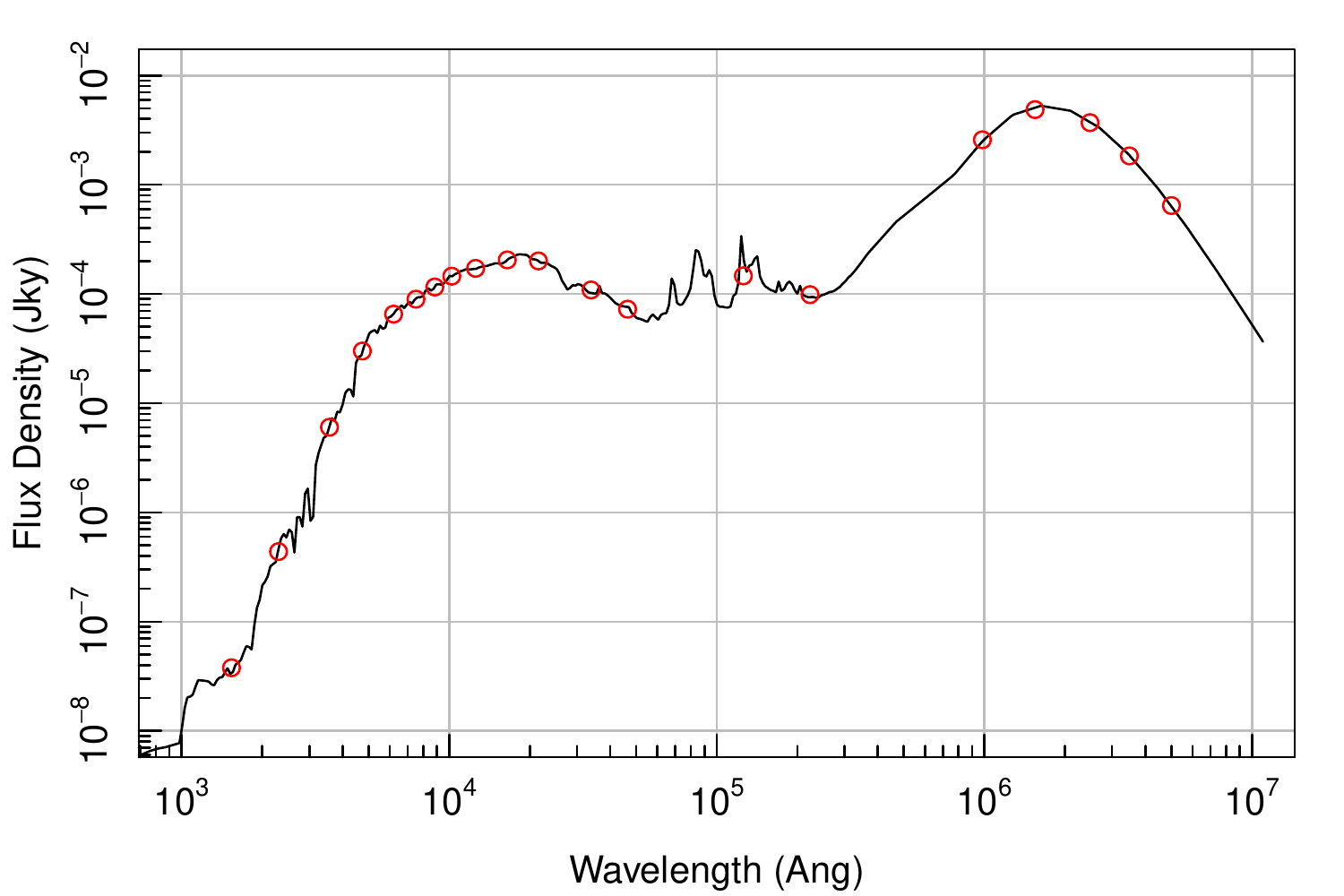}
\caption{Result of the flux density generated by running the simple example code shown in Section \ref{sec:simple_example}. The black line shows the full resolution spectrum, and the red circles show the GAMA filter responses in Jansky (which are by definition the weighted mean throughput at the pivot wavelength).}
\label{fig:simple_flux}
\end{center}
\end{figure}

\section{Nebular Lines}
\label{sec:app_lkl10}

The LKL10 nebular emission lines available in \prospect{} are presented in Table \ref{tab:lkl10}.

\begin{table}
\centering
\begin{tabular}{llr}
  \hline
Element & State & Wavelength / \rm{\AA} \\ 
  \hline
H & I & 1215.670 \\ 
  C & II & 2325.160 \\ 
  Mg & II & 2797.870 \\ 
  O & II & 3726.030 \\ 
  O & II & 3728.730 \\ 
  H & I & 3750.150 \\ 
  H & I & 3770.630 \\ 
  H & I & 3797.900 \\ 
  H & I & 3835.380 \\ 
  He & I & 3888.600 \\ 
  H & I & 3889.050 \\ 
  H & I & 3970.070 \\ 
  H & I & 4104.730 \\ 
  H & I & 4340.460 \\ 
  H & I & 4861.320 \\ 
  O & III & 4958.830 \\ 
  O & III & 5006.770 \\ 
  He & I & 5875.600 \\ 
  N & II & 6547.960 \\ 
  H & I & 6562.800 \\ 
  N & II & 6583.340 \\ 
  S & II & 6716.310 \\ 
  S & II & 6730.680 \\ 
  Ar & III & 7135.670 \\ 
  S & III & 9069.290 \\ 
  S & III & 9532.030 \\ 
  H & I & 9545.980 \\ 
  H & I & 10049.400 \\ 
  He & I & 10830.000 \\ 
  He & I & 10833.000 \\ 
  H & I & 10938.100 \\ 
  H & I & 12818.100 \\ 
  H & I & 18751.000 \\ 
  H & I & 21655.000 \\ 
  H & I & 26252.000 \\ 
  H & I & 40512.000 \\ 
  Ar & II & 69832.800 \\ 
  Ar & III & 89892.500 \\ 
  Ne & II & 128115.000 \\ 
  S & III & 186821.000 \\ 
  S & III & 336366.000 \\ 
  Si & II & 347941.000 \\ 
  O & III & 517972.000 \\ 
  O & I & 631000.000 \\ 
  O & III & 883017.000 \\ 
  N & II & 1218347.000 \\ 
  C & II & 1576366.000 \\ 
   \hline
\end{tabular}
\caption{LKL10 nebular features included in \prospect{}. Not state I means neutral, and II singly ionised etc.}
\label{tab:lkl10}
\end{table}

\section{Additional Shark Fits}
\label{sec:app_shark}

Additional example \prospect{} fits to \shark{} models are shown in Figure \ref{fig:Shark_fits_extra}.

\begin{figure*}
\begin{center}
\includegraphics[width=5.5cm]{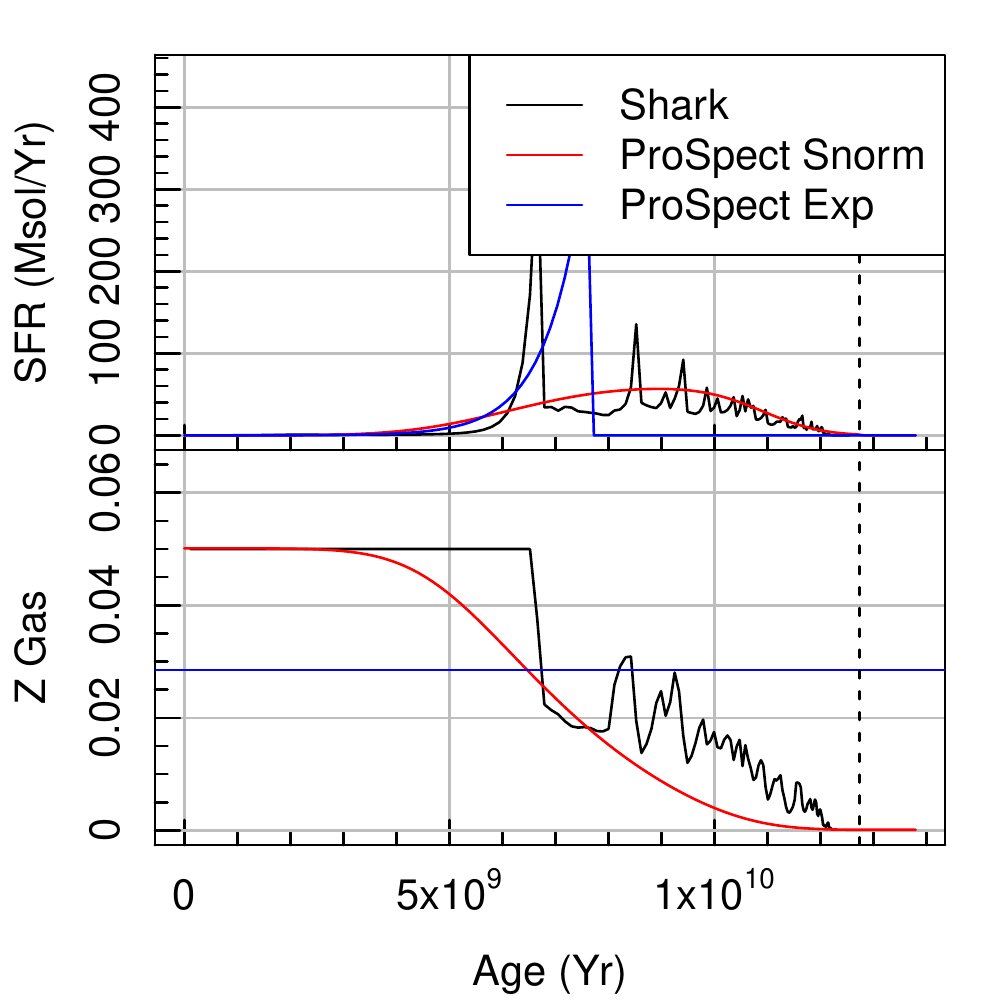}
\includegraphics[width=5.5cm]{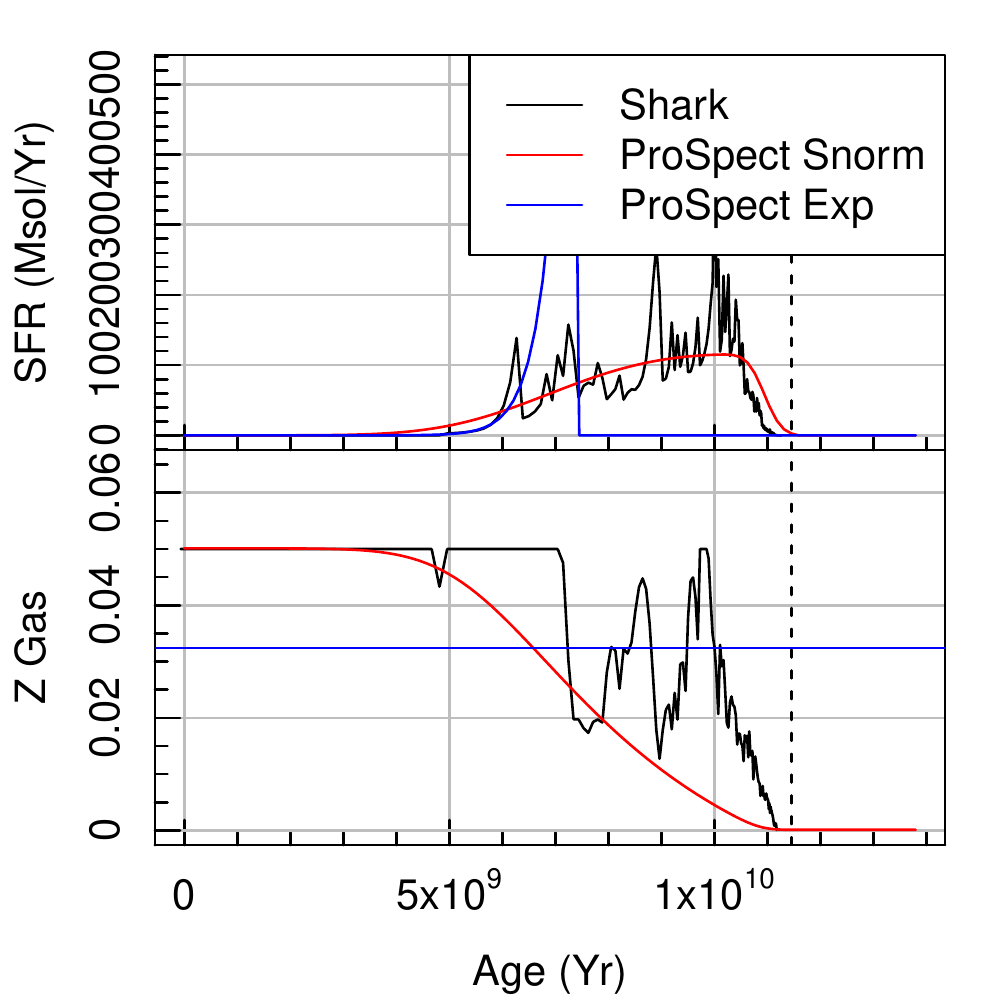}
\includegraphics[width=5.5cm]{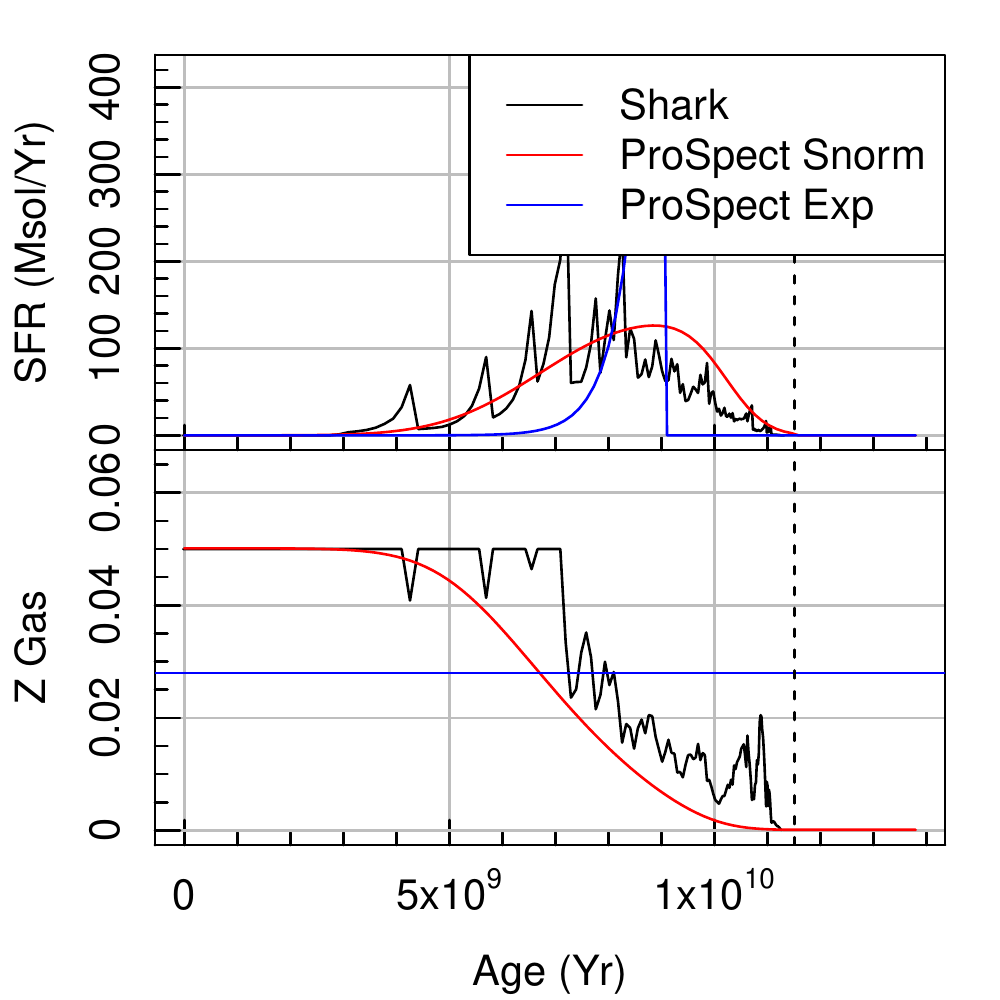}
\\
\includegraphics[width=5.5cm]{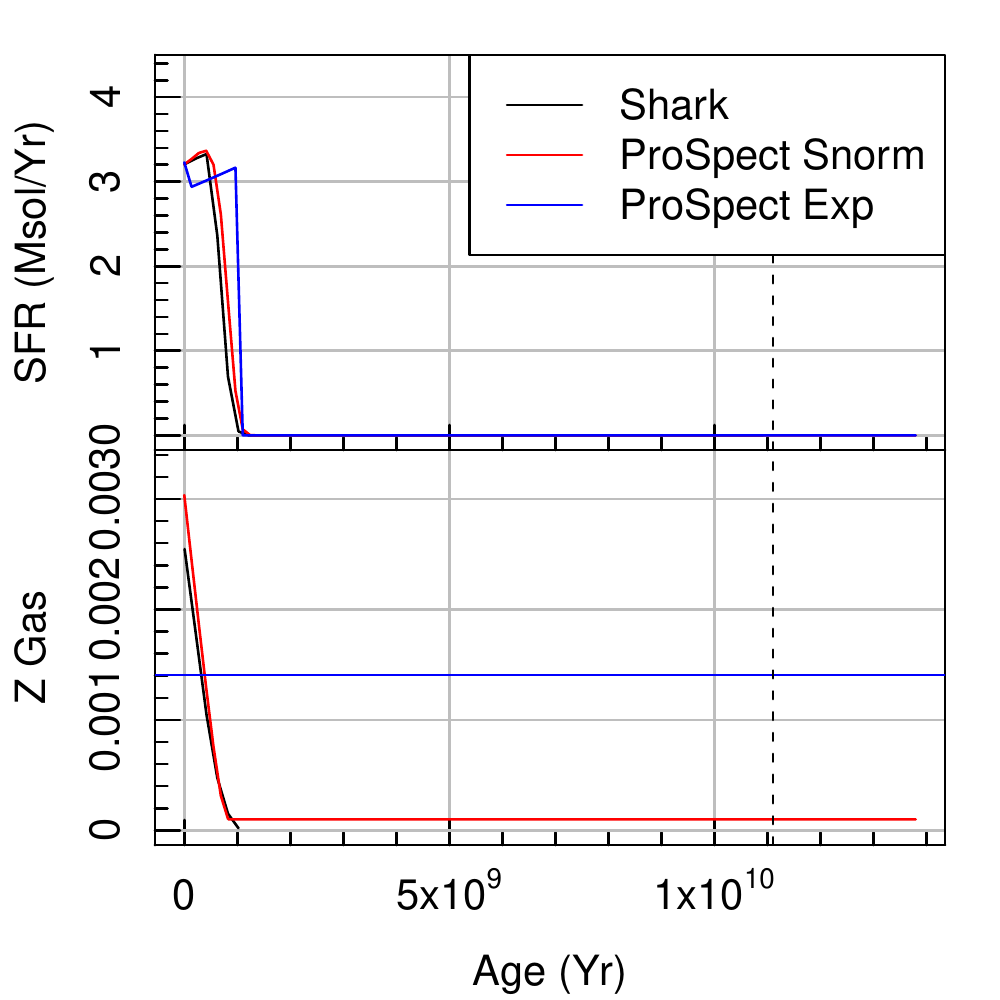}
\includegraphics[width=5.5cm]{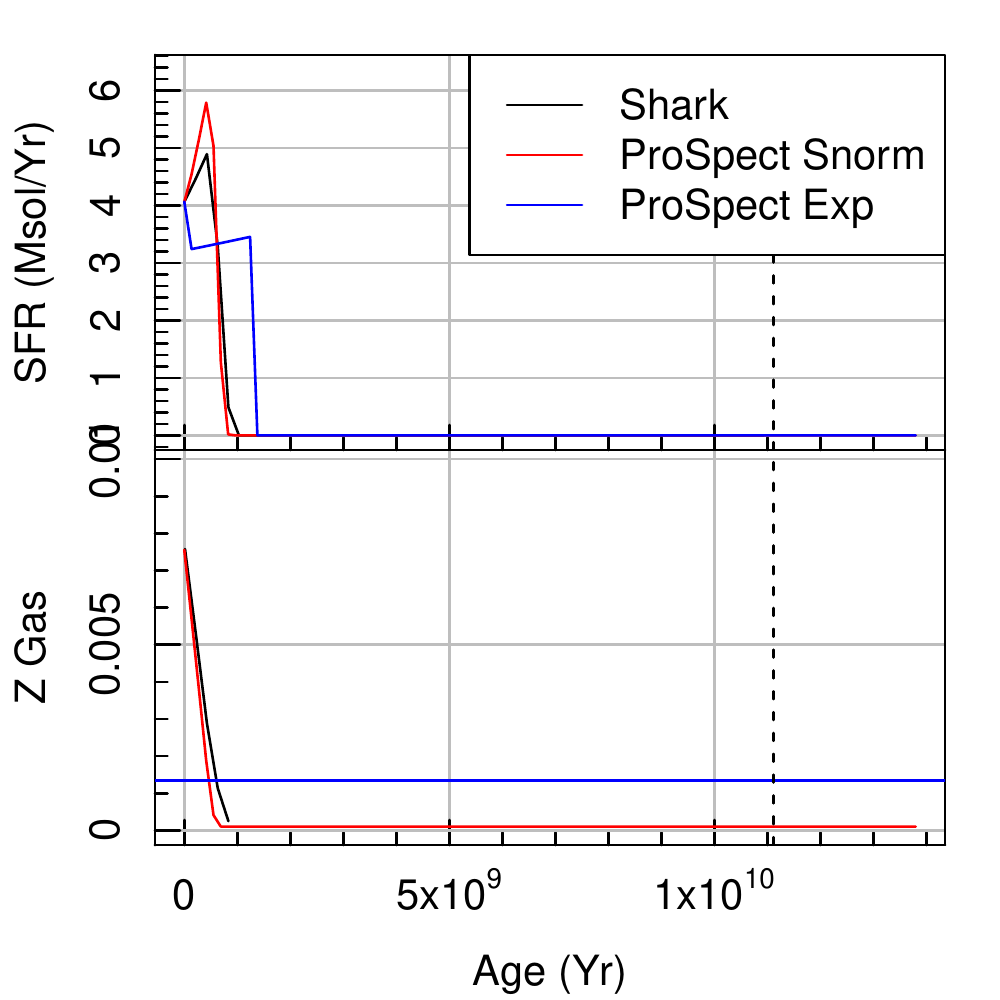}
\includegraphics[width=5.5cm]{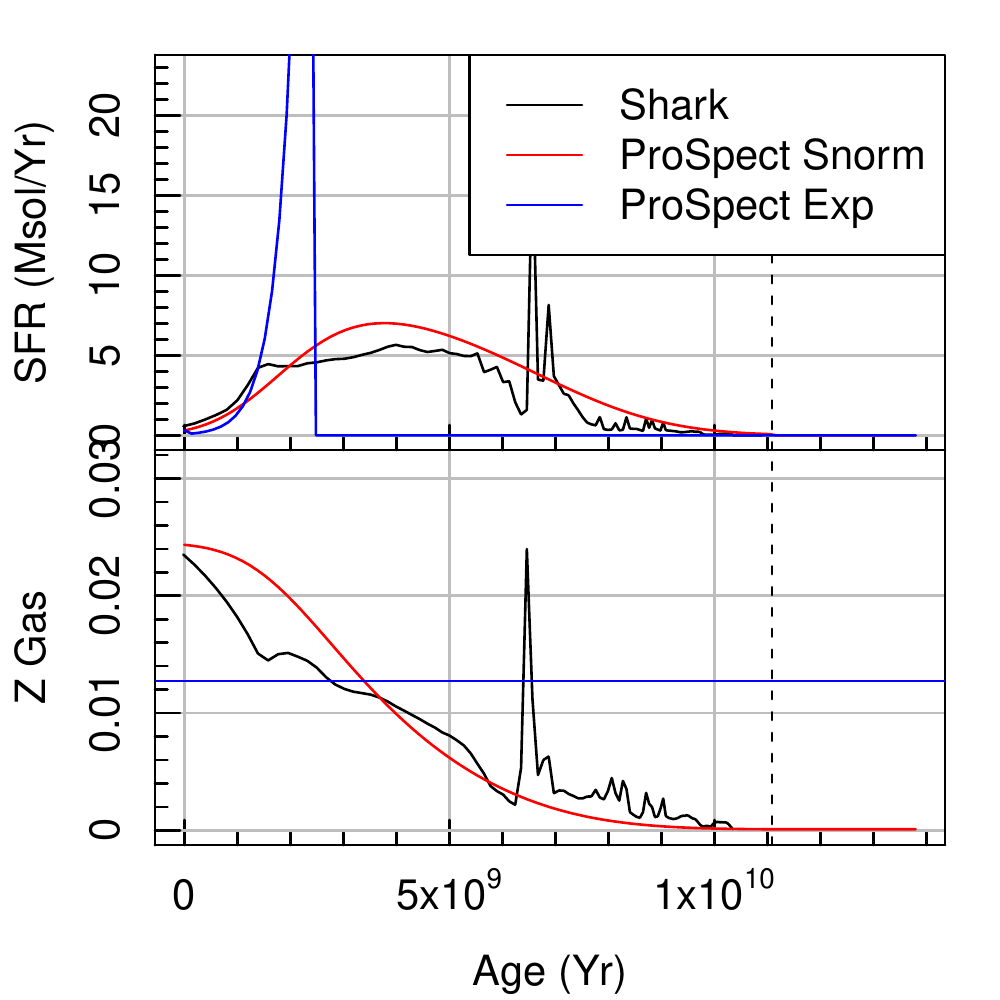}
\\
\includegraphics[width=5.5cm]{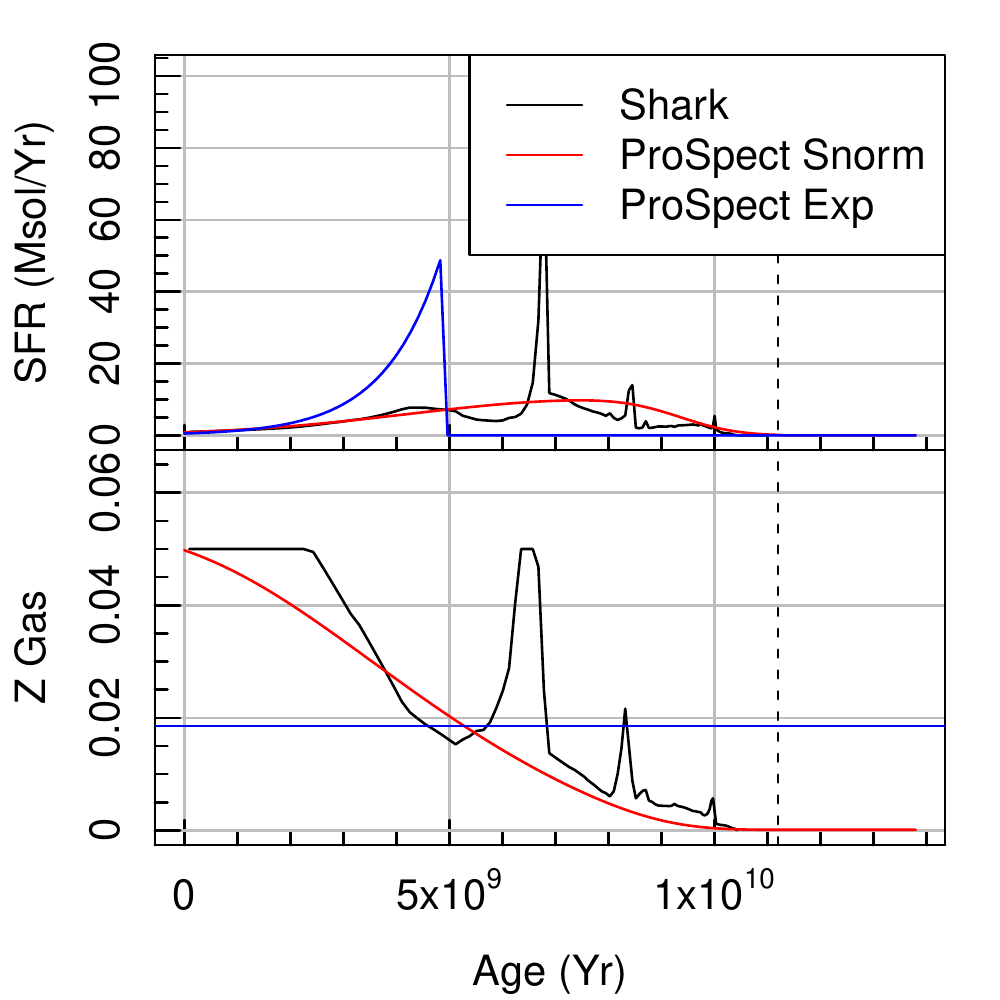}
\includegraphics[width=5.5cm]{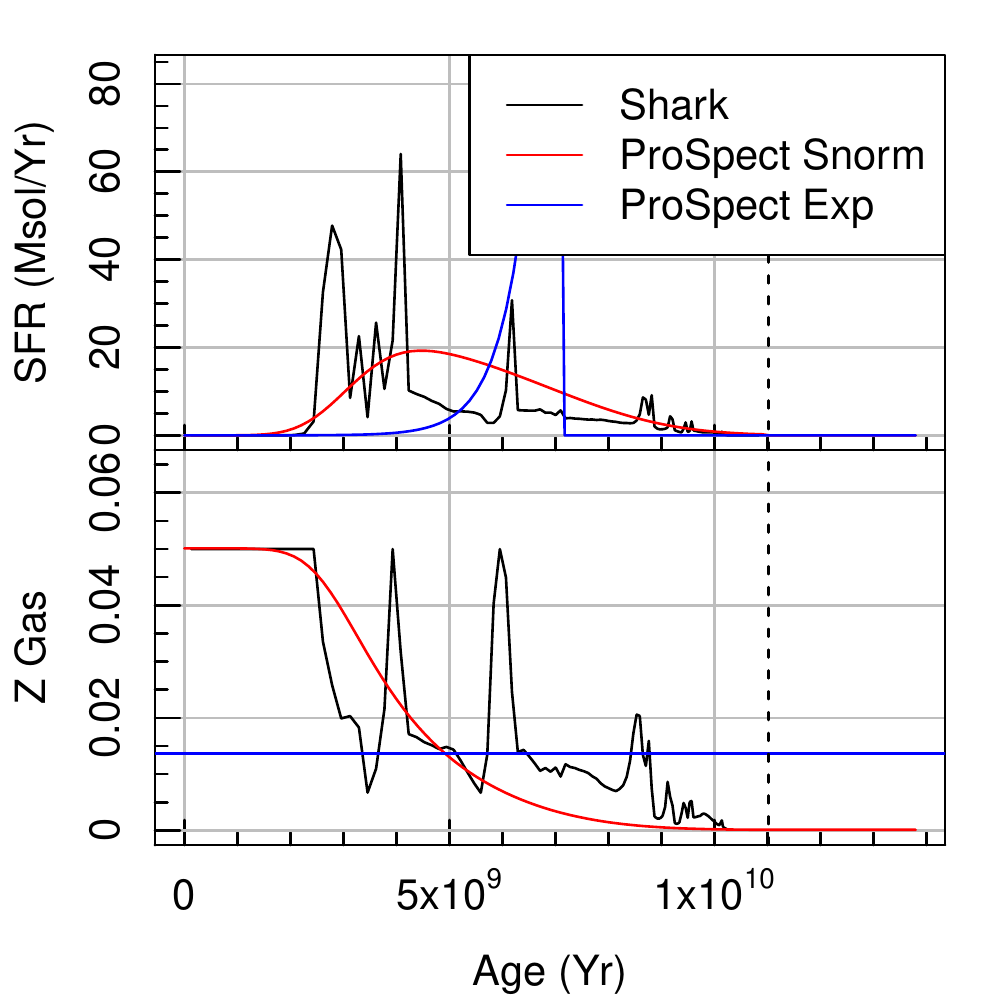}
\includegraphics[width=5.5cm]{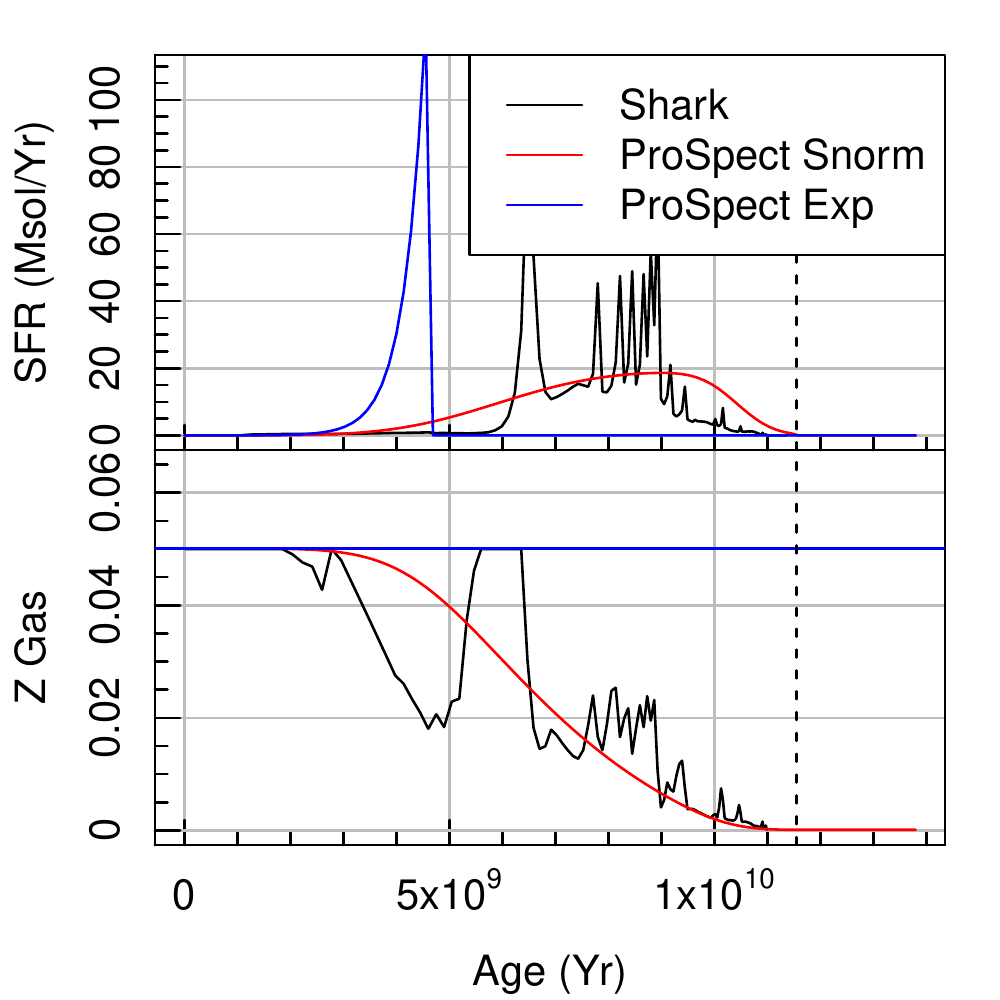}
\\
\includegraphics[width=5.5cm]{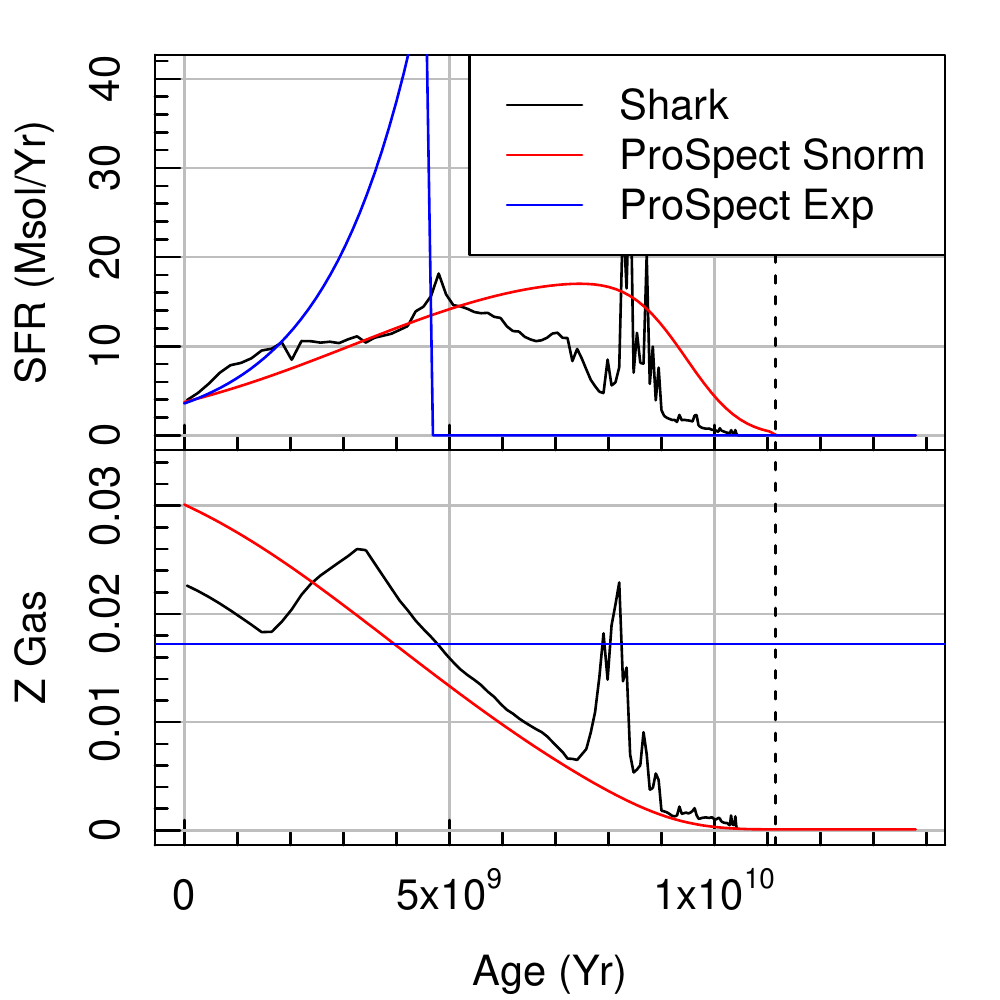}
\includegraphics[width=5.5cm]{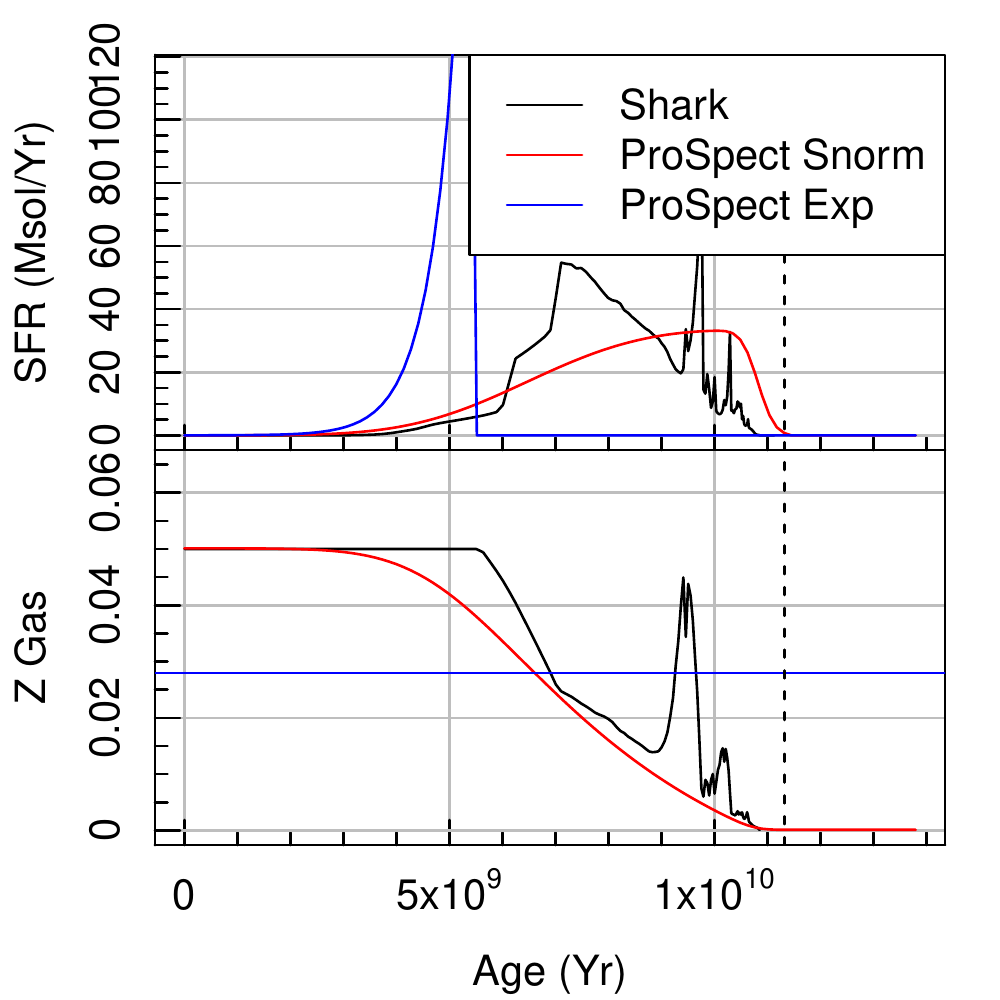}
\includegraphics[width=5.5cm]{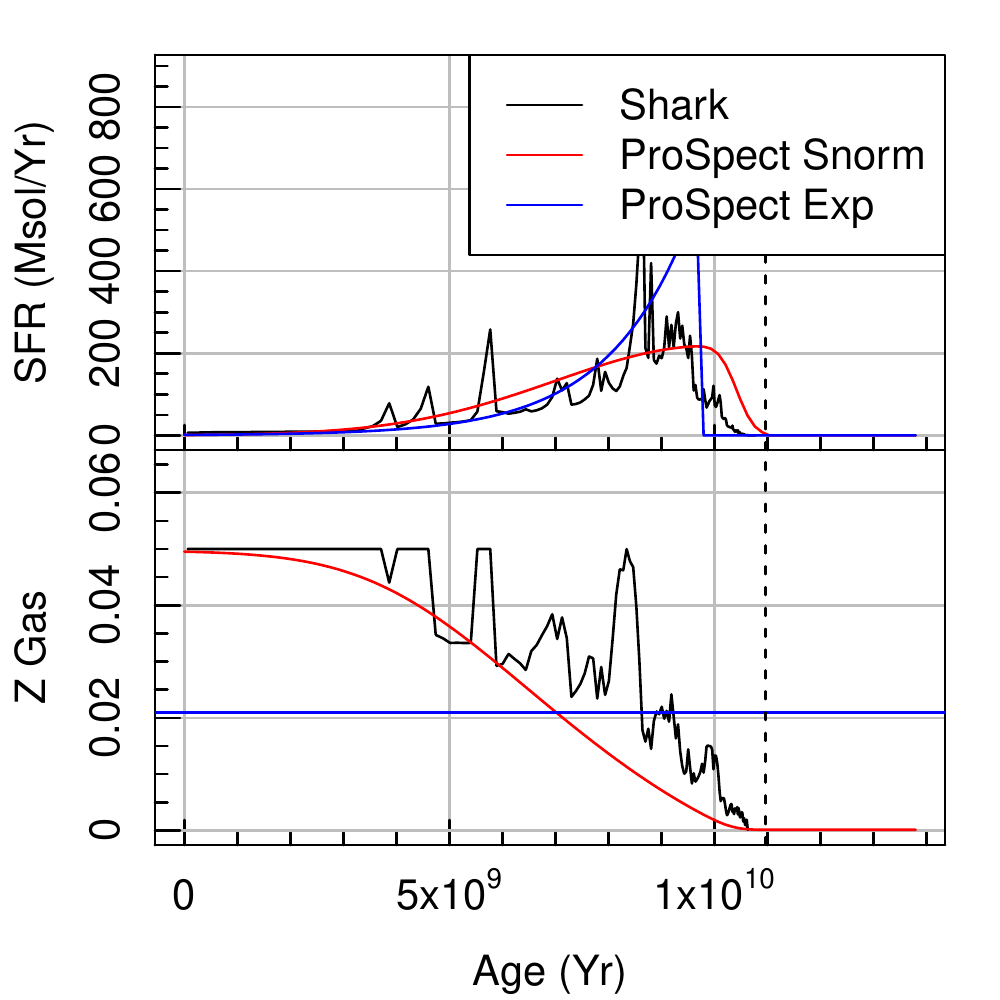}
\\
\caption{Additional example Shark SFHs and ZHs (black lines), with the inferred \prospect expectations over-plotted (red lines).}
\label{fig:Shark_fits_extra}
\end{center}
\end{figure*}

\bsp	
\label{lastpage}
\end{document}